\pdfoutput=1
\documentclass{article}     

\usepackage{float}
\usepackage{amsmath, amsthm,url,amsfonts,amssymb}
\usepackage{dsfont}
\usepackage{epsfig}

\usepackage[english]{babel}

\usepackage[left=1.75cm, top=2.7cm, bottom=2.7cm,right=1.75cm]{geometry}

\usepackage{latexsym,amssymb,amsfonts,graphicx}
\usepackage{amsmath,dsfont}
\usepackage{verbatim}
\usepackage{mathrsfs}
\usepackage{bm}
\usepackage{color}

\usepackage{url}

\usepackage{bm}
\usepackage{natbib}

\newcommand{\Px}{ \mathbb{P} }

\newcommand{\Ex}{ \mathbb{E} }
\newcommand{\gt}{\mathcal{G}_t}

\newcommand{\idc}{\mathbf{1}}

\newcommand{\Gx}{\mathbb{G}}
\newcommand{\Fx}{\mathbb{F} }

\newcommand{\F}{\mathcal{F}}
\newcommand{\G}{\mathcal{G}}
\newcommand{\R}{\mathds{R}}
\newcommand{\calY}{\mathcal{Y}}
\newcommand{\calU}{\mathcal{U}}

\newcommand{\pa}{\partial}
\newcommand{\wpost}{\underline{w}}

\newcommand{\wpre}{\bar{w}}

\newcommand{\pr}{'}

\newcommand{\loga}{\vartheta}

\def \a {{\alpha}}

\def \l {{\lambda}}

\def \R {{\mathbb {R}}}
\def \N {{\mathbb {N}}}

\def \phi {{\varphi}}

\def \tilde {\widetilde}

\def\p{\partial}

\def \rn {{\mathbb {R}}^{N-1}}

\newtheorem{theorem}{Theorem}[section]
\newtheorem{definition}{Definition}[section]

\newtheorem{proposition}[theorem]{Proposition}
\newtheorem{remark}[theorem]{Remark}
\newtheorem{lemma}[theorem]{Lemma}

\definecolor{Brown}{rgb}{0.00, 0.00, 0.00}
\newcommand{\DBrown}{\color{Brown}}

\definecolor{Br}{rgb}{0, 0, 0}

\definecolor{Green}{rgb}{0.00, 0.00, 0.00}
\newcommand{\BGreen}{\color{Green}}
\definecolor{Orange}{rgb}{0.00, 0.00, 0.00}

 \newcommand{\DMagenta}{\color{Magenta}}
\definecolor{Magenta}{rgb}{0,0,0}

 \definecolor{BRed}{rgb}{0.00, 0.00, 0.00}
    \newcommand{\BRed}{\color{BRed}}

 \definecolor{CRed}{rgb}{0.00, 0.00, 0.00}
 \newcommand{\CRed}{\color{CRed}}

 \definecolor{Red}{rgb}{0.00, 0.00, 0.00}
    \newcommand{\Red}{\color{Red}}
    \definecolor{DRed}{rgb}{0.00, 0.00, 0.00}
    
    \definecolor{Blue}{rgb}{0,0,0}
    \newcommand{\Blue}{\color{Blue}}
        \definecolor{Green}{rgb}{0.00, 0.00, 0.00}
    
\definecolor{Orange}{rgb}{0.00, 0.00, 0.00}

\definecolor{DarkBlue}{rgb}{0,0.0,0.0}

\definecolor{CadBlue}{rgb}{0.00, 0.00, 0.00}
\newcommand{\CadBlue}{\color{CadBlue}}

\definecolor{CaddBlue}{rgb}{0.00, 0.00, 0.00} 
\newcommand{\CaddBlue}{\color{CaddBlue}}

    \definecolor{PaleGrey}{rgb}{0.00, 0.00, 0.00}
    \newcommand{\PaleGrey}{\color{PaleGrey}}

    \definecolor{PPaleGrey}{rgb}{0.00, 0.00, 0.00}
    \newcommand{\PPaleGrey}{\color{PPaleGrey}}

      \definecolor{Violet}{rgb}{0.00, 0.00, 0.00}
    \newcommand{\DViolet}{\color{Violet}}

\author{Agostino Capponi\thanks{{Department of Applied Mathematics and Statistics, Johns Hopkins University, Baltimore, MD, 21218, USA ({\tt acappon1@jhu.edu}). 
}}
 \and Jos\'e E. Figueroa-L\'{o}pez\thanks{Department of Statistics, Purdue University, West Lafayette, IN, 47907,  USA ({\tt figueroa@purdue.edu}). {Research supported in part by the NSF Grant DMS-1149692.}}
 \and Andrea Pascucci\thanks{{Department of Mathematics, University of Bologna, Bologna, 40126, Italy ({\tt andrea.pascucci@unibo.it}).}}}

\date{}

\begin{document}

\title{Dynamic Credit Investment in Partially Observed Markets}

\maketitle

\begin{abstract}
We consider the problem of maximizing expected utility for a power investor who can allocate his wealth in a stock, {a} defaultable security, and {a} money market account. The dynamics of these security prices are governed by geometric Brownian motions modulated by a hidden continuous time finite state Markov chain. We reduce the partially observed stochastic control problem to a complete observation risk sensitive control problem via the filtered regime switching probabilities. We separate the latter into pre-default and post-default dynamic optimization subproblems, and obtain two coupled Hamilton-Jacobi-Bellman (HJB) partial differential equations. We prove existence and uniqueness of {a globally} bounded {classical} solution to each HJB equation, and give the corresponding verification theorem. We provide a numerical analysis 
showing that the investor increases his holdings in stock as the filter probability of being in high growth regimes increases, and decreases
his credit risk exposure when the filter probability of being in high default risk regimes gets larger. 
\end{abstract}

\vspace{0.3 cm}


\vspace{0.3 cm}



\section{Introduction}
Regime switching models constitute an appealing framework, stemming from their ability to capture the relevant features of {asset price} dynamics, which behave differently depending on the specific phase of the business cycle {in place}.
In the context of continuous time utility maximization, some studies have considered observable regimes, while others have accounted for the possibility that they are not directly visible. In the case of observable regimes, \cite{Zari} {considers} an infinite horizon investment-consumption model where the agent can invest her wealth in a stock and risk-free bond, with borrowing and stock short-selling constraints. In a similar regime switching framework, \cite{Cadenillas} {study} the infinite horizon problem of a risk averse investor maximizing regime dependent utility from terminal wealth and consumption. 
A different branch of literature has considered the case when regimes are hidden and need to be estimated from publicly available market information. \cite{NagaiRung} consider a finite horizon portfolio optimization problem, where a power investor allocates his wealth across money market account and stocks, whose price dynamics follow {a diffusion process modulated by a hidden finite-state Markov process}. \cite{Tamura} extend the analysis to the case
when the time horizon is infinite. 
\cite{ElSiu11} study the optimal investment problem of an insurer when the model uncertainty is governed by a hidden Markov chain. {\PPaleGrey \cite{Siu13} considers optimal investment problems in general non-Markovian hidden regime switching models. 
\cite{Haussmann} consider a multi-stock market model, with stochastic interest rates and drift modulated by a hidden Markov chain. Combining appropriate Malliavin calculus and filtering results from hidden Markov models, they derive explicit {representations} of the optimal strategies. In a series of two papers, \cite{NagaiRung1} and \cite{NagaiRung2} consider a regime switching framework where logarithmic and power investors optimize their terminal utility by investing in stocks at random times due to liquidity constraints. 

The literature surveyed above has considered markets consisting of securities carrying market, but not default, risk. In recent years, few studies have considered a portfolio optimization framework inclusive of defaultable securities. \cite{KS} study optimal portfolio problems with defaultable assets within a  Black-Cox framework. \cite{KS08} consider an investor who can allocate her wealth across multiple defaultable
bonds, in a model where simultaneous defaults are allowed. In the same market model, \cite{KSAll} define default as the beginning of financial distress, and discuss contagion effects on prices of defaultable bonds. \cite{BielJang} derive optimal investment strategies for a CRRA investor, allocating her wealth among a defaultable bond, risk-free bank account, and a stock. \cite{BoWang} consider a portfolio optimization problem, where a logarithmic investor can choose a consumption rate, and invest her wealth across a defaultable perpetual bond, a stock, and a money market account. \cite{JiaoPham} combine duality theory and dynamic programming to optimize the utility of a CRRA investor in a market consisting of a riskless bond and a stock subject to counterparty risk. Optimal investment under contagion risk has been considered by \cite{BoCappMF}, who construct an empirically motivated framework based on interacting intensity models, and analyze how contagion effects impact optimal allocation decisions due to abrupt changes in prices. A related study by \cite{Jiao} develop a portfolio framework where multiple default events can occur, and some of the securities may still be traded after default.

The first attempt at using regime switching within a portfolio optimization framework consisting of defaultable securities was done by \cite{CapFig1}. Such a modeling choice is also empirically supported by a study of \cite{Longstaff}, which identifies three credit regimes characterized by different levels of default intensity and recovery rates, via a historical analysis of the corporate bond market.
\cite{CapFig1} consider an investor trading in a stock and defaultable security, whose price dynamics are modulated by an observable Markov chain. Using the HJB approach, they recover the optimal investment strategies as the unique solution to a coupled system of partial differential equations.

{The present paper} considers the case where regimes are hidden, so that the power investor must decide on the optimal allocation policy using only {the observed market} prices. This improves upon the realism of the model in \cite{CapFig1}, given that in several circumstances market regimes such as inflation and recession, or credit regimes characterized by high or low credit spreads, are typically unobserved to investors. Moreover, the hidden regime feature requires a completely different analysis, and leads us to solving a partially observed stochastic control problem, where regime information must be inferred from an enlarged market filtration. {The latter is composed {\emph {both}} of a reference filtration generated by {the} observable security prices, {\emph {and}} of a credit filtration tracking the occurrence of the default event.} To the best of our knowledge, ours represents the first study in this direction.

We next describe our main contributions. First, by considering a portfolio optimization problem in a context of partial information with possibility of default, we advance earlier literature which has so far considered either one or the other aspect, but never both simultaneously. {We construct} an equivalent fully observed risk-sensitive control problem, where the new state is given by the {\PaleGrey regime filtered} probabilities, generalizing
the approach of \cite{NagaiRung} who do not deal with default event information.
We use the filter probabilities to obtain the Hamilton-Jacobi-Bellman (HJB) equation for the {dynamic} optimization problem, which we separate it into coupled pre-default and post-default optimization subproblems. This is done using the projected filter process. We remark that the decomposition of a global optimal investment problem into sub-control problems in a progressively enlarged filtration has also been considered by \cite{Jiao} and \cite{Pham}. Their approach consists in first defining the sub-control problems in the reference market filtration exclusive of default event information, and then connecting them by assuming the existence of a conditional density on the default times, previously introduced in \cite{KarouiJeanJiao}. Despite few similarities between ours and their approach arising from the fact that both consider a pre and post-default decomposition and solve backwards, there are also significant differences between the two approaches. We consider the wealth dynamics under the enlarged market filtration inclusive of default events and do not perform any pre-post default decomposition of the control problem at the level of the stochastic differential equation. It is only after deriving the HJB partial differential equations that the decomposition into pre and post-default PDEs naturally arises. Their approach instead exploits the exponential utility preference function of the investor to reduce the optimal investment problem to solving a recursive system of backward stochastic differential equations with respect to the default-free market filtration. A detailed analysis of these BSDEs including the possibility of
jump times driven by Brownian motion is provided in \cite{Khar}.

Secondly, the presence of default risk makes the HJB-PDE satisfied by the pre-default value function non-linear.
There are two sources of nonlinearity, namely quadratic growth of the gradient and exponential
nonlinearity. We first perform a suitable transformation yielding a parabolic PDE whose associated
operator is linear in the gradient and matrix of second derivatives, but nonlinear in the solution. We then provide
a rigorous analysis of the transformed PDE and prove the existence of a classical solution via { a} monotone iterative method. Since the nonlinear term is only locally, but not globally, Lipschitz {continuous} because the derivative explodes at zero, we also need to prove that the solution is bounded away from zero. In particular, we establish both a lower and upper bound for
the solution, and prove $\mathcal{C}^{2,\alpha}_P$ regularity. We then use this result to prove a verification theorem establishing the correspondence between the solution to the PDE and the value function of the control problem. {\CadBlue The proof of the theorem requires the development of a number of technical results, such as the guaranteed positivity of the filtering process.}
By contrast, the HJB-PDE corresponding to the post-default optimization problem can be linearized using a similar transformation to the one adopted by \cite{NagaiRung}, and a unique classical solution can be guaranteed as shown, for instance, in \cite{Tamura}.

Thirdly, we provide a thorough comparative statics analysis to illustrate the impact of partial information on the optimal allocation decisions. We consider a square root investor and a two-states Markov chain. We find that the fraction of wealth invested in the stock increases as the filter probability of being in the regime with the highest growth rate increases. In order to be hedged against default, the investor shorts a higher number of units of defaultable security if the filter probability of staying in the highest default risk regime increases. {Vice versa}, when the probability of being in the safest regime increases, the investor {increases} his exposure to credit risk by shorting smaller amount of units of the defaultable security. If the regime is characterized by a sufficiently low level of default intensity, the square root investor may even go long credit, and purchase units of the defaultable security. We find that lower values of price volatility induce the investor to increase the fraction of wealth invested in the risky asset. More specifically, if the stock volatility is low, the filter gain coming from received observations is higher and the investor purchases increasingly more units as the stock volatility decreases. Similarly, for a sufficiently high probability of being in the high default risk regime, the investor shorts increasingly larger number of units of defaultable security as the volatility of the latter decreases. This reflects the risk averse nature of the investor, who wants to reduce his credit risk exposure more if the filter estimate becomes more accurate due to the higher informational gain from price observations.
We also find that as observations become less informative due to higher price volatilities, the investor deposits a significant fraction of his wealth in the money market account. All this suggests that partial information tends to push the investor towards strategies reducing both market and credit risk exposure.

{The rest of the paper is organized as follows. Section \ref{sec:model} defines the market model. Section \ref{sec:utilmax} sets up the utility maximization problem. Section \ref{sec:HJB} derives the HJB equations corresponding to the risk sensitive control problem. Section \ref{sec:ver} analyzes the solutions of the HJB-PDE equations. Section \ref{sec:numanalysis} develops a numerical analysis. {Section \ref{sec:conclusions} summarizes our main conclusions. Finally, two appendices present the main proofs of the paper.}}

\section{The Market Model} \label{sec:model}
Assume $(\Omega,\G,\Gx,\Px)$ is a complete {\DMagenta filtered} probability space, where $\Px$ is the historical probability measure, {${\Gx}:=(\gt)_{t \geq 0}$} is an enlarged filtration given by {$\gt := \mathcal{F}_t \vee \mathcal{H}_t$}{, where $(\mathcal{H}_t)_{t}$ is a filtration to be introduced below}.
We take the right continuous version of $\mathbb{G}$, i.e. $\gt$ is the smallest right-continuous filtration
containing $\mathbb{F}$ and $\mathbb{H}$, with $\gt := \bigcap_{\epsilon>0} \left(\mathcal{F}_{t+\epsilon} \vee \mathcal{H}_{t+\epsilon}\right)$ {(see also \cite{Belanger})}.

Here, $\Fx:=(\F_{t})_{t}$ is a suitable filtration supporting a two dimensional Brownian motion $W_t = (W_t^{(1)},W_t^{(2)})^{\top}$, where $^{\top}$ denotes the transpose. We {also} assume that the \emph{hidden} states of the economy are modeled by a  finite-state continuous-time Markov chain  $X:=\{X_t\}_{t\geq{}0}$, which is 
 adapted to $\Gx$ and assumed to be independent of $W:=\{W_{t}\}_{t\geq{}0}$. Without loss of generality, the state space is identified by the set of unit vectors $\{e_1, e_2, \ldots, e_N \}$, where $e_i = (0,...,1,...0)^{^{\top}} \in \mathds{R}^N$.
The following semi-martingale representation is well-known {(cf. \cite{elliottb}):}
\begin{equation}
X_t = X_0 + \int_0^t {A(s)^{\top}} X_s ds + {\varphi(t)}{,}
\label{eq:MCsemim}
\end{equation}
where {$\varphi(t)=(\varphi_{1}(t),\dots,\varphi_{N}(t))^{\top}$} is a $\mathds{R}^N$-valued martingale under $\Px$, and
$A(t) := [\varpi_{i,j}(t)]_{i,j=1,\dots,N}$ is the so-called generator of the Markov process. Specifically, denoting {\DMagenta $p_{i,j}(t,s) := \Px(X_s = e_j | X_t = e_i)$}, for $s \geq t$, and $\delta_{i,j}={\bf 1}_{i=j}$, we have that
$$
    \varpi_{i,j}(t) = \lim_{h \rightarrow 0} \frac{p_{i,j}(t,t+h) - \delta_{i,j}}{h};
$$
cf.  \cite{bielecki01}. In particular, {$\varpi_{i,i}(t) = -\sum_{j \neq i} \varpi_{i,j}(t)$}.  {The following mild condition is also imposed:
\begin{equation}\label{CndOnGen}
    \sup_{t\geq{}0}\max_{i,j}\varpi_{i,j}(t)<\infty.
\end{equation}
We denote by $p^{\circ} = (p^{\circ,1},\ldots,p^{\circ,N})$} the initial distribution on the Markov chain and, throughout the paper, assume that {$p^{\circ,i} > 0$}.

We consider a {frictionless} financial market  consisting of three instruments: a risk-free bank account, a {defaultable security}, and a stock.

\medskip
\noindent\textbf{Risk-free bank account.} The instantaneous market interest rate is assumed to be constant.
The dynamics of the price process {$\{B_t\}$, which describes the risk-free bank account,} is given by
\begin{equation}
    {dB_t = r B_t d t, \qquad B_{0}=1.}
\end{equation}

\noindent\textbf{Stock Security.}
We assume that the {appreciation} rate $\{\mu_t\}$ of the stock depends on the {economic} regime $X_t$ in the following way:
$$
{\CaddBlue \mu_t:= \left<\mu, X_t\right>{,}}
$$
{\CadBlue where $\mu = {(\mu_1,\mu_2,\ldots,\mu_N)}$ is a vector {\Blue with constant} components denoting the values of the drift associated to the different economic regimes and where $\left<\cdot,\cdot\right>$ denotes the standard inner product in $\mathbb{R}^{N}$. Under the historical measure, the stock dynamics is given by
\begin{equation}\label{DfnOfS}
dS_t = \mu_t S_t dt + \sigma S_t dW^{(1)}_t, \qquad S_0 = {s^{\circ}}{.}
\end{equation}
}

\noindent\textbf{Defaultable Security.}
Before defining the vulnerable security considered in the present paper, we need to {introduce} the default process. Let $\tau$ be a nonnegative random variable, {defined on $(\Omega,\G,\Px)$}, representing the default time of the counterparty selling the security.
Let $\mathcal{H}_t = \sigma({\CaddBlue H_u}: u \leq t)$ be the filtration generated by the \emph{default process} {$H_{t}:= \idc_{\tau \leq t}$}. We use the canonical construction of the default time $\tau$ in terms of a given hazard process. The latter is defined by $\Gamma_t := -\ln\left(1-\Ex^{\Px}\left[H_t \big|  \mathcal{F}_t\right]\right)$ (see also definition 9.2.1 in \cite{bielecki01}), and postulated to have absolutely continuous sample paths with respect to the Lebesgue measure on $\mathds{R}^+$. In other words, it is assumed to admit the integral representation
$$
\Gamma_t = \int_0^t h_u du
$$
for a $\Fx$ progressively measurable, nonnegative stochastic process $\{h_t\}_{t\geq{}0}$, with integrable sample paths. The process
$h_t$ is referred to as the $\Fx$-hazard rate of $\tau$, and will be specified later.
We next give the details of the construction of the random time $\tau$.  {We assume the existence of an exponential random variable $\chi$ defined on the probability space
 {$(\Omega,\G, \Px)$}, independent of the process $(X_t)_{t}$.
We define $\tau$ by setting
\begin{equation}
{\tau:=}\inf \left \{t \in \mathds{R}^+ : \int_0^t {h_u} du \geq \chi \right \},
\label{eq:taudef}
\end{equation}
where we follow the convention $\inf\emptyset=+\infty$. It can be proven  {(see \cite{bielecki01}, Section 6.5 for details)} that
\begin{equation}\label{MrtRprDftPrcP}
    \xi_t := {\CaddBlue H_t} - \int_0^t {\CaddBlue {\bar{H}_{u^{-}}}} h_u du = {{\CaddBlue H_t} - \int_0^{t\wedge\tau} h_u du,}
\end{equation}
is a ${\Gx}$-martingale under $\Px$, where {{\CaddBlue $\bar{H}_u:=1-H_u$} and {\CaddBlue $\bar{H}_{u^{-}} := \lim_{s\uparrow u} \bar{H}_s = \idc_{\tau \geq  u}$}}. Intuitively, Eq.~\eqref{MrtRprDftPrcP} says that the single jump process needs to be compensated for default, prior to the occurrence of the event. As with the appreciation rate, we  assume that the process $h$ is driven by the hidden Markov chain as follows:
$$
    h_t:= \left<h, X_t\right>{,}
$$
where {$h = (h_1,\dots,h_N)\in(0,\infty)^{N}$} denotes the possible values that the default rate process can take depending on the economic regime in place. {\BRed We model the pre-default dynamics of the defaultable security as}
\begin{equation}
{\BRed
\frac{dP_t}{P_t} = a(t, X_t) dt + \upsilon dW_t^{(2)},\qquad {(t<\tau)},}\qquad {P_{0}=P^{\circ},}
\label{eq:pricehist}
\end{equation}
where {$P^{\circ}\in\mathbb{R}_{+}$ and} $a:\mathbb{R}_{+}\times\{e_{1},\dots,e_{N}\}\to\mathbb{R}_{+}$ is a deterministic function. {\BRed After default the security becomes worthless, i.e. $P_{t}:=0$ for any $t\geq{}\tau$} and, thus, $\{P_{t}\}_{t\geq{}0}$ follows the dynamics:
\begin{equation}\label{eq:pricehistCompl}
    d P_{t}=P_{t^{-}}\left(a(t,X_{t})dt+\upsilon d W_{t}^{(2)}-d H_{t}\right).
\end{equation}
For future reference, we also impose the following mild technical assumption:
\begin{equation}\label{CndFora}
    {\int_{0}^{T}a^{2}(t,e_{i})dt<\infty, \quad \text{for any}\quad T>0 \text{ and } i\in\{1,\dots,N\}.}
\end{equation}

\begin{remark}
{\CadBlue As usual when dealing with hidden Markov models the volatility components are assumed to be constant, see for instance \cite{NagaiRung}. If $\sigma$ and $\upsilon$ were not constant but consisting of distinct components depending on $X$, then the Markov chain $\{X_t\}_{t\geq0}$ would become observable. This is because the quadratic variation of $X_t$ would converge almost surely to the integrated volatility, see \cite{McKean}. Consequently, by inversion, the regime in place at time $t$ would become known. Further, we notice that the choice of constant volatility might also provide a fairly good fit to market data when calibrating the hidden regime switching model to market prices. This has been empirically shown by \cite{Liechty} on data from the New York Merchantile stock exchange using Markov chain Monte-Carlo methods.}
\end{remark}

\begin{remark}
The specification given in~\eqref{eq:pricehist} captures {several} relevant market models which have {been considered} in the literature:
\begin{enumerate}
\item First, {the model \eqref{eq:pricehist} may be specialized} to capture
the {\BRed pre-default} dynamics of a defaultable stock. The latter is a widely used instrument in hybrid models of equity and credit. For instance, \cite{Linetsky} and \cite{CarrMendoza} model the {pre-bankruptcy risk-neutral} dynamics of a {defaultable} stock as
$$
    d{S}_t = ({r+h_t}) S_t dt + {\upsilon} S_t dW_t,
$$
where {$\{W_t\}_{t\geq{}0}$ is a Brownian driver and $\{h_t\}_{t\geq{}0}$ is a stochastic (adapted) default intensity process}. Clearly, such a specification is a special instance of \eqref{eq:pricehist}, where we set ${\BRed a(t, X_t)}={r+ h(X_t)}$. The addition of the hazard rate in the drift ensures that the discounted stock price process is a martingale.
\item Secondly, the dynamics in Eq.~\eqref{eq:pricehist} may be used to model the time evolution of prices of credit {\BGreen sensitive} securities when {an additive type of ``microstructure or market-friction"} noise is taken into account.
In general, secondary market {investors only} observe market quotes for traded credit derivatives, such as spreads, at discrete points in {time, e.g., at times $t_k = k {\Delta}$, $k=0,\dots,N$, for a certain fixed time mesh $\Delta>0$. The corresponding observed yield spreads $z_{t_{k}}$ are then often modeled as $z_{t_k} = a_{t_k}(t_{k},X_{t_k}) + \epsilon_k$, with an i.i.d. sequence $(\epsilon_k)_k$, independent of $X$,} capturing microstructure noise due to illiquidity, {transaction costs, or transmission ``errors''. In that case, $a(t_{k},X_{t_{k}})$ represents the underlying unobserved yield spread which follows an efficient arbitrage-free model of choice.} \cite{RungFr} argue that as the interarrival time $t_k - t_{k-1}$ between consecutive observations gets smaller, the cumulative {log return process $z_{t}:=\log\left(P_{t}/P_{0}\right)$} of the defaultable security {converges, in law,} to
\begin{equation}\label{Eq:FrsDfnDynz}
\int_0^t a(s, X_s) ds + \upsilon W_t^{(2)}.
\end{equation}
Again, the dynamics of~\eqref{Eq:FrsDfnDynz} is in the form of our dynamics~\eqref{eq:pricehist}.
\end{enumerate}
\end{remark}
{For future convenience, we {introduce} the {two-dimensional} observed pre-default log-price process $Y_t = (\log(S_t), {\log(P_t)})^{\top}$, whose dynamics is given by
\begin{equation}
dY_t = {\loga(t,X_t)} dt + {\Sigma_{Y}} dW_t,
\label{eq:Yorig}
\end{equation}
where
\begin{equation}
     \Sigma_{Y} :=
\left(
\begin{array}{ccc}
\sigma  & 0 \\
 0  &  \upsilon  \\
\end{array}
\right),\qquad \loga(t,X_t):= \left[\mu_t - \frac{\sigma^2}{2}, a(t,X_t) -\frac{\upsilon^2}{2} \right]^{\top}=\left[\left<\mu,X_t\right> - \frac{\sigma^2}{2}, a(t,X_t) -\frac{\upsilon^2}{2} \right]^{\top}.
    \label{eq:defvartheta}
\end{equation}
We {also}} define two subfiltrations of {$\Gx$}, namely, the market filtration {$\Gx^{I}:=(\mathcal{G}_{t}^{I})_{t \geq 0}$ where}
$$
\mathcal{G}_t^I := \mathcal{F}_t^I  \vee \mathcal{H}_t,  \qquad  \mathcal{F}_t^I := \sigma(S_u, P_u; u \leq t),
$$
and the subfiltration {$\Fx^{X}:=(\mathcal{F}_{t}^{X})_{t \geq 0}$}, generated by the Markov chain $(X_t)_{t}$:
$$
\mathcal{F}_t^{{X}} = \sigma(X_u; u \leq t).
$$
Therefore, we may also write $\mathcal{G}_t = \mathcal{F}_t^{{X}} \vee \mathcal{G}_t^I$. From this, it is evident that {while} {$(X_t)_{t \geq 0}$} is {$(\mathcal{G}_t)_{t \geq 0}$} adapted,
it is not {$(\mathcal{G}_t^{I})_{t \geq 0}$} adapted. 
\section{The Utility Maximization Problem} \label{sec:utilmax}
 We consider an investor who wants to {maximize her expected final utility during a {trading period $[0,T]$}, by dynamically allocating her financial wealth into (1) the risk-free bank account, (2) the stock, {and} (3) the defaultable security, as defined in the previous section. Let us denote by $\nu_t^B$ the number of shares of the risk-free bank account that the investor buys (${\nu}_t^B>0$) or sells (${\nu}_t^B<0$) at time $t$. Similarly, ${\nu}_t^S$ and  {${\nu}_t^P$} denote the investor's portfolio positions in the stock and defaultable security at time $t$, respectively.
 The process $\nu:=\{({\nu}_t^B,{\nu}_t^S,{\nu}_t^P)\}_{t\geq0}$ is called a \emph{portfolio process}. We {denote $V_t({\nu})$} the wealth of the portfolio process ${\nu}=({\nu}^B,{\nu}^S,{\nu}^P)$ at time $t$, i.e.
 $$
 V_t({\nu}) = {\nu}_t^B B_t + {\nu}_t^S S_t + {\nu}_t^P {\BGreen \mathbf{1}_{\tau > t}} {P_t}.
 $$
We require the processes ${\nu}_t^B,{\nu}_t^S$, and ${\nu}_t^P$ to be $\mathcal{G}^I$-predictable. The investor does not have intermediate consumption nor capital income to support her trading of financial assets and, hence, we also assume  {the following {self-financing} condition}:
\[
    d V_{t}={\nu}_{t}^{B} dB_{t}+{\nu}^{S}_{t} dS_{t}+{\nu}_{t}^{P} {\BGreen \mathbf{1}_{\tau > t}} d{P_t}.
\]
Let
 \begin{equation}
 \pi_t^B := \frac{{\nu}_t^B B_t}{V_{t-}({\nu})}, \quad
\pi_t^S := \frac{{\nu}_t^S S_t}{V_{t-}({\nu})},
\quad
 \pi_t^P= \frac{{\nu}_t^{{\BGreen P}} {P_t}}{V_{t-}({\nu})}{\mathbf{1}_{\tau > t}} ,
  \end{equation}
if $V_{t-}({\nu})>0$, while {$\pi_t^B=\pi_t^P=\pi_t^S=0$} when $V_{t-}({\nu})=0$.
The vector $\bar\pi_{t}:=(\pi_t^B,\pi_t^S,\pi_t^P)$, called a \emph{trading strategy}, {represents} the corresponding fractions of wealth invested in each asset at time $t$.
Note that if {${\bar\pi}:=({\bar\pi}_{t})_{t}$} is admissible ({the precise definition will be given later}), then the dynamics of the resulting wealth process {in terms of ${\bar\pi}$} can be written as
\begin{equation}\label{Eq:DynWealthProc}
    {d V_{t}=V_{t^{-}}\left\{\pi_{t}^{B}\, \frac{d B_{t}}{B_{t}}+\pi^{S}_{t}\, \frac{dS_{t}}{S_{t}}+{\pi_{t}^{P}}\, {\frac{d {P}_t}{P_t}}\right\}},
\end{equation}
{under the convention that $0/0=0$. The latter convention} is needed to deal with the case when default has occurred ({$t\geq{}\tau$}), so that $P_t$=0 and we have $\pi^{P}_{t}=0$. Using that $\pi^B + \pi^P + \pi^S = 1$, and the corresponding dynamics of $B_t$, $S_t$ and $P_t$, {we can further rewrite the {dynamics (\ref{Eq:DynWealthProc})} as}
\begin{equation}\label{Eq:DynWealthProcb}
{\frac{dV^{\pi}_t}{{V^{\pi}_{t^{-}}}} = r dt + \pi_t^S \left(\mu_t - r \right) dt + \pi_t^S \sigma dW_t^{(1)} + \pi_t^P \left(a(t,X_t)-r \right) dt + \pi_t^P \upsilon dW_t^{(2)} ,\qquad V^{\pi}_{0}=v},
\end{equation}
for a given initial budget $v\in (0,\infty)$. {Above, we use $\pi_{t}:=(\pi_t^S,\pi_t^{P})^{\top}$ to denote the time-$t$ investment strategy only consisting of {positions on the} stock and defaultable security,
and write $V^{\pi}$ to emphasize the dependence of the wealth process on the strategy $\pi$.}
The objective {\PaleGrey of the power investor is then} to choose $\pi=(\pi^S,\pi^{P})^{\top}$ so to maximize the expected utility from terminal wealth
\begin{equation}
J(v,\pi,T) := \frac{1}{\gamma} \Ex^{\Px}\left[\left(V_T^{\pi}\right)^{\gamma} \right],
\label{eq:Jvpi}
\end{equation}
for a given fixed value of $\gamma \in (0,1)$. 
By It\^o's formula and Eq.~(\ref{Eq:DynWealthProcb}),} we readily obtain {that $V_{t}^{\gamma}:=\left(V^{\pi}_{t}\right)^{\gamma}$ follows the dynamics}
\begin{eqnarray*}
dV^{\gamma}_t &=& \gamma V_t^{\gamma} \left[r dt + {\pi_t^S} \left(\mu_t - r \right) dt + \pi_t^S \sigma dW_t^{(1)} + \pi_t^{P} \left( {a(t,X_t) -r} \right) dt + \pi_t^P \upsilon dW_t^{(2)} \right] \\
 &  & + \frac{1}{2} \gamma (\gamma-1) V_t^{\gamma} \left[ (\pi_t^S)^2 \sigma^2 dt + (\pi_t^P)^2 \upsilon^2 dt  \right].
\end{eqnarray*}
Next, recalling that $W_{t}:=(W^{(1)}_{t},W^{(2)}_{t})^{\top}$, $\pi_t = (\pi_t^S,\pi_t^P)^{\top}$, and the definition of $\Sigma_{Y}$ given in (\ref{eq:defvartheta}),} we may rewrite the above SDE as
\begin{equation}\label{eq:PowerSDEV}
dV^{\gamma}_t = V_t^{\gamma} \left[- \gamma {\eta(t,X_t,\pi_t)} dt + \gamma \pi_t^{\top} {\Sigma_{Y}} dW_t \right],
\end{equation}
where
\begin{equation}
{\eta(t,X_t,\pi_t)} = -r + \pi_t^S(r-{\left<\mu,X_{t}\right>}) + \pi_t^P \left({\BRed r- {a(t,X_t)}} \right) + \frac{1-\gamma}{2} \pi_t^{\top} {\Sigma^{\top}_{Y}} {\Sigma_{Y}} \pi_t.
\label{eq:defeta}
\end{equation}
It is then clear that the solution to the {stochastic differential equation (\ref{eq:PowerSDEV}) with initial condition $V_0=v$} is given by
\begin{equation}\label{SimplExpV}
V_t^{\gamma} = v^{\gamma} {\exp\left( \gamma \int_0^t \pi_s^{\top} {\Sigma_{Y}} dW_s-\gamma \int_0^t {\eta(s,X_s,\pi_s)} ds  - \frac{\gamma^2}{2} \int_0^T \pi_s^{\top} {\Sigma_{Y}} {\Sigma^{\top}_{Y}} \pi_s ds\right)}.
\end{equation}

{\PaleGrey
From Eq.~\eqref{eq:Jvpi} and~\eqref{SimplExpV}, we can see that we need to solve a maximization problem with partial information since the regime $X_t$ is not directly observable and investment strategies can only be based on past information of security prices. Our approach is to transform it into a fully observed risk sensitive control problem. Such a reduction is accomplished through two main steps. First, in Section~\ref{sec:equivform} we show equivalence to a complete observation control problem with finite dimensional Markovian state. Then, in Section~\ref{sec:risksens} we transform the complete observation control problem into a risk-sensitive stochastic control problem.
}

\subsection{{\PaleGrey An Equivalent Formulation as {a} Complete Observation Control Problem}} \label{sec:equivform}
{\PaleGrey The objective of this section is to show how the partially observed control problem in~\eqref{eq:Jvpi} may be reduced
to a complete observation control problem. This is accomplished by showing that the criterion~
\eqref{eq:Jvpi} may be equivalently rewritten as an expectation, taken with respect to a suitably chosen probability measure,
{\DBrown of an exponential function of the (observable) regime filtered probabilities}.
Next, we start developing the change to the new measure, chosen so that the underlying chain {$(X_t)_{t}$} becomes independent of the investor filtration {$\Gx^I$} under such a measure.}
First, we introduce some needed notation and terminology. Given two semimartingales $L$ and $M$, we denote {by} $[L]$ and $[L,M]$ the quadratic variation of $L$ and the quadratic covariation of $L$ and $M$, respectively.  We also denote the stochastic exponential of $L$ by $\mathcal{E}(L)$. If $L$ is of the form $L_t = \int_0^t \theta^{\top}_s dY_s$, where $Y_s$ is a $\mathbb{R}^{d}$-valued {continuous It\^o} process, and $\{\theta_s\}_{s\geq0}$ is $\Gx$ predictable, then
\begin{equation}
\mathcal{E}_t(L) = {\exp\left(\int_0^t \theta^{\top}_u dY_u - \frac{1}{2} \int_0^t \theta^{\top}_u  {\theta_u d[Y]_u}\right).}
\label{eq:stocexpcont}
\end{equation}
If $Z$ is of the form $Z_t = \int_0^t \iota_s d \xi_s$, where $\xi_s$ has been defined in \eqref{MrtRprDftPrcP}, and $\{\iota_s\}_{s\geq0}$ is $\Gx$-predictable, with $\iota > -1$, then
\begin{equation}
\mathcal{E}_t(Z) = {\exp\left(\int_0^t \log(1+\iota_{s}) dH_s - \int_0^{t \wedge \tau} \iota_s h_s ds\right).}
\label{eq:stocexpdisc}
\end{equation}
It is well known (see \cite{bielecki01}, {Section 4.3}) that $R_t := \mathcal{E}_t({L}) \mathcal{E}_t(Z)$ follows the SDE
\begin{equation}
R_t = 1 + \int_{{(0,t]}} R_{s-} \left(\theta^{\top}_s dY_s + \iota_s d \xi_s \right).
\label{eq:stocexpdyn}
\end{equation}

{\PaleGrey We now proceed to introduce the new measure {$\hat\Px$} on ($\Omega, \Gx)$.} 
Such a measure is {defined} in terms of its density process as follows:
\begin{equation}\label{Eq:DfnRho}
    {\frac{d\hat{\Px}}{d\Px}\bigg|_{\mathcal{G}_t}:={\mathcal{E}_{t}\left(\int_0^{\cdot} -{\loga(s,X_s)^{\top}} {\Sigma_{Y}^{-1}}  d{W_s} \right) \mathcal{E}_{t}\left(\int_0^{\cdot} {\frac{1-h_{s^{-}}}{h_{s^{-}}}} d\xi_s \right)  =: \rho^{(1)}_t \rho^{(2)}_t}}.
\end{equation}
In particular, using Eqs.~\eqref{eq:stocexpcont}
and~\eqref{eq:stocexpdisc}, $\rho^{(1)}_t$ and $\rho^{(2)}_t$ above are given by
\begin{align*}
\rho^{(1)}_t &= \exp\left(-\int_0^t {\loga(s,X_s)}^{\top} {\Sigma_{Y}^{-1}} dW_s - \frac{1}{2}\int_0^T \loga^{\top} ({\Sigma_{Y}} {\Sigma^{\top}_{Y}})^{-1} {\loga(s,X_s)} ds\right),\\
\rho^{(2)}_t &= \exp\left(-\int_0^t \log(h_{u^{-}}) dH_u -\int_0^{t \wedge \tau} (1-h_u) du\right)  = h_{\tau^{-}}^{-\idc_{\{\tau \leq  t\}}} \exp\left(-\int_0^{t \wedge \tau} (1-h_u) du\right).
\end{align*}
Moreover, from Eq.~\eqref{eq:stocexpdyn}, {the density process $\rho_t=\rho^{(1)}_{t}\rho^{(2)}_{t}$} admits the following representation
\begin{equation}\label{SDEForrho}
\rho_t = 1 + {\int_{0}^{t}} \rho_{s-} \left({-}{\loga(s,X_s)}^{\top} {\Sigma_{Y}^{-1}} dW_s +{\frac{1-h_{s^{-}}}{h_{s^{-}}}} d \xi_s \right).
\end{equation}
{In order to show that $\hat\Px$ is well-defined, one must verify that $\mathbb{E}^{\Px}\left(\rho_{T}\right)=1$.  To this end, we use a general version of Novikov's condition, as proved in \cite{ProtterShimbo} (see Theorem 9 therein), which states that the stochastic exponential $\mathcal{E}(M)$ of a locally square integrable martingale $M$ is a martingale on $[0,T]$ if
\begin{equation}\label{GeneralNovikov}
     {\Ex^{\Px}\left[e^{\frac{1}{2}\left<M^{c},M^{c}\right>_{T}+\left<M^{d},M^{d}\right>_{T}}\right]<\infty},
\end{equation}
where $M^{c}$ and $M^{d}$ are the continuous and purely discontinuous martingale parts of $M$. {Here, $\left<M^{c},M^{c}\right>_{T}$ and $\left<M^{d},M^{d}\right>_{T}$ denote the compensators of the  quadratic variations of $M^c$ and $M^d$ at time $T$, respectively (see \cite{Protter}, Page 70). From (\ref{SDEForrho}), $\rho_{t}=\mathcal{E}_{t}(M)$ with
\[
    M_{t}= -\int_{0}^{t} \loga(s,X_s)^{\top} \Sigma_{Y}^{-1} dW_s +\int_{0}^{t}\frac{1-h_{s^{-}}}{h_{s^{-}}} d \xi_s.
\]
Therefore, we have
\begin{align*}
\left<M^{c},M^{c}\right>_{T}&=\int_{0}^{T}{\loga^{\top}(s,X_s)(\Sigma_{Y}\Sigma_{Y})^{-1}\loga(s,X_s)} ds,\\
\left<M^{d},M^{d}\right>_{T}&= \int_0^T\left[\frac{(1-h_s)^2}{h_s}\right]\bar{H}_s ds.
\end{align*}
Clearly, $\left<M^{c},M^{c}\right>_{T}$ is bounded in view of the condition (\ref{CndFora}). It remains to prove that $\left<M^{d},M^{d}\right>_{T}$ is also bounded. We have
\begin{equation*}
\left<M^{d},M^{d}\right>_{T}
= \int_0^T\left(\frac{1}{h_s}+h_s-2\right)\bar{H}_s ds
\leq \int_0^T\left(\frac{1}{h_s}+h_s\right) ds.
\end{equation*}
Since $h_i >0$  for all $i \in \{1,\ldots,N\}$ {and $h_{s}=\left<h,X_{s}\right>$}, we obtain that
\[
\max_{i \in\{1,\ldots,N\}} \left(\frac{1}{h_i} + h_i \right)<C,
\]
for some constant $C>0$. Thus, we conclude that $\left<M^{d},M^{d}\right>_{T}$ is also bounded.
}


Under {the probability measure $\hat\Px$}, by {Girsanov's theorem} ({see, e.g.,} \cite{bielecki01}, Section 5.3), we have that
$$
\hat{W}_t = W_t + \int_0^t {\Sigma_{Y}^{-1}} {\loga(s,X_s)} ds
$$
is a Brownian motion, and
\begin{equation}
\hat{\xi}_t = \xi_t - \int_0^{t \wedge \tau} (1-h_u) du = H_t - \int_0^{t \wedge \tau} du= {H_t - \int_0^{t} {\CaddBlue \bar{H}_{u^{-}}} du}
\label{eq:zteq}
\end{equation}
is a {$\Gx^I$-martingale}. Note also that, {from Eq.~(\ref{eq:Yorig}),}  the observed pre-default log-price process $Y_t = (\log(S_t), {\log(P_t)})^{\top}$ possesses the dynamics $dY_t = {\Sigma_{Y}} d\hat{W}_t$ {under $\hat{\Px}$}. Furthermore, the inverse density process,
$$
U_t := \frac{d\Px}{d\hat{\Px}} \bigg|_{{\mathcal{G}_t}},
$$
{can} be written as $U_t = U_t^{(1)} U_t^{(2)}$, where
\begin{align*}
U_t^{(1)} &:= {\exp\left(\int_0^t {\loga(s,X_s)}^{\top} {\Sigma_{Y}^{-1}}  dW_s + \frac{1}{2} \int_0^t \loga^{\top} \left( {\Sigma_{Y}} {\Sigma^{\top}_{Y}} \right)^{-1} {\loga(s,X_s)} ds \right)}\\
U_t^{(2)} &:= {h_{\tau^{-}}^{\idc_{\{\tau \leq t\}}}} \exp\left(\int_0^{t \wedge \tau} (1-h_u) du\right)= {\mathcal{E}_{t} \left(\int_0^{\cdot} ({h_{s^{-}}}-1) d\hat{\xi}_s \right)
          .}
\end{align*}

{Using the previous probability measure $\hat{\Px}$ together with the representation (\ref{SimplExpV}), Eq.~\eqref{eq:Jvpi}} may be rewritten as
\begin{eqnarray}
\nonumber \frac{1}{\gamma} {\Ex^{\Px}} \left[V_T^{\gamma} \right] &=& \frac{v^{\gamma}}{\gamma} \Ex^{{\hat{\Px}}}\left[e^{-\gamma \int_0^T {\eta(s,X_s,\pi_s)} ds + \gamma \int_0^T \pi_s^{\top} {\Sigma_{Y}} dW_s - \frac{\gamma^2}{2} \int_0^T \pi_s^{\top} {\Sigma_{Y}} {\Sigma^{\top}_{Y}} \pi_s ds} U_T \right] \\
 &=& \frac{v^{\gamma}}{\gamma} \Ex^{{\hat{\Px}}} \left[L_T  \right],
\label{eq:gammaeq}
\end{eqnarray}
where
\begin{align} \label{eq:lteq}
 &L_t :={\mathcal{E}_{t}\left(\int_0^{\cdot} {Q(s,X_s,\pi_s)^{\top}}  {\BRed \Sigma_Y d\hat{W}_s} \right) U_t^{(2)} \exp\left(-\gamma \int_0^t {\eta(s,X_s,\pi_s)} ds\right)}, \\
&{Q(s,{e_{i}},\pi_s)} := ({\Sigma_{Y}} {\Sigma^{\top}_{Y}})^{-1} {\loga(s,{e_{i}})} + \gamma \pi_s= {\left(\frac{1}{\sigma^{2}}\left(\mu_{i}-\frac{\sigma^{2}}{2}\right)+\gamma\pi^{S}_{s},\frac{1}{\upsilon^{2}}\left(a(t,e_{i})-\frac{\upsilon^{2}}{2}\right)+\gamma\pi^{P}_{s}\right)^{\top}.}
 \label{eq:defQ}
\end{align}

Next, we proceed to give the filter probabilities {and some useful related relationships, for which we first need to introduce some notation.} Throughout, the {unit} $N$-simplex in $\mathbb{R}^{N}$ is denoted by
$$
\Delta_{N-1} = \left \{{(d^1,d^2,\ldots,d^N)}:\; d^1 + d^2 + \ldots d^N = 1, \; d^i \geq  0, \; i = 1,\ldots,N \right \}.
$$
{\CadBlue Let $g: D \to \mathbb{R}$, where $D = D_1 \times \{e_{1},\dots,e_{N}\} \times D_2$, with $D_1$ and $D_2$
arbitrary, possibly empty, domains. The {mapping} $\hat{g}: D_1 \times {\Delta}_{N-1} \times D_2 \to \mathbb{R}$ is defined as}
\begin{equation}
{\CadBlue \hat{g}(y,d,z) = \sum_{i=1}^N g(y,e_i,z) {d^i},}
\label{eq:optproj}
\end{equation}
{for each $y \in D_1$, $d \in {\Delta}_{N-1}$, and $z \in D_2$.
Similarly, given a vector $l=(l_{1},\dots,l_{N})\in\mathbb{R}^{N}$, we define the associated mapping $\hat{l}:\Delta_{N-1}\to\mathbb{R}$ as}
\begin{equation}\label{eq:optproj2}
    {\hat{l}(d):=\sum_{i=1}^{N}l_{i}d^{i}}.
\end{equation}
Throughout, the filter probability that the regime $X_t$ is $e_i$ at time $t$, conditional on the filtration $\mathcal{G}_t^I$, is denoted by
\begin{equation}
p_t^i :=\Px \left(X_t = e_i \big| \mathcal{G}_t^I\right),\qquad{i=1,\dots,N}.
\label{eq:filtprob}
\end{equation}
In particular, note the following useful relationships in terms of the transformations introduced in {\CadBlue  Eqs.~(\ref{eq:optproj})-(\ref{eq:optproj2}):
\begin{equation}
     {\Ex^{\Px} \left[g(y, X_t, z) \big| \mathcal{G}_t^I \right]=\hat{g}(y,{p_{t}},z),\qquad
     \Ex^{\Px} \left[\left<l,X_t\right> \big| \mathcal{G}_t^I \right]=\hat{l}(p_{t}),}
\label{eq:optprojb}
\end{equation}
where} hereafter $p_{t}:=(p_{t}^{1},\dots,p_{t}^{N})^{\top}$. We then have the following fundamental result, {\PaleGrey proven} in \cite{FreyS}:
\begin{proposition}[Proposition 3.6 in \cite{FreyS}] \label{prop:filterprob}
The normalized filter probabilities {$p_{t}:=(p_{t}^{1},\dots,p_{t}^{N})^{\top}$} are governed by the {SDE}
\begin{align}\nonumber
dp_t^i &= \sum_{\ell=1}^{N}\varpi_{\ell,i}(t) p^{\ell}_{t} dt + p_t^i ({\loga(t,e_i)^{\top}}-{\hat\loga}(t,p_t)^{\top}) (\Sigma_{Y} \Sigma_{Y}^{\top})^{-1} {\BRed (dY_t - {\hat\vartheta(t,p_t)} dt )}\\
&\quad  + {p_{t^{-}}^i \frac{h_i - {\hat{h}(p_{t^{-}})}}{{\hat{h}(p_{t^{-}})}} \left({d H_t} -  {\hat{h}(p_{t^{-}})} {\CaddBlue \bar{H}_{t^{-}}}dt \right)},
\label{eq:dynpti}
\end{align}
{with initial condition $p_{0}^{i}=p^{\circ,i}$}.
\nonumber
\end{proposition}
{Note that since $\hat{h}(p_{t})\geq\min_{i}\{h_{i}\}>0$, there is no singularity arising in the filtering equation (\ref{eq:dynpti}).}
{\PaleGrey We remark that \cite{RungFS}, Section 4.1, also consider filter equations for finite-state Markov chains in the presence of multiple default events. However, they provide the dynamics of the unnormalized filter probabilities using a Zakai-type SDE, and then construct an algorithm to compute the filter probabilities.}

\begin{remark}\label{Step1Const}
Uniqueness of a strong solution to the system (\ref{eq:dynpti}) was also established in \cite{FreyS} (see the discussion following Eqs.~(A2)-(A-3) therein), building on results of \cite{Kliemannetal}.
\end{remark}
{\PaleGrey
We are now ready to give the main result of this section.
Define
\begin{equation}
    {\hat{L}_t = \mathcal{E}_{t}\left(\int_0^{\cdot} {\hat{Q}(s,p_s,\pi_s)^{\top}} {\BRed \Sigma_Y d\hat{W}_s} \right) \mathcal{E}_{t}\left(\int_0^{\cdot} ({\hat{h}(p_{s^{-}})} - 1) d \hat{\xi}_s\right) e^{-\gamma \int_0^t {\hat{\eta}(s,p_s,\pi_s)} ds}},
\label{eq:hatLt}
\end{equation}
where,
using the notation (\ref{eq:optproj}) and recalling the definitions of $\eta$ and $Q$ given in Eqs.~(\ref{eq:defeta}) and (\ref{eq:defQ}), respectively,
\begin{align*}
    \hat\eta(t,p_t,\pi_t)&=\sum_{i=1}^{N}\eta(t,e_{i},\pi_{t})p_{t}^{i}=-r + \pi_t^S(r-{\hat\mu(p_{t})}) + \pi_t^P \left(r- {\hat{a}(t,p_{t})} \right) + \frac{1-\gamma}{2}\left(\sigma^{2}(\pi^{S}_{t})^{2}+\upsilon^{2}(\pi^{P}_{t})^{2}\right)\\
         \hat{Q}(t,p_t,\pi_t)&=\sum_{i=1}^{N}Q(t,e_{i},\pi_{t})p_{t}^{i}=\left(\frac{1}{\sigma^{2}}\left({\hat\mu(p_{t})}-\frac{\sigma^{2}}{2}\right)+\gamma\pi^{S}_{t},\frac{1}{\upsilon^{2}}\left({\hat{a}(t,p_{t})}-\frac{\upsilon^{2}}{2}\right)+\gamma\pi^{P}_{t}\right)^{\top}.
\end{align*}
Using the definitions of the stochastic exponentials given in~\eqref{eq:stocexpcont} and~\eqref{eq:stocexpdisc}, it follows from a direct application of {It\^o{'}s} formula that
\begin{equation}
d\hat{L}_t = \hat{L}_{{t^{-}}} \left({\hat{Q}(t,p_t,\pi_t)^{\top}} dY_t + ({\hat{h}(p_{{t^{-}}})} -1) d\hat{\xi}_t \right) - \gamma {\hat\eta(t,p_t,\pi_t)} \hat{L}_t dt.
\label{eq:ltdyn}
\end{equation}
Then, we have the following crucial result, whose proof is reported in Appendix \ref{app:utilmax}.}
\begin{proposition} \label{prop:equivform}
It holds {that}
\begin{equation}
J(v,\pi,T) = \frac{v^{\gamma}}{\gamma} \Ex^{\hat{\Px}} \left[ \hat{L}_T \right].
\label{eq:stocontr}
\end{equation}
\end{proposition}
The representation in~\eqref{eq:stocontr} establishes the correspondence between the original partially observed control problem
(criterion~\eqref{eq:Jvpi} depending on the hidden state $X_t$) and {a} complete observation control problem (criterion~\eqref{eq:stocontr} depending on the observed (filter) probabilities $p_t$).

\subsection{The Risk-Sensitive Control Problem} \label{sec:risksens}
{\PaleGrey
The objective of this section is to show how the complete observation control problem~\eqref{eq:stocontr} may be reduced to
a risk-sensitive stochastic control problem. Such a representation proves to be useful for analyzing the control problem via the HJB approach in the next section. The reduction is obtained building on the approach of \cite{NagaiRung} who do not consider the defaultable security. Next, we develop the change to the new measure $\tilde{\Px}$, so to write the criterion~\eqref{eq:stocontr} in the risk sensitive form. The measure change from $\hat{\Px}$ to $\tilde{\Px}$ is defined via {its Radon-Nikodym density as follows:}}
\begin{equation}\label{TrnsFromHatPtoTildeP}
\frac{d\tilde{\Px}}{d\hat{\Px}} \bigg|_{{\mathcal{G}_t^I}} := \zeta_t := \mathcal{E}_{t}\left(\int_0^{\cdot} {\hat{Q}(s,p_s,\pi_s)^{\top}} {\BRed \Sigma_Y d\hat{W}_s} \right) \mathcal{E}_{t}\left(\int_0^{\cdot} ({\hat{h}(p_{s^{-}})} - 1) d \hat{\xi}_s\right).
\end{equation}
{\BGreen Note that the probability measure $\tilde{\Px}$ depends, through $\zeta_t$, on the strategy $\pi_t$. Hence, in order for
$\tilde{\Px}$ to be a probability measure, we need to require that the set of admissible strategies satisfies the condition}
\begin{equation}
    {\BGreen \Ex^{\hat{\Px}}\left[\zeta_T\right] = \Ex^{\Px}\left[\rho_T  \zeta_T \right] = 1.}
\label{Eq:AdPrbMes}
\end{equation}
{In order to impose (\ref{Eq:AdPrbMes}), we again use the general Novikov's condition (\ref{GeneralNovikov}). In this case, it is easy to check $\rho_t  \zeta_t =\mathcal{E}_{t}(M)$ with
\[
    M_{t}=\int_0^{t} \left(\hat{Q}(s,p_s,\pi_s)^{\top} \Sigma_Y -\loga(s,X_s)^{\top} {\Sigma_{Y}^{-1}}\right) d{W}_s +\int_0^{t} \frac{\hat{h}(p_{s^{-}}) -h_{s^{-}}}{h_{s^{-}}}d {\xi}_s,
\]
and, thus,
\begin{align*}
    \left<M^{c},M^{c}\right>_{T}&\leq 2
    \int_{0}^{T}\hat{Q}^{\top} \Sigma_Y \Sigma^{\top}_Y \hat{Q}(s,p_s,\pi_s)ds+2\int_{0}^{T}\loga^{\top}(\Sigma_{Y}\Sigma_{Y})^{-1}\loga(s,X_s)ds\\
    \left<M^{d},M^{d}\right>_{T}&=\int_{0}^{\tau\wedge{}T}\left(\frac{\hat{h}(p_{s}) -h_{s}}{h_{s}}\right)^{2}h_{s}ds.
\end{align*}
The second term in the expression of $\left<M^{c},M^{c}\right>_{T}$ is uniformly bounded in view of the condition (\ref{CndFora}), while $\left<M^{d},M^{d}\right>_{T}$ is also  bounded since the integrand therein is bounded by $2\max_{i}\{h^{2}_{i}\}/\min_{i}\{h_{i}\}$. Therefore, we only need to require that $\Ex^{\Px} \left[e^{\frac{1}{2}\int_{0}^{T}\hat{Q}^{\top}\Sigma_{Y}\Sigma_{Y}^{\top}\hat{Q}(s,p_{s},\pi_{s})ds}\right]<\infty$, for which it suffices that $\pi=(\pi^{S},\pi^{P})^{\top}$ meets the integrability condition:
\begin{equation}\label{NovCnd2}
    \Ex^{\Px}\left[e^{\frac{\sigma^{2}\gamma^{2}}{2}\int_{0}^{T}\left(\pi_{s}^{S}\right)^{2}ds+
     \frac{\upsilon^{2}\gamma^{2}}{2}\int_{0}^{T}\left(\pi_{s}^{P}\right)^{2}ds}\right]<\infty.
\end{equation}
Once we have established {conditions for} the validity of the probability transformation (\ref{TrnsFromHatPtoTildeP}), we can then apply Girsanov's theorem  (cf. \cite{bielecki01}) to conclude that}
$$
\tilde{W}_t = \hat{W}_t - \int_0^t \Sigma_Y^{\top} {\hat{Q}(s,p_s,\pi_s)} ds
$$
is a {$\mathcal{\Gx}^I$-}Brownian motion under $\tilde{\Px}$, {while}
\begin{equation}\label{Eq:DfnTildexi}
\tilde{\xi}_t = \hat{\xi}_t - \int_0^{t \wedge \tau} ({\hat{h}(p_s)}-1) ds = H_t - \int_0^{t \wedge \tau} {\hat{h}(p_s)} ds=
 {H_t - \int_0^{t} {\hat{h}(p_s)} {\CaddBlue \bar{H}_{s^{-}}} ds,}
\end{equation}
is a {$\mathcal{\Gx}^I$} martingale under $\tilde{\Px}$. It then follows immediately that
\begin{equation}
J(v,\pi,T) = \frac{v^{\gamma}}{\gamma} \Ex^{\tilde{\Px}}\left[\hat{L}_T \zeta_T^{-1} \right] = \frac{v^{\gamma}}{\gamma} \Ex^{\tilde{\Px}}\left[e^{-\gamma \int_0^T {\hat{\eta}(s,p_s,\pi_s)} ds} \right],
\label{eq:probreform}
\end{equation}
where the dynamics of $p_t^i$ in Eq.~\eqref{eq:dynpti} may be rewritten under the measure $\tilde{\Px}$ as
\begin{align}
\nonumber dp_t^i &= p_t^i \left(\loga(t,e_i)^{\top} - \hat\loga(t,p_t)^{\top} \right) {\Sigma_Y^{-1}} d\tilde{W}_t \\
 & \quad+ \left( \sum_{\ell=1}^{N}\varpi_{\ell,i}(t) p^{\ell}_{t} + \gamma p_t^i \left(\loga(t,e_i)^{\top} - \hat\loga(t,p_t)^{\top} \right) \pi_t \right) dt + {p_{t^{-}}^i \frac{{h}_{i} - \hat{h}(p_{t^{-}})}{{\hat{h}(p_{t^{-}})}} {d \tilde{\xi}_t}},\quad t\geq{}0.
\label{eq:ptiptilde}
\end{align}
Hence, the overall conclusion is that the original problem is {reduced} to a risk sensitive control problem of the form:
\begin{equation}\label{RSCPDf}
    \sup_{\pi} J(v;\pi;T)
 = \frac{v^{\gamma}}{\gamma}\sup_{\pi} \Ex^{\tilde{\Px}}\left[e^{-\gamma \int_0^T {\hat{\eta}(s,p_s,\pi_s)} ds} \right],
\end{equation}
where {the} maximization is done across {suitable} strategies $(\pi_t)_{t}$. {\BGreen We shall specify later on the precise class of trading strategies $\pi$ on which the portfolio optimization problem is defined.

\begin{remark}\label{Exist2}
    As customary with Markovian optimal control problems, we will solve the risk-sensitive control problem starting at time $t=0$ by embedding it into a dynamical control problem starting at any time $t\in [0,T]$. {Roughly speaking}, the latter problem can be seen as the original problem (\ref{RSCPDf}) but starting at time $t\in(0,T]$ instead of $0$. In order to formally define the dynamical problem, we consider a family of SDEs indexed by $t$ of the form:
    \begin{align}
\nonumber d{p}_s^{t,i} &= {p}_s^{t,i} \left(\loga(s,e_i)^{\top} - \hat\loga(s,{p}_s^{t})^{\top} \right) {\Sigma_Y^{-1}} d\widetilde{W}^{t}_s \\
 & \quad+ \left( \sum_{\ell=1}^{N}\varpi_{\ell,i}(s) {p}_s^{t,\ell} + \gamma {p}_s^{t,i} \left(\loga(s,e_i)^{\top} - \hat\loga(s,{p}_s^{t})^{\top} \right) {\pi^{t}_s} \right) ds + {{p}^{t,i}_{s^{-}} \frac{{h}_{i} - \hat{h}({p}_{s^{-}}^{t})}{{\hat{h}({p}_{s^{-}}^{t})}} {d \widetilde{\xi}^{t}_s}},
 \quad {s\in (t,T]},
\label{eq:ptiptildeb}
\end{align}
with initial condition ${p}^{t,i}_{t}=p^{\circ,i}$, defined on a suitable space $(\Omega^{t},\G^{t},\Gx^{t},\widetilde{\Px}^{t})$, equipped with a Wiener process $\{\widetilde{W}^{t}_{s}\}_{s\geq{}t}$ starting at $t$ and, an independent, one-point counting process $\{H^{t}_{s}\}_{s\geq{}t}$ such that $H^{t}_{t}=z^{\circ}\in\{0,1\}$ and
\begin{equation}\label{Eq:DfnTildexiFict}
\tilde{\xi}^{t}_s :=
 H^{t}_s - \int_{t}^{s} {\hat{h}({p}^{t}_u)} {\CaddBlue \bar{H}^{t}_{u^{-}}} du,\quad s\geq{}t,
\end{equation}
is a $\widetilde{\Px}^{t}$-martingale. {Hereafter, the construction of the process $\{p^{t}_{s}\}_{s\in[t,T]}:=\{(p^{t,1}_{s},\dots,p^{t,N}_{s})^{\top}\}_{s\in[t,T]}$ is} carried out in a similar way as the construction of the solution to (\ref{eq:ptiptilde}). Concretely,  start defining the process $\{{p}^{t}_{s}\}_{s\geq{}0}:=\{({p}^{t,1}_{s},\dots,{p}^{t,N}_{s})^{\top}\}_{s\geq{}0}$ via the representation (\ref{eq:filtprob}), which, {analogously to $\{p_{s}\}_{s\geq{}0}$ follows the SDE (\ref{eq:dynpti})}, is the solution of a system of SDE's of the form:
    \begin{align}\nonumber
d{p}_s^{t,i} &= \sum_{\ell=1}^{N}{\varpi}_{\ell,i}(s){p}^{t,\ell}_{s} ds + {p}_s^{t,i} (\loga(s,e_i)^{\top}-\hat{\loga}(s,{p}^{t}_s)^{\top}) (\Sigma_{Y} \Sigma_{Y}^{\top})^{-1} {\left(d{Y}^{t}_s - \hat{{\vartheta}}(s,{p}^{t}_s) ds \right)}\\
&\quad  + p_{s^{-}}^{t,i}\, \frac{h_i - {\hat{h}({p}^{t}_{s^{-}})}}{{\hat{h}({p}^{t}_{s^{-}})}} \left(d {H}^{t}_s -  \hat{h}(p^{t}_{s^{-}})\bar{H}_{s^{-}}^{t} ds\right),\qquad s\geq{}t,
\label{eq:dynptib}
\end{align}
on a probability space $(\Omega^{t},\G^{t},\Gx^{t},\Px^{t})$. Here, ${Y}^{t}_{s}:=(\log({S}^{t}_{s}),\log({P}^{t}_{s}))^{\top}$, $s\geq{}t$,  with $\{{S}^{t}_{s}\}_{s\geq{}t}$ and $\{P^{t}_{s}\}_{s\geq{}t}$ defined analogously to (\ref{DfnOfS}) and (\ref{eq:pricehistCompl}):
\begin{align*}
    dS^{t}_s &= \left<\mu,X_{s}^{t}\right> S_s^{t} ds + \sigma S^{t}_s dW^{(1,t)}_s, \quad s>{}t,\quad S^{t}_t = s^{\circ},\\
    d P_{s}^{t}&=P_{s^{-}}^{t}\left(a(s,X^{t}_{s})ds+\upsilon d W_{s}^{(2,t)}-d H^{t}_{s}\right),\quad s>{}t,\quad P_{t}^{t}=P^{\circ}.
\end{align*}
The hidden Markov chain $\{X^{t}_{s}\}_{s\geq{}t}$ has initial distribution ${p_{t}^{t,i}}=\Px^{t}\left(X^{t}_{t}=e_{i}\right)=p^{\circ,i}$ and generator $A(s):= [\varpi_{i,j}(s)]_{i,j=1,\dots,N}$, $s\geq{}t$. Once we have defined the process $\{p_{s}^{t}\}_{s\geq{}t}$, we proceed to define $\tilde{\Px}^{t}$ in terms of a suitable trading strategy $\{\pi^{t}_{s}\}_{s\in[t,T]}:=\{(\pi^{t,S}_{s},\pi^{t,P}_{s})^{\top}\}_{s\in[t,T]}$ analogously to $\tilde{\Px}$, and processes $\{\widetilde{W}^t\}_{s\geq{}t}$ and $\{\widetilde\xi^{t}_{s}\}_{s\geq{}t}$ analogously to $\tilde{W}$ and $\tilde{\xi}$ so that, under $\tilde{\Px}^{t}$, the process ${p}^{t}=({p}^{t,1},\dots,{p}^{t,N})^{\top}$ satisfies (\ref{eq:ptiptildeb}).
Note that the existence of the measure transformation $\widetilde{\Px}^{t}$ and, hence, of the solution to the SDE (\ref{eq:ptiptildeb}) is guaranteed provided that $\pi^{t}$ satisfies the analogous of (\ref{NovCnd2}):
\begin{equation}\label{NovCnd2bbc}
    \Ex^{\Px^{t}}\left[e^{\frac{\sigma^{2}\gamma^{2}}{2}\int_{t}^{T}\left(\pi_{s}^{t,S}\right)^{2}ds+
     \frac{\upsilon^{2}\gamma^{2}}{2}\int_{t}^{T}\left(\pi_{s}^{t,P}\right)^{2}ds}\right]<\infty.
\end{equation}
\end{remark}

\section{HJB formulation}\label{sec:HJB}
This section is devoted to {formulating} the HJB equation.
Given that the filter probability process $p_s=(p_s^{1},\dots,p_s^{N})$ is degenerate in $\mathds{R}^N$, we consider the projected $N-1$ dimensional process
\[
     \tilde{p}_{s}:=(\tilde{p}^{1}_{s},\dots,\tilde{p}^{N-1}_{s})^{\top}:=(p^{1}_{s},\dots,p^{N-1}_{s})^{\top},
\]
as opposed to the actual filtering process. Next, we rewrite the problem (\ref{eq:probreform}) in terms of the above process, which now lies in the space
$$
    {\tilde{\Delta}_{N-1} = \left \{(d^1,\dots,d^{N-1}): \; {d^1 + \cdots + d^{N-1} < 1}, \; {d^i > 0} \right \}},
$$
in view of the Lemma \ref{Rem:TryPrvRmnPs} below.
Let us start with some notation needed to write the SDE of $\tilde{p}$ in matrix form.
{First, similarly to {(\ref{eq:optproj}) and (\ref{eq:optproj2}), given a vector  $l=(l_{1},\dots,l_{N})\in\mathbb{R}^{N}$ and a} function $g: D \to \mathbb{R}$, where $D = D_1 \times \{e_{1},\dots,e_{N}\} \times D_2$, with $D_1$ and $D_2$ arbitrary, possibly empty domains, define {the mappings
$\tilde{g}: D_1 \times\tilde{\Delta}_{N-1}\times D_2 \to\mathbb{R}$ and $\tilde{l}:\tilde\Delta_{N-1}\to\mathbb{R}$ as follows:
\begin{equation}\label{eq:HndNtWtpv2}
    \tilde{g}(y,d,z) =g(y,e_N,z)+ \sum_{i=1}^{N-1} [g(y,e_i,z)-g(y,e_N,z)] d^{i}, \qquad
    \tilde{l}(d):=l_{N}+\sum_{i=1}^{N-1}[l_{i}-l_{N}]d^{i},
\end{equation}
for $d=(d^1,\dots,d^{N-1}) \in\tilde{\Delta}_{N-1}$ and $y \in D_1, z \in D_2$.}}
{\CadBlue The following relationships are useful in what follows. For $y \in D_1, z \in D_2$,
\begin{equation}\label{UsflRel2}
    {\hat{g}(y,p_{t},z)=\tilde{g}(y,\tilde{p}_{t},z), \qquad \hat{l}(p_{t})=\tilde{l}(\tilde{p}_{t}).}
\end{equation}
Throughout,} the projection of a vector $l=(l_{1},\dots,l_{N})$ on the first $N-1$ coordinates is denoted by $l^{\pr}:=(l_{1},\dots,l^{N-1})$. {\CadBlue Similarly, for a given matrix $B$, we use $B^{\pr}$ to denote
the projection on the submatrix consisting of the first $N-1$ columns. Hence,
\begin{align*}
    {\loga(t)^{\pr}} &:= (\loga(t,e_1),\ldots,\loga(t,e_{N-1}))={\left(\begin{array}{ccc}
    \mu_{1} - \frac{\sigma^2}{2},&\dots&,\mu_{N-1} - \frac{\sigma^2}{2}\\
     a(t,e_{1}) -\frac{\upsilon^2}{2},
     &\dots &,
      a(t,e_{N-1}) -\frac{\upsilon^2}{2}\end{array}\right)}.
\end{align*}
We} use ${{\rm Diag}}({\bf b})$ to denote the diagonal matrix, whose {$i^{th}$} diagonal element is the $i^{th}$ component of the vector ${\bf b}$.
{\CadBlue
Further, let $\beta_{\varpi}(t,\tilde{p}_{t})$ be the $(N-1)\times 1$ vector defined by}
\begin{align}
    \beta_{\varpi}(t,\tilde{p}_t) &= \left( \varpi_{N,1}(t)+\sum_{\ell=1}^{N-1} [\varpi_{\ell,1}(t) -\varpi_{N,1}(t)]
    \tilde{p}^{\ell}_{t},\dots, \varpi_{N,N-1}(t)+\sum_{\ell=1}^{N-1} [\varpi_{\ell,N-1}(t) -\varpi_{N,N-1}(t)]
    \tilde{p}^{\ell}_{t}\right)^{\top}.
\label{eq:tildeBdef}
\end{align}
Finally, we also use $\mathbf{1}$ to denote the $N-1$ dimensional column vector whose entries are all ones.

In what follows, we work with the collection of processes  $\{p^{t}_{s}\}_{t\leq{}s\leq{}T}=\{(p^{t,1}_{s},\dots,p^{t,N}_{s})^{\top}\}_{t\leq{}s\leq{}T}$ {constructed on a suitable
probability space $(\Omega^{t},\G^{t},\Gx^{t},\Px^{t})$ as described in Remark \ref{Exist2}}.
 Using Eq.~\eqref{eq:ptiptildeb}, the dynamics of the vector process $\tilde{p}^{t}_s:=(\tilde{p}^{t,1}_{s},\dots,\tilde{p}^{t,N-1}_{s}):=
({p}^{t,1}_{s},\dots,{p}^{t,N-1}_{s})$ under $\widetilde{\Px}^{t}$ may be rewritten as
\begin{align}
\nonumber d\tilde{p}^{t}_{s} &= {\rm Diag}(\tilde{p}^{t}_s) \left({\loga(s)^{\pr}} - \mathbf{1} \tilde\loga(s,\tilde{p}^{t}_s) \right)^{\top} {\Sigma_Y^{-1}}d\tilde{W}^{t}_s + {{\CadBlue \beta_{\varpi}}(s,\tilde{p}^{t}_{s}) ds}  \\
\nonumber & \quad+ \gamma {\rm Diag}(\tilde{p}^{t}_s) \left(\loga(s)^{\pr} - {\mathbf{1}} \tilde{\loga}(s,\tilde{p}^{t}_s) \right)^{\top}  {\pi_{s}^{t}}ds + {{\rm Diag}}(\tilde{p}^{t}_{s^{-}}) \frac{1}{\tilde{h}(\tilde{p}^{t}_{s^{-}})} \left(h^{\pr} - {\mathbf{1}} \tilde{h}(\tilde{p}^{t}_{s^{-}}) \right) {d\tilde\xi^{t}_{s}},\quad t<s\leq{}T,\\
\tilde{p}^{t}_{t}&=\tilde{p}^{\circ},\nonumber
\end{align}
where the initial value of the process is {$\tilde{p}^{\circ}:=(\tilde{p}^{\circ,1},\dots,\tilde{p}^{\circ,N-1})^{\top}=({p}^{\circ,1},\dots,{p}^{\circ,N-1})^{\top}$}.
Next, let us define
\begin{align*}
    {\kappa}(s,\tilde{p}^{t}_s) &:= 
{{\rm Diag}}(\tilde{p}^{t}_s) \left({\loga(s)^{\pr}} - \mathbf{1} \tilde\loga(s,\tilde{p}^{t}_s) \right)^{\top} \Sigma_Y^{-1}
\\
    {\beta}_{\gamma}(s,\tilde{p}^{t}_s,\pi_{s}^{t}) &:= {\beta_{\varpi}}(s,\tilde{p}^{t}_s) + \gamma {\kappa}(s,\tilde{p}^{t}_s) \Sigma_Y^{\top} \pi_s^{t},\\
    {\varrho}(\tilde{p}^{t}_{s^{-}})&:={{\rm Diag}}(\tilde{p}^{t}_{s^{-}}) \frac{1}{\tilde{h}(\tilde{p}^{t}_{s^{-}})} \left(h^{\pr} - {\mathbf{1}} \tilde{h}(\tilde{p}^{t}_{s^{-}}) \right).
\end{align*}
Then,} the dynamics of {$\tilde{p}^{t}_s=(\tilde{p}^{t,1}_{s},\dots,\tilde{p}^{t,N-1}_{s})^{\top}$ for $s \in [t,T]$, under the probability measure {$\tilde{\Px}^{t}$}, is given by
\begin{equation}\label{eq:DynPortArIP}
 {d\tilde{p}^{t}_s = {\beta}_{\gamma}(s,\tilde{p}^{t}_s,\pi^{t}_s) ds + {\kappa}(s,\tilde{p}^{t}_s) d\tilde{W}^{t}_s + {\varrho}(\tilde{p}^{t}_{s^{-}}) {d\tilde\xi^{t}_{s}}}\quad (t<s\leq T),\qquad
\tilde{p}^{t}_t = \tilde{p}^{\circ}\in\tilde\Delta_{N-1}.
\end{equation}
A similar expression may be written for the vector {$\tilde{p}_s:=(\tilde{p}^{1}_{s},\dots,\tilde{p}^{N-1}_{s})^{\top}:=
({p}^{1}_{s},\dots,{p}^{N-1}_{s})^{\top}$ solving Eq.~\eqref{eq:ptiptilde}, which lives in the ``real world" space $(\Omega,\G,\Gx,\Px)$ and starts at time $0$.}}

Note that {$\pi^{t}$} affects the evolution of {$\{\tilde{p}^{t}_s\}_{t\leq{}s\leq{}T}$} through the drift ${\beta}_{\gamma}$, and also through the measure {$\tilde{\Px}^{t}$} due to an
admissibility constraint analogous to (\ref{Eq:AdPrbMes}).
{As explained in Remark \ref{Exist2}, the condition~\eqref{Eq:AdPrbMes} is satisfied provided that~\eqref{NovCnd2bbc} holds true. In light of these observations, the following class of admissible controls is natural.}
\begin{definition}\label{def:admiss}
The class of admissible strategies $\mathcal{A}(t,T;\tilde{p}^{\circ},z^{\circ})$ consists of locally bounded feedback trading strategies
$$
\pi_{s}^{t}:= \left(\pi_{s}^{t,S}, \pi_{s}^{t,P}\right) = \left( \pi^{t,S}(s,\tilde{p}^{t}_{s^{-}},H^{t}_{s^{-}}), \pi^{t,P}(s,\tilde{p}^{t}_{s^{-}},H^{t}_{s^{-}}) \right)
$$
for $t<s\leq T$ and $\pi_{t}^{t}:=\left(\pi_{t}^{t,S}, \pi_{t}^{t,P}\right) = \left(\pi^{t,S}(t,\tilde{p}^{t}_{t},H^{t}_{t}),\pi^{t,P}(t,\tilde{p}^{t}_{t},H^{t}_{t})\right)
 = \left(\pi^{S}(t,\tilde{p}^{\circ},z^{\circ}),\pi^{P}(t,\tilde{p}^{\circ},z^{\circ}) \right)$, satisfying
\begin{equation}\label{KCnAdm}
    {\Ex^{\Px^{t}}\left[\exp\left(\frac{\sigma^{2}\gamma^{2}}{2}\int_{t}^{T}\left(\pi^{t,S}\left(s,\tilde{p}^{t}_{s-},H^{t}_{s-} \right)\right)^{2}ds+
     \frac{\upsilon^{2}\gamma^{2}}{2}\int_{t}^{T}\left(\pi^{t,P}\left(s,\tilde{p}^{t}_{s-}, H^{t}_{s-} \right)\right)^{2}ds\right)\right]<\infty}.
\end{equation}
so to guarantee that the measure change defined by~\eqref{TrnsFromHatPtoTildeP} is well defined.
\end{definition}


{Let us now define the dynamic programming problem associated with our original utility maximization problem. For each $t\in[0,T)$, $\tilde{p}^{\circ}\in\tilde{\Delta}_{N-1}$, $z^{\circ}\in\{0,1\}$, and Markov strategy {$\pi^{t}\in\mathcal{A}(t,T;\tilde{p}^{\circ},z^{\circ})$}, we set
\begin{equation}
G(t,\tilde{p}^{\circ},z^{\circ},\pi^{t}) := \Ex^{{\tilde{\Px}^{t}}} \left[e^{-\gamma \int_t^T \tilde{\eta}\left(s,\tilde{p}^{t}_s,\pi^{t}_s\right) ds} \right],
\end{equation}
where we recall that by construction $H^{t}_{t}=z^{\circ}$, $\tilde{p}^{t}$ is given as in (\ref{eq:DynPortArIP}), and $\tilde{\eta}$ is defined from $\eta$ in accordance to (\ref{eq:HndNtWtpv2}) as}
\[
    {\tilde{\eta}(s,\tilde{p}_{s}^{t},\pi_{s}^{t})=\eta(s,{e_{N},\pi_{s}^{t}})+\sum_{i=1}^{N-1}(\eta(s,{e_{i},\pi_{s}^{t}})-\eta(s,{e_{N},\pi_{s}^{t}}))\tilde{p}_{s}^{t,i}.}
\]
{Next, we} define the value function
\begin{equation}
{w(t,{\tilde{p}^{\circ},z^{\circ}}) := \sup_{{\pi^{t} \in \mathcal{A}(t,T;{\tilde{p}^{\circ},z^{\circ}})}} \log\left(G\left(t,{\tilde{p}^{\circ},z^{\circ}},\pi^{t}\right)\right)}.
\label{eq:valuefn}
\end{equation}
{The crucial step to link the above dynamic programming problem with our original problem is outlined next:
\begin{align*}
\sup_{{\pi \in \mathcal{A}(0,T;\tilde{p}^{\circ},z^{\circ})}} J(v;\pi;T)
& = \frac{v^{\gamma}}{\gamma}\sup_{\pi \in \mathcal{A}(0,T;\tilde{p}^{\circ},z^{\circ})} \Ex^{\tilde{\Px}}\left[e^{-\gamma \int_0^T {\hat{\eta}(s,p_s,\pi_s)} ds} \right]\\
&= \frac{v^{\gamma}}{\gamma}\sup_{\pi \in \mathcal{A}(0,T;\tilde{p}^{\circ},z^{\circ})} \Ex^{\tilde{\Px}}\left[e^{-\gamma \int_0^T {\tilde{\eta}(s,\tilde{p}_s,\pi_s)} ds} \right]\\
&=\frac{v^{\gamma}}{\gamma}\sup_{\pi^{0} \in \mathcal{A}(0,T;\tilde{p}^{\circ},z^{\circ})} \Ex^{{\widetilde{\Px}^{0}}} \left[e^{-\gamma \int_0^T \tilde{\eta}(s,\tilde{p}^{0}_s,\pi_s^{0}) ds} \right]\\
&=\frac{v^{\gamma}}{\gamma} e^{w(0,\tilde{p}^{\circ,1},\dots,\tilde{p}^{\circ,N-1},z^{\circ})},
\end{align*}
where the first and second equalities follow from Eqs.~\eqref{eq:probreform} and \eqref{UsflRel2}, while the third equality follows from the uniqueness of the strong solution to the system
(\ref{eq:dynptib}), which can be established {as noticed in Remark \ref{Step1Const}}.

{We now proceed to} derive the HJB equation corresponding to the value function {in Eq.~\eqref{eq:valuefn}.} Before doing so, we need to compute the generator $\mathcal{L}$ of the Markov process $s\in[t,T] \to {(s, \tilde{p}^{t}_s, H^{t}_s)}$. This is done in the following lemma. {Below and hereafter, we use ${\bf y} \cdot {\bf h}$ to denote componentwise multiplication of two vectors ${\bf y}$ and ${\bf h}$.}
\begin{lemma} \label{lem:geneKS}
Let $({\tilde{p}^{t}_{s}})_{s\in[t,T]}$ be the process in (\ref{eq:DynPortArIP}) with {$\pi^{t}$} of the form
\[
    {\pi^{t}_{s}:=\pi(s,\tilde{p}^{t}_{s^{-}},H^{t}_{s^{-}})\quad (t<s\leq{}T),\qquad
    {\pi_{t}^{t}}:=\pi(t,\tilde{p}^{\circ},z^{\circ}),}
\]
 for a suitable function $\pi(s,\tilde{p},z)$ such that (\ref{eq:DynPortArIP}) admits a unique {strong} solution. Then, {for any $f(t,\tilde{p},z)$ such that $f(t,\tilde{p},1)$ and $f(t,\tilde{p},0)$ are both $C^{1,2}$-functions, we} have
\begin{equation}\label{Eq:SemiMrtDcm1}
    {f(s,\tilde{p}^{t}_s,H^{t}_s) = f(t,\tilde{p}^{\circ},z^{\circ}) + \int_t^s \tilde{\mathcal{L}} f(u,\tilde{p}^{t}_u,H^{t}_u) du +  \tilde{M}_s(f),\qquad s\in (t,T],}
\end{equation}
where, denoting $\nabla_{\tilde{p}}f(t,\tilde{p},z):=(\frac{\partial f}{\partial \tilde{p}^{1}},\dots,\frac{\partial f}{\partial \tilde{p}^{N-1}})$, $f_{t}(t,\tilde{p},z):=\frac{\partial f}{\partial t}$, and $D^{2} f:=\left[\frac{\partial^{2} f}{\partial \tilde{p}^{i}\partial \tilde{p}^{j}}\right]_{i,j=1}^{N-1}$ and recalling the
notation $\tilde{h}(\tilde{p}):={h^{N}}+\sum_{i=1}^{N-1}(h_{i}-h_{N})\tilde{p}^{i}$ and $h^{\pr}:=(h_{1},\dots,h_{N-1})^{\top}$,
\begin{align*}
    \tilde{\mathcal{L}} f(t,\tilde{p},z) &:= f_t(t,\tilde{p},z) + \nabla_{\tilde{p}} f\, {\beta}_{\gamma}(t,\tilde{p},\pi(t,\tilde{p},z)) +
    \frac{1}{2} \text{tr}({\kappa} {\kappa}^{\top} D^2 f(t,\tilde{p},z))\\
    &\quad +  (1-z)\left(f\left({t} ,  \frac{1}{\tilde{h}(\tilde{p})}(\tilde{p}\cdot h^{\pr}),1\right)-f\left({t}, \tilde{p},0\right)\right)  \tilde{h}(\tilde{p}).
\end{align*}
Moreover, the {$\widetilde{\Px}^{t}$-local} martingale component is
\begin{equation}
{\tilde{M}_s(f) =  \int_t^s \nabla_{\tilde{p}}f \, {\kappa}(u,\tilde{p}^{t}_u) d \tilde{W}^{t}_u +{\int_{t}^{s}}
\left(f\left(u,  \frac{1}{\tilde{h}(\tilde{p}^{t}_{u-})}(\tilde{p}^{t}_{u^{-}}\cdot h^{\pr}),1\right)-f\left(u, \tilde{p}^{t}_{u-},0\right)\right) {d\tilde{\xi}^{t}_u}.}
\end{equation}
\end{lemma}
\proof
{For simplicity, throughout the proof we drop the superscript $t$ in the processes $\tilde{p}^{t}$, $\pi^{t}$, $\tilde{W}^{t}$, $\tilde{\xi}^{t}$, and $H^{t}$.}
Let $\tilde{p}^{c,i}$ denote the continuous component of $\tilde{p}^{i}$, determined by the first two terms on the right-hand side of Eq.~(\ref{eq:DynPortArIP}).
Using {It\^o{'}s} formula, we have
\begin{align}
\nonumber f(s,\tilde{p}_s,H_s) &= f(t,\tilde{p}_t,H_t) + \int_t^{s} f_{\CRed u}(u,\tilde{p}_u,H_u) du + \sum_{i=1}^{N-1} \int_t^s \frac{\pa f}{\pa \tilde{p}^i} d\tilde{p}_u^{c,i} + \frac{1}{2} \sum_{i,j=1}^{N-1} \int_{t}^{s}\frac{\pa^2 f}{\pa \tilde{p}^i \tilde{p}^j} d\left<\tilde{p}^{c,i}, \tilde{p}^{c,j}\right>_u \\
& \quad + \sum_{t< u \leq s} \left(f(u,\tilde{p}_u,H_u) - f(u,\tilde{p}_{u-},H_{u-}) \right).
\label{eq:fito}
\end{align}
Note that {the} size of the jump of ${\CRed \tilde{p}_t^i}$ at the default time $\tau$ is given by
\begin{equation}
    \tilde{p}_{\tau}^i - \tilde{p}_{\tau-}^i = \tilde{p}_{\tau-}^i \frac{h_i - \tilde{h}(\tilde{p}_{\tau-})}{\tilde{h}(\tilde{p}_{\tau-})},
\end{equation}
{thus implying} that $\tilde{p}_{\tau}^i = \tilde{p}_{\tau-}^i h_i/\tilde{h}(\tilde{p}_{\tau-})$ and $\tilde{p}_{\tau}= (1/\tilde{h}(\tilde{p}_{\tau-}))(\tilde{p}_{\tau^{-}}\cdot h^{\pr})$. {For $t<\tau\leq s$}, this leads to
\begin{align*}
\nonumber \sum_{t< u \leq s}\left( f(u,\tilde{p}_u,H_u) - f(u,\tilde{p}_{u-},H_{u-})\right) &= \left[f\left(\tau,  \frac{1}{\tilde{h}(\tilde{p}_{\tau-})}(\tilde{p}_{\tau^{-}}\cdot h^{\pr}),1\right)-f\left(\tau, \tilde{p}_{\tau-},0\right)\right] (H_{s}-H_{t}) \\
&= \int_t^s \left(f\left(u,  \frac{1}{\tilde{h}(\tilde{p}_{u-})}(\tilde{p}_{u^{-}}\cdot h^{\pr}),1\right)-f\left(u, \tilde{p}_{u-},0\right)\right) dH_{u} \\
&= \int_t^s \left(f\left(u,  \frac{1}{\tilde{h}(\tilde{p}_{u-})}(\tilde{p}_{u^{-}}\cdot h^{\pr}),1\right)-f\left(u, \tilde{p}_{u-},0\right)\right)  \left({d\tilde{\xi}_u} + {\CaddBlue \bar{H}_{u^{-}}} \tilde{h}(\tilde{p}_{u^{-}}) du\right),
\end{align*}
where in the last equality we have used Eq.~(\ref{Eq:DfnTildexi}) and the fact that $\hat{h}_{u}=\sum_{i=1}^{N}h_{i}p_{u}^{i}=h_{N}+\sum_{i=1}^{N-1}(h_{i}-h_{N})p_{u}^{i}=\tilde{h}(\tilde{p}_{u})$.
From this, we deduce that Eq.~\eqref{eq:fito} may be rewritten as
\begin{align}
\nonumber f(s,\tilde{p}_s,H_s) &= f(t,\tilde{p}_t,H_t) + \int_t^s f_{u}(u,\tilde{p}_u,H_u) du +
\int_{t}^{s}\nabla_{\tilde{p}} f\, {{\beta}_{\gamma}(u,\tilde{p}_u,\pi_u)} du + \frac{1}{2} \sum_{i,j=1}^{N-1} \int_t^s({\kappa} {\kappa}^{\top})_{ij} \frac{\pa^2 f}{\pa \tilde{p}^i \tilde{p}^j}(u,\tilde{p}_u,H_u) du \\
\nonumber &\quad+  \int_t^s \nabla_{\tilde{p}}f \, {{\kappa}(u,\tilde{p}_u)} d \tilde{W}_u +\int_{s}^{t}
\left(f\left(u,  \frac{1}{\tilde{h}(\tilde{p}_{u-})}(\tilde{p}_{u^{-}}\cdot h^{\pr}),1\right)-f\left(u, \tilde{p}_{u-},0\right)\right) {d\tilde{\xi}_u}  \\
&\quad+ \int_t^s \left(f\left(u,  \frac{1}{\tilde{h}(\tilde{p}_{u})}(\tilde{p}_{u}\cdot h^{\pr}),1\right)-f\left(u, \tilde{p}_{u},0\right)\right)   (1-H_{u})\tilde{h}(\tilde{p}_{u}) du,
\label{eq:fitorew}
\end{align}
which proves the lemma.
\endproof

We are now ready to derive the HJB equation associated to the control problem. We first obtain it based on standard heuristic arguments, and then in the subsequent section we provide
rigorous verification theorems for the solution. In light of the dynamic programming principle, we expect that, for {any $s\in(t,T]$},
\begin{equation}
    {w(t,\tilde{p}^{\circ},z^{\circ}) = \sup_{\pi^{t} \in \mathcal{A}(t,T;\tilde{p}^{\circ},z^{\circ})} \log \Ex^{{\widetilde{\Px}^{t}}}\left[e^{w(s,\tilde{p}^{t}_s,H^{t}_s) - \gamma \int_t^s \tilde{\eta}(u,\tilde{p}^{t}_u,\pi^{t}_u) du} \right]}.
\label{eq:vf}
\end{equation}
{with $\tilde{p}^{t}_t = \tilde{p}^{\circ}$, and $H^{t}_t = z^{\circ}$.}
Next, define $\varepsilon({s},\tilde{p},z) = e^{w({s},\tilde{p},z)}$ and note that, in light of  Lemma \ref{lem:geneKS},
\begin{equation*}
    {\varepsilon(s,\tilde{p}^{t}_s,H^{t}_s) = \varepsilon(t,\tilde{p}^{\circ},z^{\circ}) + \int_t^s \tilde{\mathcal{L}} \varepsilon(u,\tilde{p}^{t}_u,H^{t}_u) du +
\tilde{M}_{s}(\varepsilon)},
\end{equation*}
where the last term $\tilde{M}_{s}(\varepsilon)$ represents the local martingale component of $\varepsilon(s,\tilde{p}^{t}_s,H^{t}_s)$.
Plugging the previous equation into \eqref{eq:vf}, we {expect the following relation to hold:}
\begin{align}
\nonumber {0  =  \sup_{\pi^{t} \in \mathcal{A}(t,T;\tilde{p}^{\circ},z^{\circ})} \Ex^{{\widetilde\Px^{t}}} \bigg[\varepsilon(t,\tilde{p}^{\circ},H^{\circ}) \left(e^{-\gamma \int_t^s \tilde{\eta}(u,\tilde{p}^{t}_u,\pi^{t}_u)du} -1 \right)
                + e^{-\gamma \int_t^s \tilde{\eta}(u,\tilde{p}^{t}_u,\pi^{t}_u) du} \int_t^s \tilde{\mathcal{L}} \varepsilon(u,\tilde{p}^{t}_u,H^{t}_u) du \bigg]},
\label{eq:rewhjb}
\end{align}
{assuming} that the local martingale component is a true martingale. Dividing by $s-t$ and taking the limit of the above expression as $s \rightarrow t$ {leads us to the HJB equation:}
\begin{equation}\label{Eq:GenHJBEpsilon}
{0=\sup_{\pi} \left[\left(\tilde{\mathcal{L}} - \gamma \tilde{\eta}(t,\tilde{p}^{\circ},\pi)\right) \varepsilon(t,\tilde{p}^{\circ},z^{\circ}) \right]}.
\end{equation}
Let us write (\ref{Eq:GenHJBEpsilon}) in terms of $w$.
%
To this end, let us denote the differential component of $\tilde{\mathcal{L}}$ as $\tilde{\mathcal{D}}$; i.e.,
\[
    \tilde{\mathcal{D}} f(t,\tilde{p},z):= f_t(t,\tilde{p},z) + \nabla_{\tilde{p}} f(t,\tilde{p},z)\, {{\beta}_{\gamma}(t,\tilde{p},\pi(t,\tilde{p},z))} +
    \frac{1}{2} \text{tr}({\kappa} {\kappa}^{\top} D^2 f(t,\tilde{p},z)).
\]
Then, we note that
\begin{align}\nonumber
     \tilde{\mathcal{L}}\varepsilon(t,\tilde{p},z)&=\tilde{\mathcal{D}}\varepsilon(t,\tilde{p},z)+(1-z) \tilde{h}(\tilde{p})\left( e^{w \left(t,\frac{1}{\tilde{h}(\tilde{p})}\tilde{p}\cdot h^{\pr},1\right)}-e^{w \left(t,\tilde{p},0\right)} \right)\\
    &=e^{w(t,\tilde{p},z)}\left(\tilde{\mathcal{D}}w+ \frac{1}{2}  \|{\nabla_{\tilde{p}} w \, \kappa} \|^2+(1-z) \tilde{h}(\tilde{p}) \left[ e^{w \left(t,\frac{1}{\tilde{h}(\tilde{p})}\tilde{p}\cdot h^{\pr},1\right)-w \left(t,\tilde{p},0\right)}-1 \right]\right).\label{Eq:ApplGenexpw}
\end{align}
Thus, Eq.~(\ref{Eq:GenHJBEpsilon}) takes the form:
\begin{equation}
{0 = \sup_{\pi} \left[\varepsilon(t,\tilde{p}^{\circ},z^{\circ}) \left(\tilde{\mathcal{D}}w+ \frac{1}{2} \|{\nabla_{\tilde{p}} w \, \kappa}\|^2+(1-z^{\circ}) \tilde{h}(\tilde{p}^{\circ})\left[ e^{w \left(t,\frac{1}{\tilde{h}(\tilde{p}^{\circ})}\tilde{p}^{\circ}\cdot h^{\pr},1\right)-w \left(t,\tilde{p}^{\circ},0\right)}-1 \right] - \gamma \tilde{\eta}(t,\tilde{p}^{\circ},z^{\circ},\pi)\right) \right]}.
\label{eq:HJBalm}
\end{equation}
In order to get a more explicit form, let us recall that
\[
    {\eta(t,e_{i},\pi)} = -r + \pi^S(r-\left<\mu,e_{i}\right>) + \pi^P \left({r-a(t,e_{i})} \right) + \frac{1-\gamma}{2} \pi^{\top} {\Sigma^{\top}_{Y}} {\Sigma_{Y}} \pi{,}
\]
 and note that
\begin{align}
    {\tilde{\eta}(t,\tilde{p},\pi)}&=\eta(t,e_{N},\pi(t,\tilde{p},z))+\sum_{i=1}^{N-1}\left(\eta(t,e_{i},\pi(t,\tilde{p},z))-\eta(t,e_{N},\pi(t,\tilde{p},z))\right)\tilde{p}^{i} \nonumber \\
    &=-r + \pi^S(r-\tilde{\mu}(\tilde{p})) +  {\pi^P\left({r- \tilde{a}(t,\tilde{p})} \right)}+ \frac{1-\gamma}{2} \pi^{\top} {\Sigma^{\top}_{Y}} {\Sigma_{Y}} \pi.
    \label{Eq:DfnSmptldeta}
\end{align}
We can now rewrite Eq.~\eqref{eq:HJBalm} {as
\begin{eqnarray}
\nonumber & & \frac{\pa w}{\pa t} + \frac{1}{2} \text{tr}({\kappa} {\kappa}^{\top} D^2 w) + \frac{1}{2} {(\nabla_{\tilde{p}} w ){\kappa} {\kappa}^{\top} ({\nabla_{\tilde{p}}} w)^{\top}} + \gamma r +
(1-z^{\circ}) \tilde{h}(\tilde{p}^{\circ})\left[ e^{w \left(t,\frac{1}{\tilde{h}(\tilde{p}^{\circ})}\tilde{p}^{\circ}\cdot h^{\pr},1\right)-w \left(t,\tilde{p}^{\circ},0\right)}-1 \right]  \\
& & +\sup_{\pi} \bigg \{{(\nabla_{\tilde{p}} w) {\beta}_{\gamma}} - \gamma \pi^S (r-\tilde{\mu}(\tilde{p}^{\circ})) - (1-z^{\circ}) \gamma \pi^P {\BRed( r - \tilde{a}(t,\tilde{p}^{\circ}))} - \frac{1}{2} \gamma (1-\gamma) \pi^{\top} \Sigma_Y^{\top} \Sigma_Y \pi\bigg \} = 0,
\label{eq:finalHJB}
\end{eqnarray}
with} terminal condition $w(T,\tilde{p}^{\circ},z^{\circ}) = 0$ {and where all the derivatives of $w$ are evaluated at $(t,\tilde{p}^{\circ},z^{\circ})$.}

Depending on whether or not default has occurred, we will have two separate optimization problems to solve. Indeed, after default has occurred, the investor cannot invest in the {\BRed defaultable security} and only allocates his wealth in the stock and risk-free asset. {The next section analyzes in detail the two cases.}

\section{Solution to the Optimal Control Problem} \label{sec:ver}
We analyze the control {problems} developed in the previous section. We first decompose it into two related optimization {subproblems:} the post and the pre-default problems. As we will demonstrate, in order to solve the pre-default optimization subproblem, we need the solution of the post-default one.
{Before proceeding further, we recall some functional spaces which will be needed for the following proofs. {\Red We set $\mathcal{C}_{P}^{2,\alpha}$ as the set of functions locally in $C_{P}^{2,\a}\left((0,T)\times\tilde{\Delta}_{N-1}\right)\cap {C\left([0,T]\times\tilde{\Delta}_{N-1}\right)}$,}
where we recall that for a given domain $D$ of {\CRed $\rn$} and $\a\in(0,1)$,
the parabolic H\"older space $C_{P}^{2,\a}(D)$ is defined by the following norms
\begin{equation}\label{DfnC2alpah}
    \|\psi\|_{C_{P}^{2,\a}(D)}=\|\psi\|_{C_{P}^{\a}(D)} +{\|\p_{t}\psi\|_{C_{P}^{\a}(D)}}+
    \sum_{i=1}^{N-1}\|\p_{\tilde{p}_i}\psi\|_{C_{P}^{\a}(D)} +
    \sum_{i,j=1}^{N-1}\|\p_{\tilde{p}_i
    \tilde{p}_j}\psi\|_{C_{P}^{\a}(D)}
\end{equation}
with
 \begin{equation*}
    {\|\psi\|_{C_{P}^{\a}(D)}:=\sup_{(t,\tilde{p})\in D}|\psi(t,\tilde{p})|+ \sup_{\stackrel{(t,\tilde{p}),(t',\tilde{p}') \in D}{(t,\tilde{p}) \neq (t',\tilde{p}')}}
    \frac{|\psi(t,\tilde{p}) - \psi(t',\tilde{p}')|}{\left(|\tilde{p}-\tilde{p}'|^{2}+|t-t'|\right)^{\frac{\a}{2}}}}.
 \end{equation*}
Further, we denote
\begin{equation}\label{DfnC1alpah}
    \|\psi\|_{C_{P}^{1,\a}(D)}=\|\psi\|_{C_{P}^{\a}(D)} +
    \sum_{i=1}^{N-1}\|\p_{\tilde{p}_i}\psi\|_{C_{P}^{\a}(D)}{\Blue .}
\end{equation}

\subsection{Post Default Optimization Problem}
Assume that default has already occurred; i.e., we are at a time $t$ so that $\tau < t$. In particular, this means that $\pi_t^{P}=0$.
Let us denote by {$\wpost(t,\tilde{p}):=w(t,\tilde{p},1)$} the value function in the post-default optimization problem. Then, we may rewrite Eq.~\eqref{eq:finalHJB} as follows:
\begin{eqnarray}
\nonumber 0 &=& {\wpost_t} + \frac{1}{2} \text{tr}(\underline{\kappa} \underline{\kappa}^{\top} D^2  \wpost) + \frac{1}{2} {(\nabla_{\tilde{p}} \wpost) \underline{\kappa} \underline{\kappa}^{\top} (\nabla_{\tilde{p}} \wpost)^{\top}} + \gamma r \\
\label{eq:HJBpost} & & + \sup_{{\pi^{S}}} \left[{(\nabla_{\tilde{p}} w)\underline{\beta}_{\gamma}} - \gamma {\pi^S} (r- {\tilde{\mu}(\tilde{p})}) - \frac{{\sigma^2}}{2} \gamma (1-\gamma) ({\pi^S})^2 \right], \\
\label{eq:termpost} \wpost(T,{\tilde{p}}) &=& 0.
\end{eqnarray}
Here, $\underline{\kappa}(\tilde{p})$ is a $(N-1) \times 1$ vector determined by the first column of  {${\kappa}(t,\tilde{p})$} (the second column of {${\kappa}(t,\tilde{p})$} consists of all zeros). Concretely,
\begin{equation}\label{Eq:DfnUnderAlpha}
    \underline{\kappa}(\tilde{p})  := \left[{{\rm Diag(\tilde{p})}} \left( {\underline{\loga}^{\top}} - {\mathbf 1} \underline{\loga}(\tilde{p}) \right) \right] \frac{1}{\sigma},
\end{equation}
where $\underline{\loga} = (\mu_1 - \frac{1}{2} \sigma^2, \ldots, \mu_{N-1} - \frac{1}{2} \sigma^2)$ is the first row of {$\loga(t)^{\pr}$} (the second row consists of all zeros) and, correspondingly,
{$\underline{\loga}(\tilde{p}) = \tilde{\mu}(\tilde{p}) - \frac{1}{2} \sigma^2$} is a scalar with $\tilde{\mu}(\tilde{p})=\mu_{N}+\sum_{i=1}^{N-1}(\mu_{i} -  \mu_{N}) \tilde{p}^{i}$.
Similarly, $\underline{\beta}_{\gamma}(t,\tilde{p},\pi)$ in (\ref{eq:HJBpost}) is defined as
\begin{equation*}
\underline{\beta}_{\gamma}(t,\tilde{p},\pi) := {\beta_{\varpi}}(t,\tilde{p}) + \gamma  \sigma \pi^S {\underline{\kappa}(\tilde{p})} ,
\end{equation*}
{\CadBlue where we recall that {\CadBlue $\beta_{\varpi}(t,\tilde{p})$} has been defined in Eq.~\eqref{eq:tildeBdef}.}
It can easily be checked that the maximizer of Eq.~\eqref{eq:HJBpost} is given by
\begin{equation}
    {{\pi^S}(t,\tilde{p}) = \frac{1}{\sigma^{2}(1-\gamma)} \left\{\tilde{\mu}(\tilde{p})- r + \sigma{(\nabla_{\tilde{p}}  \wpost{\CadBlue(t,\tilde{p})}) {\CadBlue \underline{\kappa}(\tilde{p})}}  \right\}}.
\label{eq:pits}
\end{equation}
Plugging the {maximizer~\eqref{eq:pits}} in~\eqref{eq:HJBpost}, we obtain
\begin{eqnarray}\label{Eq:PstDfltDP}
{\wpost_t} + \frac{1}{2} \text{tr}(\underline{\kappa} \underline{\kappa}^{\top} D^2 \wpost) + \frac{1}{2(1-\gamma)} {(\nabla_{\tilde{p}} \wpost) \underline{\kappa}\, \underline{\kappa}^{\top} (\nabla_{\tilde{p}} \wpost)^{\top}}+ {(\nabla_{\tilde{p}} \wpost)\underline{\Phi}} + \underline{\Psi} &=& 0 ,  \\
{\wpost(T,\tilde{p})} &=& 0, \label{Eq:PstDfltDPBC}
\end{eqnarray}
where
\begin{align*}
\underline{\Phi}(t,\tilde{p}) &= {\CadBlue \beta_{\varpi}}(t,\tilde{p}) + \frac{\gamma}{1-\gamma} \frac{\tilde{\mu}(\tilde{p})-r}{\sigma}\, {\underline{\kappa}(\tilde{p})} ,  \\
{\CadBlue \underline{\Psi}}(\tilde{p}) &= \gamma r + \frac{\gamma}{2 (1-\gamma)} \left(\frac{\tilde{\mu}(\tilde{p})-r}{\sigma} \right)^2.
\end{align*}
Next we state, without proof, a useful result as a lemma.
\begin{lemma}\label{lem:classpost}[Theorem 3.1 in \cite{Tamura}]
For any ${\BRed T \geq 0}$, there exists a classical solution $\underline{w}$ that solves the Cauchy problem
(\ref{Eq:PstDfltDP})-(\ref{Eq:PstDfltDPBC}).
\end{lemma}

\begin{remark}\label{rem:sol}
\cite{Tamura} show the existence of a classical solution on the extension of the simplex to $\R^{N-1}$. Then, they prove that that if $\tilde{p} \in \tilde{\Delta}_{N-1}$,
the solution coincides with the solution to the Cauchy problem (\ref{Eq:PstDfltDP})-(\ref{Eq:PstDfltDPBC}). It is well known from standard results, see \cite{Friedman}, Theorem 1, pag.92, that such {\Blue a} solution is $\mathcal{C}_{P}^{2,\alpha}$. We will use this fact in our subsequent proofs.
\end{remark}
We then have
\begin{theorem}\label{MntPstDftVal}
The following assertions hold true:
\begin{enumerate}
    \item[{\rm (1)}] The solution ${\underline{w}}(t,\tilde{p})$ coincides with the value function $w(t,\tilde{p},1)$ introduced in (\ref{eq:valuefn}).
     \item[{\rm (2)}] The optimal feedback control $\{\pi_s^S\}_{s\in[t,T)}$, denoted by
$\widetilde{\pi}_s^S$, can be written as $\widetilde{\pi}_s^S = {\widetilde{\pi}^{S}(s,{\tilde{p}^{t}_{s}})}$ with
\begin{equation}
    \widetilde{\pi}^S(s,\tilde{p}):= \frac{1}{\sigma^{2}(1-\gamma)} \left(\tilde{\mu}(\tilde{p})- r + \sigma{\nabla_{\tilde{p}}  \wpost(s,\tilde{p}){\underline{\kappa}(\tilde{p})}}  \right).
\label{eq:optpi}
\end{equation}
{\CRed Moreover, the feedback trading strategy $\widetilde{\pi}_{s}:=(\tilde{\pi}^{S}_{s},\tilde{\pi}^{P}_{s})^{\top}$, $\tilde{\pi}^{P}_{s} := 0$, is admissible{\Blue ; i.e.,} it
satisfies the conditions of Definition~\ref{def:admiss}}.
\end{enumerate}
\end{theorem}
The  proof of Theorem \ref{MntPstDftVal} is reported in Appendix \ref{app:verifications}. For now, let us mention a few useful remarks about
{the solution $\underline{w}$ of (\ref{Eq:PstDfltDP})-(\ref{Eq:PstDfltDPBC}). The existence of the solution $\underline{w}$} follows from the Feynman-Kac
formula as outlined in, e.g., the proof of Theorem 3.1 in \cite{Tamura} (see also \cite{NagaiRung}). {More specifically,} the idea therein is to transform the problem into a
linear PDE via the Cole-Hopf transformation:
\begin{equation}\label{HpfTran0}
    \underline{\psi}(t,\tilde{p}) = e^{\frac{1}{1-\gamma} \underline{w}(t,\tilde{p})}.
\end{equation}
Then, it follows that $\underline{w}(t,\tilde{p})$ solves Eq.~(\ref{Eq:PstDfltDP})-(\ref{Eq:PstDfltDPBC}) if and only if {$\underline{\psi}(t,\tilde{p})$} solves the linear PDE
\begin{eqnarray}
\nonumber \frac{\pa {\underline{\psi}}}{\pa t} + \frac{1}{2} \text{tr}(\underline{\kappa} \underline{\kappa}^{\top} D^2 \underline{\psi} ) + \underline{\Phi}^{\top} \nabla_{\tilde{p}} {\underline{\psi}} + \frac{\underline{\Psi}}{1-\gamma} \underline{\psi} &=& 0, \\
\underline{\psi}(T,\tilde{p}) &=& 1.
\label{eq:transpost}
\end{eqnarray}

\subsection{Pre Default Optimization Problem} \label{sec:pre}
Assume that $\tau > t$, i.e. default has not occurred by time $t$. Let us denote by {$\wpre(t,\tilde{p}):=w(t,\tilde{p},0)$} the value function {\CadBlue corresponding to the} pre-default optimization problem. Then, we may rewrite
Eq.~\eqref{eq:finalHJB} as
\begin{eqnarray}
\nonumber & & \wpre_t + \frac{1}{2} \text{tr}(\bar{\kappa} \bar{\kappa}^{\top} D^2 \wpre) + \frac{1}{2} {(\nabla_{\tilde{p}} \bar{w} ){\bar{\kappa}} {\bar{\kappa}}^{\top} ({\nabla_{\tilde{p}}} \bar{w})^{\top}} + \\
 \nonumber & & \sup_{\pi=(\pi^{S},\pi^{P})} \bigg \{{(\nabla_{\tilde{p}} \bar{w}) {\CadBlue \bar{\beta}_{\gamma}}} - \gamma {\pi^S} (r-\tilde{\mu}(\tilde{p})) - \gamma {\pi^P} \left( {\BRed r - \tilde{a}(t,\tilde{p})} \right) - \frac{1}{2} \gamma (1-\gamma) {\pi^{\top}} \Sigma^{\top}_Y \Sigma_Y {\pi} \bigg \}  \\
 \label{eq:preHJB} & & + {\tilde{h}(\tilde{p})} {\left[ e^{\wpost \left(t, \frac{1}{\tilde{h}(\tilde{p})} \tilde{p} \cdot h^{\pr}\right) -\wpre \left(t,\tilde{p}\right)}-1 \right]} + \gamma r = 0, \\
 \nonumber & & \wpre(T,\tilde{p}) = 0.
\end{eqnarray}
 Above, {$\bar{\kappa}(t,\tilde{p})$} is a $(N-1) \times 2$ matrix given by
\begin{equation*}
\bar{\kappa}(t,\tilde{p}) := {\kappa(t,\tilde{p})} = \left[{\rm Diag}(\tilde{p}) \left( \bar{\loga}(t)^{\top} - {\mathbf 1} \bar{\loga}(t,\tilde{p})^{\top} \right) \right] \left( \Sigma_Y \Sigma_Y^{\top} \right)^{-1} \Sigma_Y,
\end{equation*}
 with $\bar{\loga}(t) = \loga(t)^{\pr}$ being a $2 \times (N-1)$ matrix and
{$\bar{\loga}(t,\tilde{p}):=\tilde{\loga}(t,\tilde{p})$}. Further,
\begin{equation*}
{\CadBlue \bar{\beta}_{\gamma}(t,\tilde{p},\pi) := \beta_{\gamma}(t,\tilde{p},\pi)} = {\CadBlue \beta_{\varpi}}(t,\tilde{p}) + \gamma \bar{\kappa}(t,\tilde{p}) \Sigma_Y^{\top} \pi.
\end{equation*}
{It is important to point out the explicit appearance of the solution $\wpost$ to the HJB post default equation in the PDE (\ref{eq:preHJB}) satisfied by the pre-default HJB equation $\wpre$.
This establishes the required relationship between pre and post-default optimization subproblems.}

Next, define
$$
\Upsilon(t,\tilde{p}) = {\left(r-\tilde{\mu}(\tilde{p}), {\BRed r- \tilde{a}(t,\tilde{p})} \right)^{\top}}.
$$
Then, we can rewrite Eq.~\eqref{eq:preHJB} as
\begin{eqnarray}
\nonumber & & \wpre_t + \frac{1}{2} \text{tr}(\bar{\kappa} \bar{\kappa}^{\top} D^2 \wpre ) + \frac{1}{2} {(\nabla_{\tilde{p}} \wpre) \bar{\kappa}\bar{\kappa}^{\top} (\nabla_{\tilde{p}} \wpre)^{\top}} + \\
\label{eq:HJBrewr}  & & \sup_{\pi} \bigg \{{(\nabla_{\tilde{p}} \wpre) {\CadBlue \bar{\beta}_{\gamma}}}- \gamma \pi^{\top} \Upsilon - \frac{1}{2} \gamma (1-\gamma) \pi^{\top} \Sigma^{\top}_Y \Sigma_Y \pi \bigg \} +
 {\tilde{h}({\CRed \tilde{p}})} {\left[ e^{\wpost \left(t, \frac{1}{\tilde{h}({\CRed \tilde{p}})} {\CRed \tilde{p}} \cdot h^{\pr}\right) -\wpre \left(t,{\CRed \tilde{p}}\right)}-1 \right]} + \gamma r= 0 \\
\nonumber & & \wpre(T,\tilde{p}) = 0.
\end{eqnarray}
Using the first order condition, we obtain {that} the {maximal point} {$\pi^{*}$} is the solution of the following equation{:}
$$
\gamma {\Sigma_Y} \bar{\kappa}^{\top} {(\nabla_{\tilde{p}} \bar{w})^{\top}} - \gamma \Upsilon - \gamma (1-\gamma) \Sigma_Y^{\top} \Sigma_Y {\pi^{*}} = 0{.}
$$
Solving the previous equation  for $\pi^*$ yields:
\begin{equation}
{\pi^{*}} = \frac{1}{1-\gamma} (\Sigma_Y^{\top} \Sigma_Y)^{-1} \left({-\Upsilon} + \Sigma_Y \bar{\kappa}^{\top} {(\nabla_{\tilde{p}} \bar{w})^{\top}} \right){.}
\label{eq:pimaximizer}
\end{equation}
After plugging {$\pi^{*}$} into~(\ref{eq:HJBrewr}), and performing algebraic simplifications {(see Appendix \ref{app:verifications} for details)}, we obtain
\begin{eqnarray}\label{eq:finHJB0}
 \wpre_t + \frac{1}{2} \text{tr}(\bar{\kappa} \bar{\kappa}^{\top} D^2 \wpre) + \frac{1}{2(1-\gamma)} {(\nabla_{\tilde{p}} \wpre) \bar{\kappa}\bar{\kappa}^{\top} (\nabla_{\tilde{p}} \wpre)^{\top}} + {(\nabla_{\tilde{p}} \wpre) \bar{\Phi}} + {\tilde{h}( {\CRed \tilde{p}}) e^{\wpost \left(t, \frac{{\CRed \tilde{p}} \cdot h^{\pr}}{\tilde{h}({\CRed \tilde{p}})} \right)}e^{-\wpre(t,{\CRed \tilde{p}} )}} + \bar{\Psi} = 0, & &  \\
 \wpre(T,{\CRed \tilde{p}} ) = 0, & &
\label{eq:finHJB0in} 
\end{eqnarray}
where
\begin{eqnarray}
\nonumber \bar{\Phi}(t,\tilde{p}) &=& {\CadBlue \beta_{\varpi}}(t,\tilde{p}) - \frac{\gamma}{1-\gamma} \bar{\kappa} \Sigma_Y^{-1} {\CadBlue \Upsilon(t,\tilde{p})},  \\
\nonumber \bar{\Psi}(t,\tilde{p}) &=& {\frac{1}{2}} \frac{\gamma}{1-\gamma} \Upsilon^{\top} (\Sigma_Y^{\top} \Sigma_Y)^{-1} {\CadBlue \Upsilon(t,\tilde{p})} + \gamma r - \tilde{h}(\tilde{p}).
\end{eqnarray}

Next, we prove the existence of a classical solution to the above Cauchy problem. We first
perform a similar transformation as in the post-default case, and obtain that the function
$\bar{w}(t,\tilde{p})$ solves the problem (\ref{eq:finHJB0}) if and only if the function
\begin{equation}\label{HpfTran1}
    {\bar{\psi}}(t,\tilde{p}) = e^{\frac{1}{1-\gamma} \bar{w}(t,\tilde{p})},
\end{equation}
solves the Cauchy problem
\begin{eqnarray}\label{eq:finHJB2}
 \nonumber \bar{\psi}_t(t,{\tilde{p}}) + \frac{1}{2} \text{tr}(\bar{\kappa} \bar{\kappa}^{\top} D^2 \bar{\psi}(t,{\tilde{p}})) +  \bar{\Phi}(t,{\tilde{p}}) \nabla_{\tilde{p}} \bar{\psi}(t,{\tilde{p}}) + \bar{\Psi}(t,{\tilde{p}})\frac{\bar\psi(t,{\tilde{p}})}
 {1-\gamma} +\tilde{h}({\tilde{p}}) e^{\wpost \left(t, \frac{{\tilde{p}} \cdot h^{\pr}}{\tilde{h}({\tilde{p}})} \right)}\frac{\bar\psi^{\gamma}(t,{\tilde{p}})}{1-\gamma} = 0 \\
 \bar\psi(T,{\tilde{p}}) =1.
\end{eqnarray}
{\CRed Notice that problem \eqref{eq:finHJB2} is {\it non-linear}. Hence, proving the existence of a
classical solution is not as direct as in the post-default case where the transformed HJB-PDE
given by Eq.~\eqref{eq:transpost} turned out to be linear. We establish this result by applying a
monotone iterative method used in \cite{DifrancescoPascucciPolidoro2007} for the study of obstacle
problems for American options. There are, however, significant differences between the two problems, mainly arising
from the appearance of the non-linear term $\bar\psi^{\gamma}$ in our PDE~\eqref{eq:finHJB2}. This
term is not globally Lipschitz continuous, while all PDE coefficients in
\cite{DifrancescoPascucciPolidoro2007} satisfy this condition. For this reason, it is crucial for
us to prove that $\bar{\psi}$ is bounded away from zero, while for their obstacle problem
\cite{DifrancescoPascucciPolidoro2007} only need to show that the solution is bounded, i.e.
$|\bar{\psi}(t,\tilde{p})| \leq c e^{c t}$ for some positive constant $c$.}
Below, we make use of the parabolic H\"older space $C^{2,\alpha}_P$ defined by the norm (\ref{DfnC2alpah}).
We then have
\begin{theorem}\label{t3}
Problem \eqref{eq:finHJB2} admits a classical solution {\CRed $\bar{\psi} \in
{\mathcal{C}_{P}^{2,\a}}$} for any $\a\in(0,1)$. Moreover, there exists a
constant $C\ge 1$, only depending on the $L^{\infty}$-norms of the coefficients of the PDE, such
that, for each $(t,\tilde{p})$,
\begin{equation}\label{stimaLinfinito1}
 \frac{1}{C}\le \bar{\psi}(t,\tilde{p}) \leq e^{C T}.
\end{equation}
\end{theorem}
The  proof of Theorem \ref{t3} is reported in Appendix \ref{app:verifications}.
{Let us remark that also
\begin{equation}\label{PrpHolderw}
	\bar{w}(t,\tilde{p}) \in
{\CRed {\mathcal{C}_{P}^{2,\a}}}
\end{equation}
in view of the relation~\eqref{HpfTran1} and the estimate~\eqref{stimaLinfinito1}, yielding that $\bar{w}$ has the same properties of $\bar{\psi}$ in the previous theorem.}
The following result {shows} a verification theorem for the pre default optimization problem.
\begin{theorem}\label{MntPreDftVal2}
Suppose  that the conditions of Theorem \ref{MntPstDftVal} are satisfied and, in particular, let $\underline{w}\in {\mathcal{C}_P^{2,\alpha}}$ be the solution of
(\ref{Eq:PstDfltDP}) with terminal condition (\ref{Eq:PstDfltDPBC}).
Additionally, let $\bar{w} \in \mathcal{C}_P^{2,\alpha}$ be the solution to the Cauchy problem (\ref{eq:finHJB0})-(\ref{eq:finHJB0in}) established in Theorem \ref{t3}.
Then, the following assertions hold true:
\begin{enumerate}
    \item[{\rm (1)}] The solution {${\bar{w}}(t,\tilde{p})$} coincides with the optimal value function {$w(t,\tilde{p},0)$} introduced in (\ref{eq:valuefn}).

    \item[{\rm(2)}] The optimal feedback controls {$\{\pi^{t}_{s}\}_{s\in[t,T)}:=\{(\pi_s^S,\pi^{P}_{s})^{\top}\}_{s\in[t,T)}$}, denoted by
$\widetilde{\pi}:=(\widetilde{\pi}^S,\widetilde{\pi}^P)^{\top}$, can be written as {$\widetilde{\pi}_s^S = {\widetilde{\pi}^{S}(s,\tilde{p}^{t}_{s^{-}},H^{t}_{s^{-}})}$ and $\widetilde{\pi}_s^P = {\widetilde{\pi}^{P}(s,\tilde{p}^{t}_{s^{-}},H^{t}_{s^{-}})}$} with
\begin{align}
    (\widetilde{\pi}^S(s,\tilde{p},0),\widetilde{\pi}^P(s,\tilde{p},0))^{\top}:=
    \frac{1}{1-\gamma} (\Sigma_Y^{\top} \Sigma_Y)^{-1} \left({\DViolet -\Upsilon(s,\tilde{p})} + \Sigma_Y \bar{\kappa}^{\top} {\nabla_{\tilde{p}} \bar{w}(s,\tilde{p})^{\top}} \right),\label{eq:optpiPrD1}\\
    (\widetilde{\pi}^S(s,\tilde{p},1),\widetilde{\pi}^P(s,\tilde{p},1))^{\top}:=
    \left(\frac{1}{\sigma^{2}(1-\gamma)}
    \left(\tilde{\mu}(\tilde{p})- r + \sigma{\underline{\kappa}^{\top}\nabla_{\tilde{p}}  \wpost(s,\tilde{p})^{\top}}  \right),0\right)^{\top}.
    \label{eq:optpiPrD2}
\end{align}
\end{enumerate}
\end{theorem}
The proof of Theorem \ref{MntPreDftVal2} is reported in Appendix \ref{app:verifications}.

\section{Numerical Analysis} \label{sec:numanalysis}
We provide a numerical analysis of the optimal strategies and value functions derived in the previous sections. We set $N=2$, i.e. we consider two regimes, thus the
vector $\tilde{p} := p$ becomes one dimensional with $p$ denoting the filter probability that the Markov chain is in regime ``1''. Unless otherwise specified, throughout the section we use the following benchmark parameters: $\sigma = 0.4$, $\upsilon = 0.5$, $r = 0.03$, $\mu_1 = 0.5$, $\mu_2 = 0.2$, $h_1 = 1$, $h_2 = 0.2$, $\varpi_{11} = -0.5$, and $\varpi_{22} = -1$.
We fix the time horizon to $T = 10$. We set $\gamma = 0.5$, i.e. we consider a square root investor.  We describe the numerical setup in
Section \ref{sec:setup}, and give an economic analysis of the strategies in Section \ref{sec:analysis}.

\subsection{Setup} \label{sec:setup}
Since the solution to the pre-default HJB-PDE depends on the solution to the post-default HJB PDE, we first need to solve Eq.~\eqref{eq:transpost}. In case of two regimes,
the PDE~\eqref{eq:transpost} becomes two dimensional with $t \in \mathds{R}^+$ and $0 \leq p \leq 1$. More specifically, Eq.~\eqref{eq:transpost} reduces to
\begin{eqnarray}
\nonumber \frac{\partial \underline{\psi}(t,p)}{\partial t} + \frac{1}{2} \underline{\kappa}(t,p)^2 {\frac{\partial^{2} \underline{\psi}(t,p)}{\partial p^{2}}}  +
\underline{\Phi}(t,p) {\frac{\partial \underline{\psi}(t,p)}{\partial p}} + \underline{\Psi}(t,p) \frac{\underline{\psi}(t,p)}{1-\gamma} &=& 0, \\
\nonumber \underline{\psi}(T,p) &=& 1,
\end{eqnarray}
where
\begin{eqnarray}
\nonumber \underline{\kappa}(t,p) &=& \sigma^{-1} p \left(\mu_1 - \left(\mu_1 p + \mu_2 (1-p) \right)\right) \\
\nonumber \beta_{\varpi}(t,p) &=& \varpi_{11} p + \varpi_{21} (1-p)  = \varpi_{11} p - \varpi_{22} (1-p)  \\
\nonumber \underline{\Phi}(t,p) &=& {\CRed \beta_{\varpi}(t,p)} + \frac{\gamma}{1-\gamma} \frac{\mu_1 p + (1-p) \mu_2 -r}{\sigma} \underline{\kappa}(t,p) \\
\underline{\Psi}(t,p) &=& \gamma r + \frac{\gamma}{2 (1-\gamma)} \left( \frac{\mu_1 p + (1-p) \mu_2 - r}{\sigma} \right)^2
\end{eqnarray}
We numerically solve the above derived PDE using a standard Crank-Nicolson method. From the transformation (\ref{HpfTran0}), we obtain that the post-default value function is given
by $\underline{w}(t,{p}) = (1-\gamma) \log (\underline{\psi}(t,{p}))$. The latter is then used into the pre-default PDE, computed as described next.
On $(t,p)$, with $t \in \mathds{R}^+$, and $0\leq p \leq{}1$, the PDE~\eqref{eq:finHJB2} satisfied by ${\bar{\psi}}(t,{p})$ reduces to
\begin{align*}
\frac{\partial {\bar{\psi}(t,p)}}{\partial t} + \frac{1}{2}\bar\kappa(t,p)\bar\kappa(t,p)^{\top}\frac{\partial^2 {\bar{\psi}}(t,p)}{\partial {p}^{2}}+ \bar{\Phi}(t,p) \frac{\partial {\bar{\psi}}}{\partial {p}}+ \bar{\Psi}(t,p) \frac{{\bar{\psi}}(t,p)}{1-\gamma} +
(h_2 + (h_{1}-h_2)p) e^{{\wpost \left(t, \frac{p h_1}{h_2 + (h_{1}-h_2)p}\right)}} \frac{{\bar{\psi}}(t,p)^{\gamma}}{1-\gamma} &=0,\\
\nonumber {\bar{\psi}}(T,p) &= 1,
\end{align*}
{\CadBlue where}
\begin{eqnarray*}
\nonumber {\bar{\kappa}}(t,p) &=& p (1-p)\left[ \frac{\mu_1 - \mu_2}{\sigma}, \; \frac{{\CadBlue a}(t,e_1) - {\CadBlue a}(t,e_2)}{\upsilon} \right]=: [\bar{\kappa}_{11}(t,p), \ \bar{\kappa}_{12}(t,p)] \\
\nonumber \bar{\Phi}(t,p) &=& \varpi_{21} + (\varpi_{11}-\varpi_{21} )p + \frac{\gamma}{1-\gamma} \left( \frac{\bar{\kappa}_{11}(t,p)}{\sigma}(\tilde{\mu}(p)-r) + \frac{\bar{\kappa}_{12}(t,p)}{{\upsilon}} \left({\BRed {\CadBlue \tilde{a}}(t,p)-r} \right)\right) \\
\bar{\Psi}(t,p) &=& \frac{1}{2} \frac{\gamma}{1-\gamma} \left(
{\CadBlue \frac{\left(\tilde{\mu}(p)-r \right)^2}{\sigma^2}} + {\CadBlue \frac{\left(\tilde{a}(t,p)-r  \right)^2}{{\upsilon}^2}} \right) + \gamma r - h_{2}+\left(h_1 - h_2 \right)p.
\end{eqnarray*}
In the following analysis, we set $a(t, e_1) = -(r + h_1)$, and $a(t, e_2) = -(r + h_2)$. In agreement with the notation in Section \ref{sec:model},
$a(t, e_1)$ and $a(t, e_2)$ denote the risk adjusted returns of the defaultable security when the current regime is ``1'', respectively ``2''.
Similarly to the post-default case, we employ a standard Crank-Nicolson method to solve the above derived nonlinear PDE.
The solution to the pre-default PDE is then obtained as $\underline{w}(t,p) = (1-\gamma) \log (\underline{\psi}(t,p))$.

\subsection{Analysis of Strategies}\label{sec:analysis}
Figure \ref{fig:prestrategy} shows that the stock investment strategy is increasing in the filter probability of the hidden chain being in the first regime. This happens because under our parameter choices, the growth rate of the stock is higher in regime ``1'', while the volatility stays unchanged. Consequently, if the filter estimate of being in the most profitable regime gets higher, the risk averse investor would prefer to shift a larger amount of wealth in the stock. On the other hand, as the probability of being in regime ``1'' increases, the risk averse investor
shorts a higher amount of {the defaultable security}. This happens because the default intensity in regime ``1'' is the highest, and thus the risk averse investor wishing to decrease his
exposure to default risk goes short in the vulnerable security. Notice also the key role played by the stock volatility $\sigma$. As the volatility gets lower, the investor shorts more units of the vulnerable security and invest the resulting proceeds in the stock security. Indeed, since the stock volatility is low while the default risk unchanged, the risk averse investor prefers to invest larger fraction of wealth in the stock security, and does so by raising cash via short-selling of the vulnerable security. The latter action also results in him having reduced exposure to credit risk. Since all model parameters depend on time only through the underlying hidden regime,
investment strategies are not very sensitive to passage of time.

\begin{figure}
\centering
\begin{tabular}{cc}
\epsfig{file={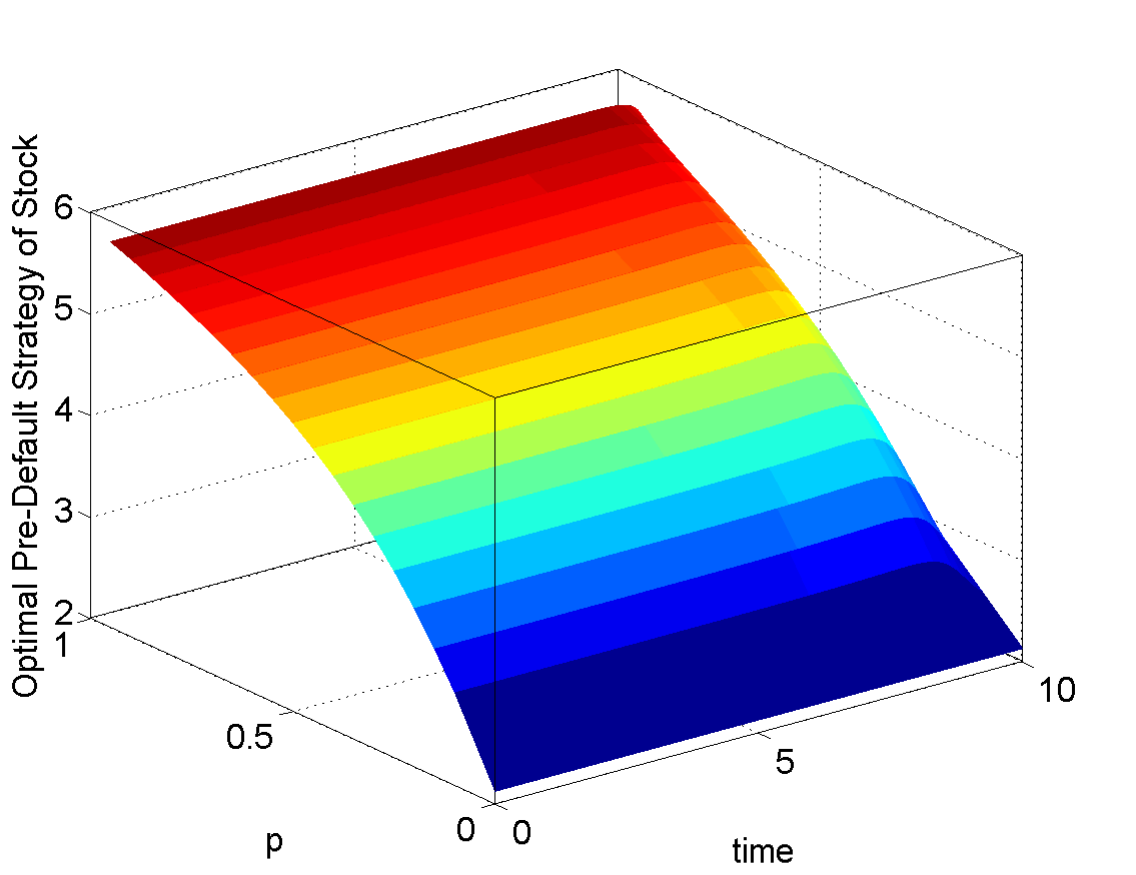},width=0.4\linewidth,clip=}
\epsfig{file={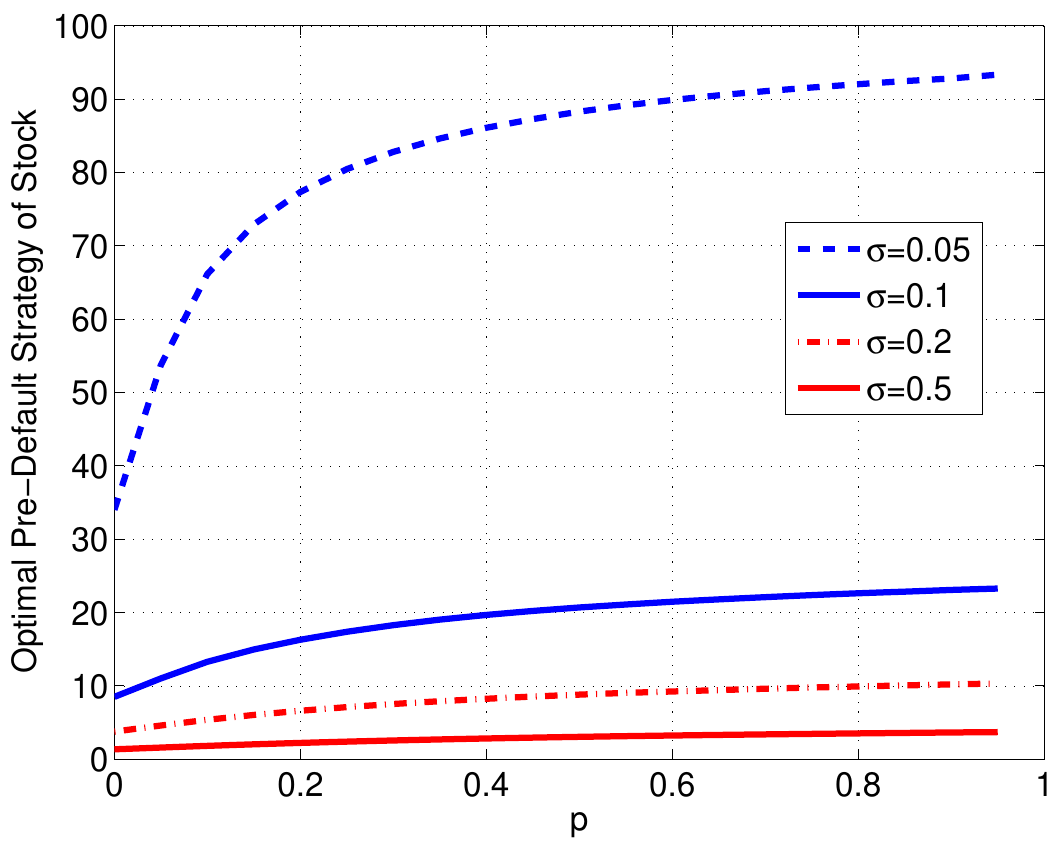},width=0.4\linewidth,clip=} \\
\epsfig{file={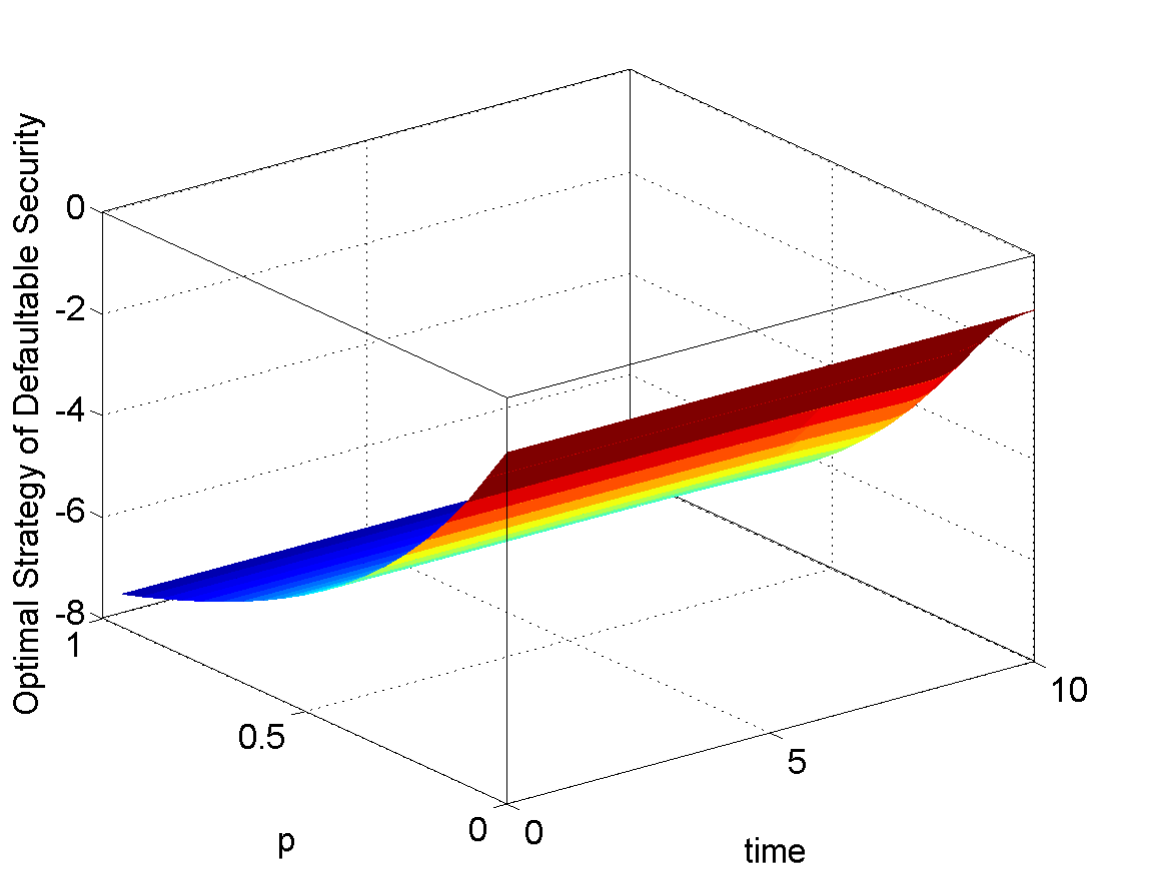},width=0.4\linewidth,clip=}
\epsfig{file={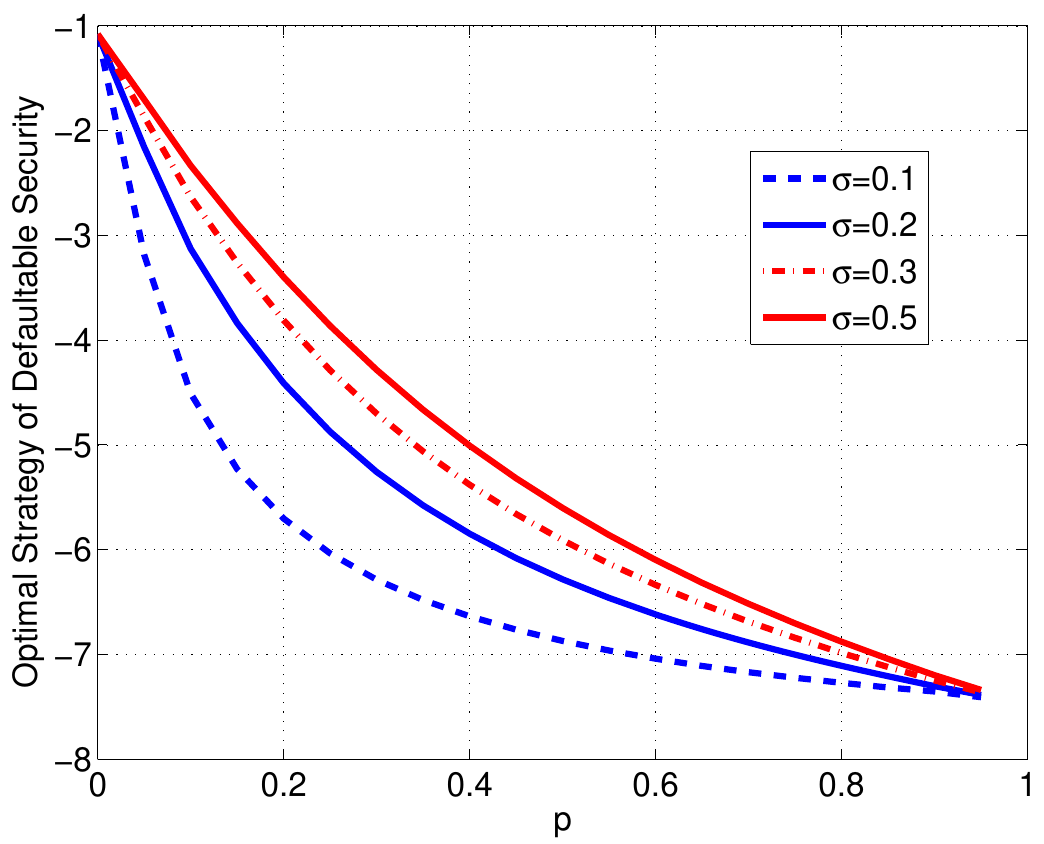},width=0.4\linewidth,clip=} \\
\end{tabular}
\caption{The top panels report the stock investment strategy. The bottom panels report the investment strategy in the vulnerable security. In the right panels, we set $t=0$.}
\label{fig:prestrategy}
\end{figure}

From figure~\ref{fig:valuefn}, we can also see that both the pre-default and post-default value functions are decreasing in time, and increasing in the filter probability $p$. Moreover, as the filter probability of being in regime ``1'' increases, the investor extracts more utility given that he realizes higher gains by simultaneously shorting the vulnerable security and purchasing the stock security.
\begin{figure}
\centering
\begin{tabular}{cc}
\epsfig{file={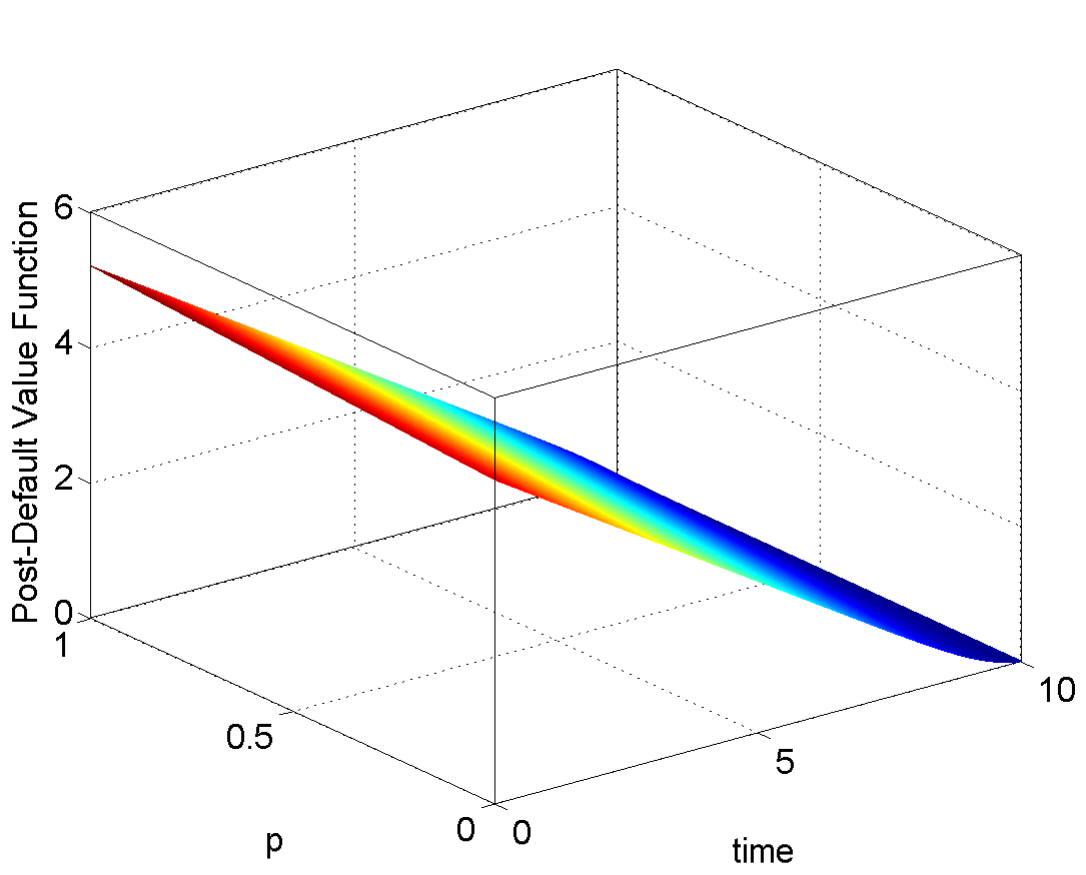},width=0.4\linewidth,clip=}
\epsfig{file={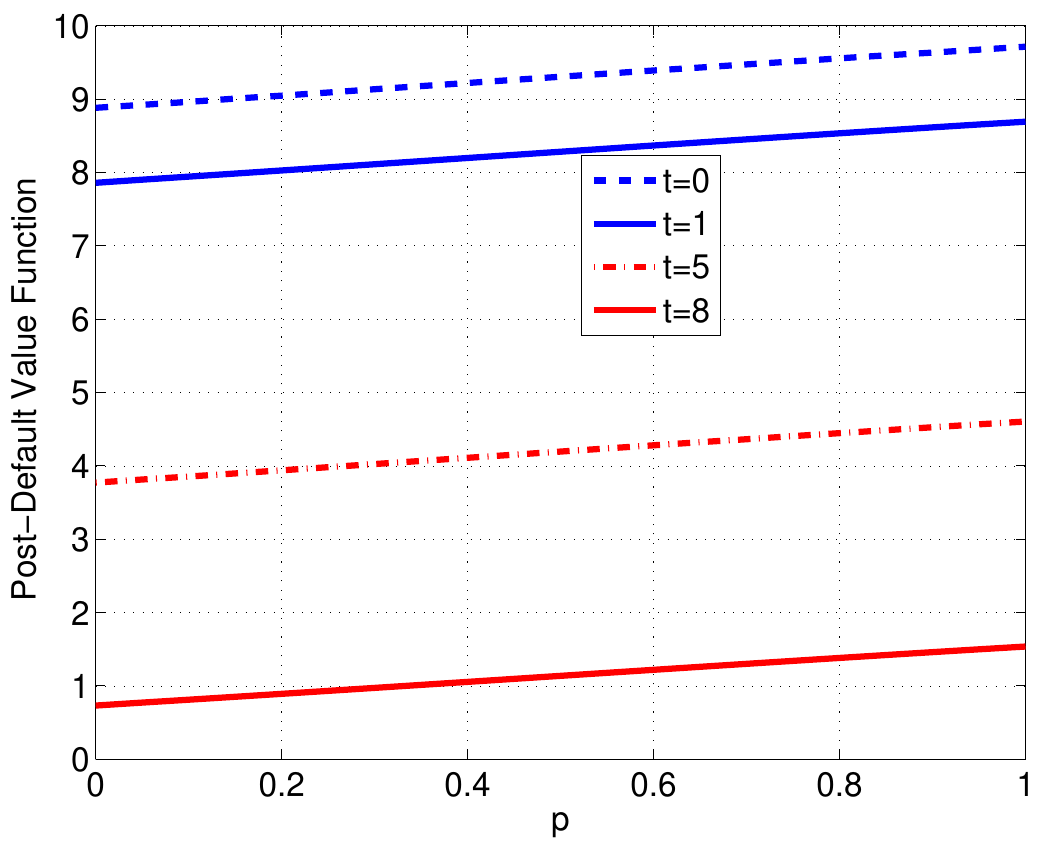},width=0.4\linewidth,clip=}\\
\epsfig{file={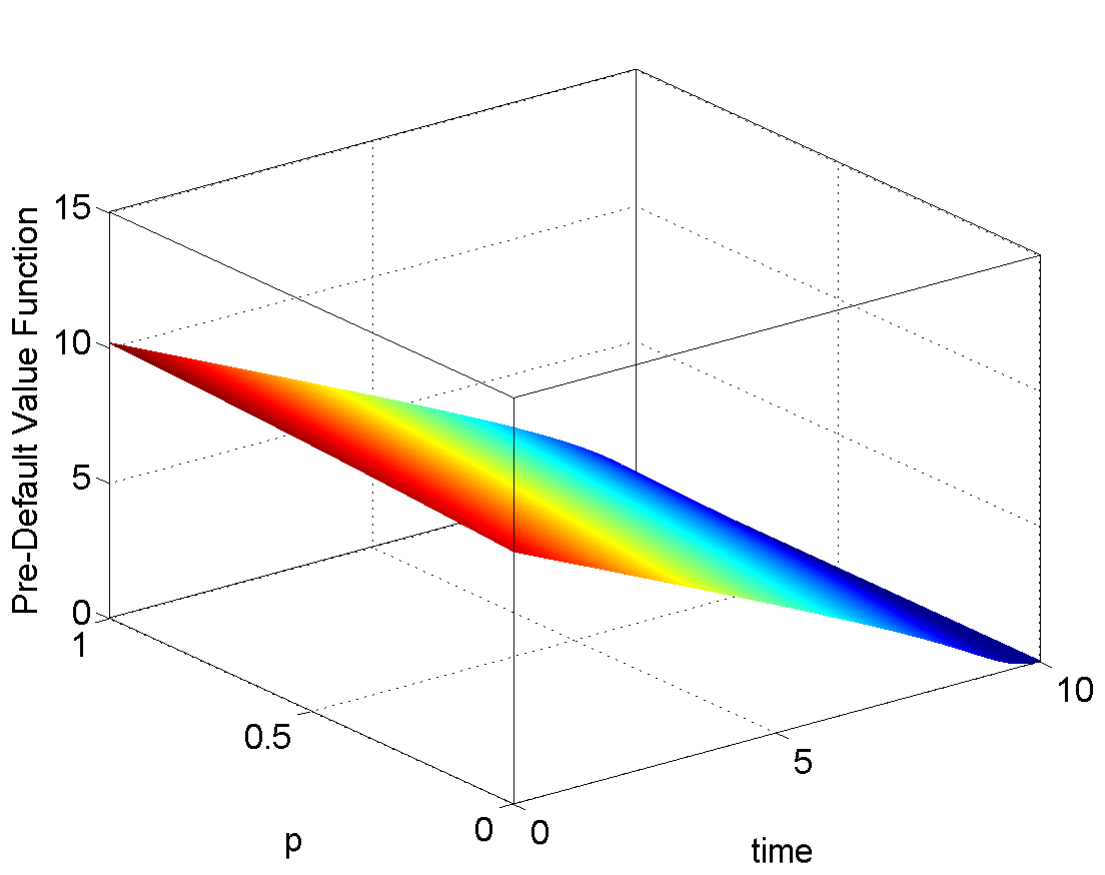},width=0.4\linewidth,clip=}
\epsfig{file={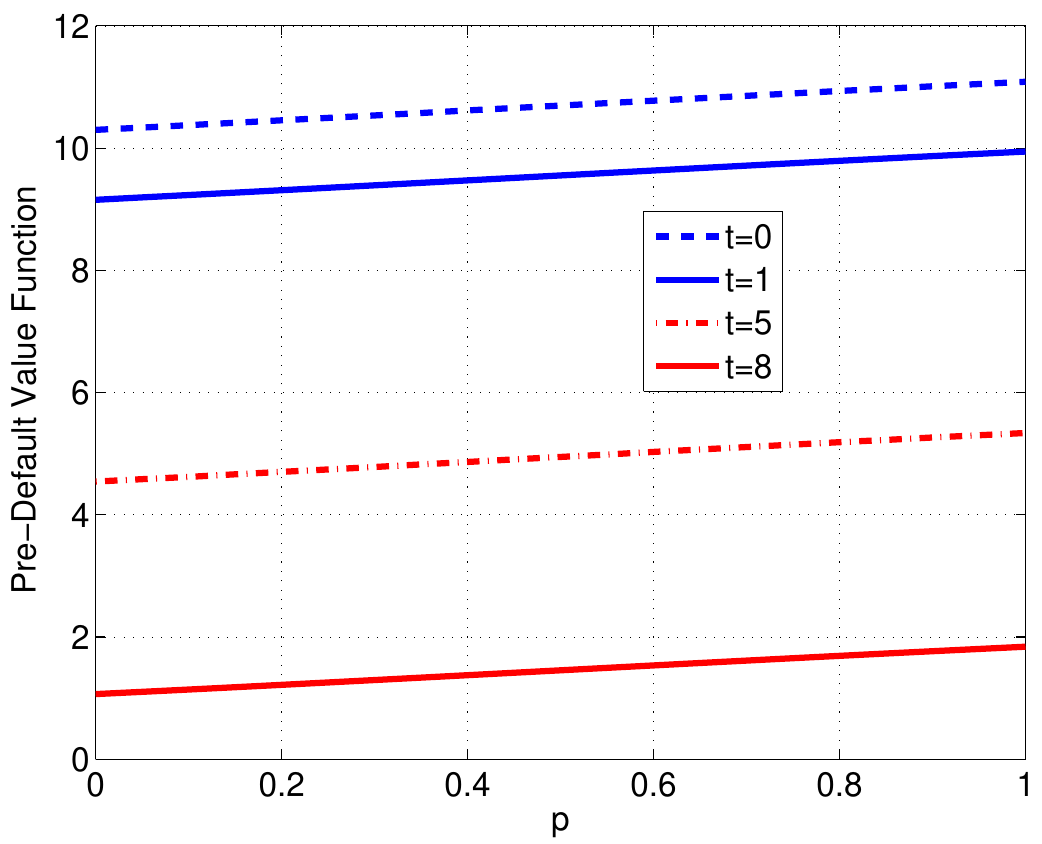},width=0.4\linewidth,clip=}\\
\end{tabular}
\caption{The pre-default and post-default value functions plotted versus time $t$ and filter probability $p$.}
\label{fig:valuefn}
\end{figure}

The investor increases the amount of shorted units of the vulnerable security if the filter probability of staying in the regime characterized by high default intensity gets larger. By contrast, if the filter probability of staying in the regime with low default intensity increases he takes higher credit risk exposure by purchasing more bond units. This is clearly illustrated in the left panel of figure \ref{fig:predefintens}.
Moreover, when $h_1 < h_2$ smaller amount of units of the vulnerable security are shorted if the probability $p$ of staying in the low risk regime gets higher. However, when the default intensity $h_1$ in regime ``1'' exceeds the default intensity $h_2$ in regime ``2'' ($h_2 = 0.2$), the opposite effect is observed. As $h_1$ gets higher, a larger amount of units of {the vulnerable security} are shorted if the filter probability of staying in regime ``1'' increases. We also notice that, ceteris paribus, the pre-default value function is increasing in $h_1$ for a fixed $p$ value, and also, for a given value of $h_1$ the pre-default value function increases in $p$.

\begin{figure}
\centering
\begin{tabular}{cc}
\epsfig{file={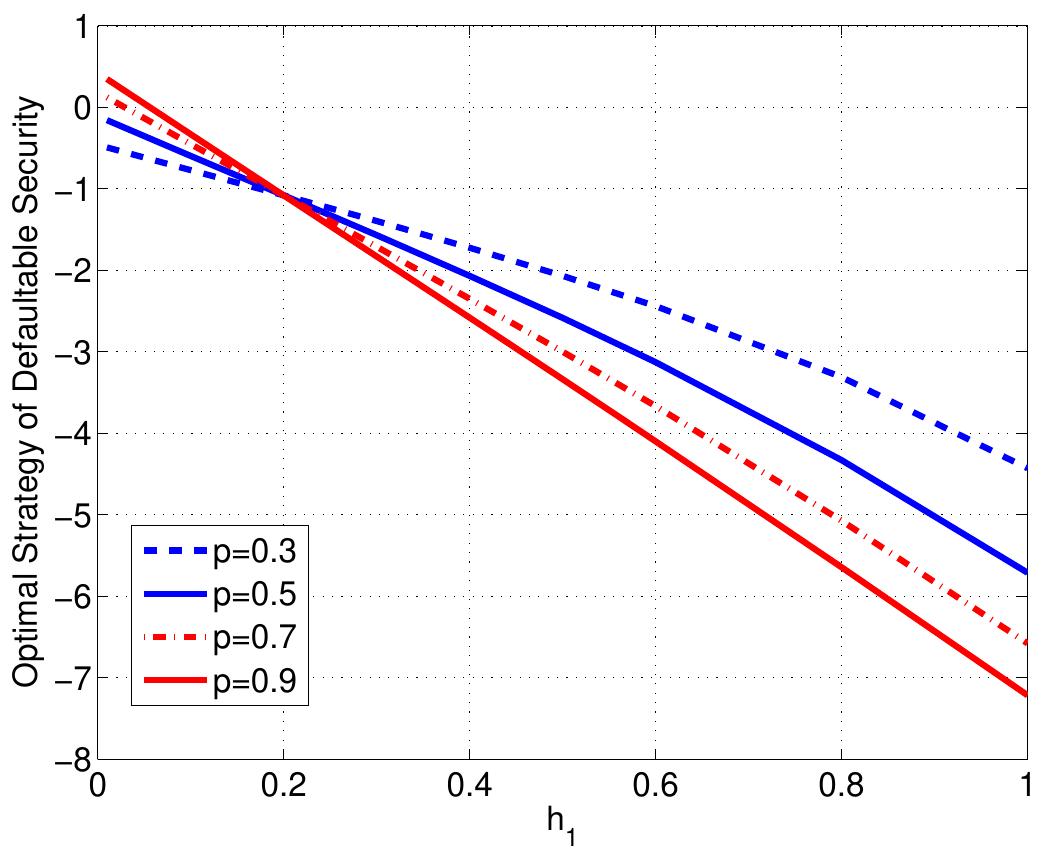},width=0.4\linewidth,clip=}
\epsfig{file={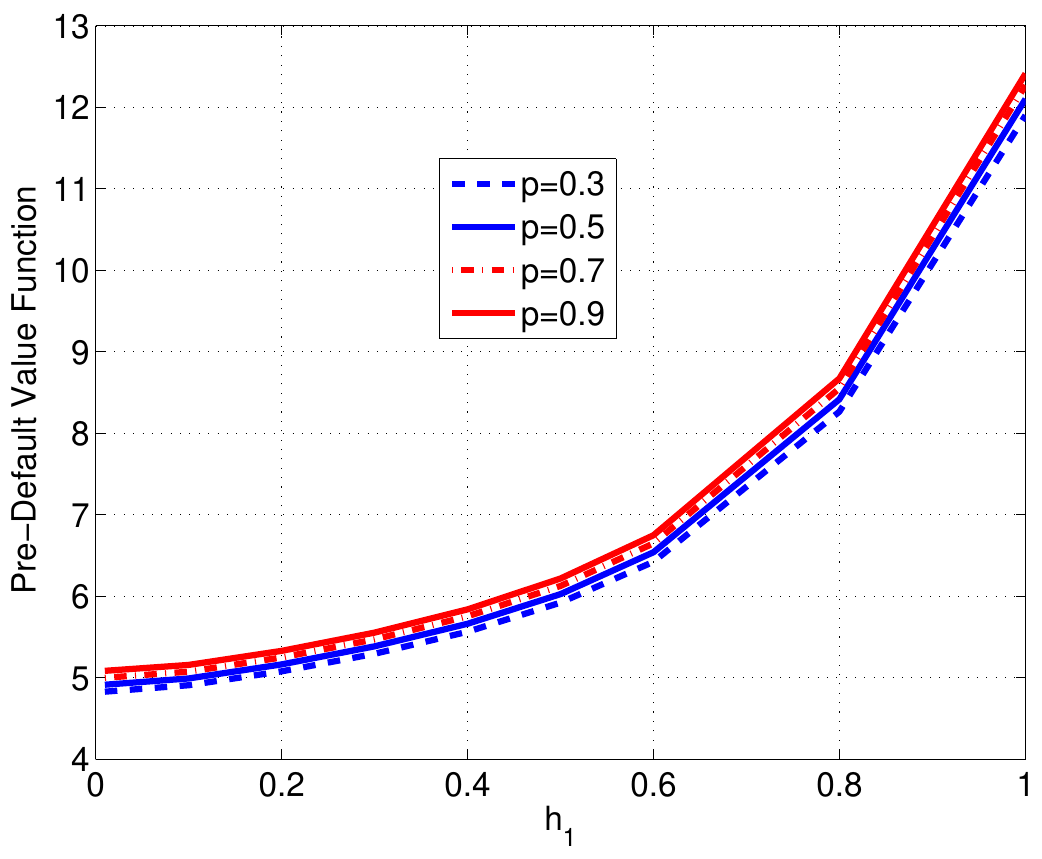},width=0.4\linewidth,clip=} \\
\end{tabular}
\caption{The left panel reports the optimal investment strategy in the {defaultable security}. The right panel reports the pre-default value function. We set $t=0$.}
\label{fig:predefintens}
\end{figure}

We next analyze the dependence of the optimal stock and vulnerable security investment strategies on the volatilities $\sigma$ and $\upsilon$.
Figure \ref{fig:stockbondsnu} shows that when the stock volatility is low, the investor puts a large fraction of wealth in the stock
security. This happens because the filter gain coming from receiving stock observations is the highest.
Since observations are more informative, the risk averse investor deposits a larger fraction of wealth in the stock, especially if the filter probability of being in the high growth regime (regime ``1'' with $\mu_1 = 0.5$) is high. As the volatility gets larger, stock price observations become less informative leading the investor to decrease the wealth fraction invested in stock. When the volatility exceeds a certain threshold, regardless of the filter probability the investor always puts a small amount of wealth in the stock. We also notice that a similar role is played by the volatility $\upsilon$ of the defaultable security. From the right panel of figure \ref{fig:stockbondsnu}, we notice that when $\upsilon$ is low, i.e. price observations of the vulnerable security are very informative, the investor wants to reduce more his exposure to default risk. Hence, he shorts more units of the vulnerable security especially if the filter probability of being in the highest credit risk regime (regime ``1'' with $h_1 = 0.05$) is large. This reflects the risk averse nature of the investor who dislikes default risk and uncertainty. As for the stock, when $\upsilon$ gets larger the investment strategy in the defaultable security becomes less sensitive to the filter probability and for large values of $\upsilon$ the investor may even find it optimal to purchase the defaultable security. This happens when the potential loss incurred by the investor when he is long credit and default occurs (hence making the vulnerable security worthless) is outweighed by the risk adjusted return resulting from holding the defaultable security. 

We conclude by relating partial to full information settings. As price volatilities become smaller, the regime switching model becomes more observable. This is because price observations become more informative and allow the investor to build more accurate estimates of the regime in place. Consequently, the above analysis outlines the important role played by regime uncertainty in determining the optimal strategies of risk averse investors. Compared to the case of fully observed regimes studied in \cite{CapFig1}, the presence of incomplete information induces the risk averse investor to decrease the wealth amount invested in the risky securities. As clearly illustrated in figure \ref{fig:stockbondsnu}, when the price volatilities are sufficiently high ($\sigma \approx 0.8$ for the stock and $\upsilon \approx 0.6$ for the defaultable security), the investor deposits almost entire amount of wealth in the money market account.
\begin{figure}
\centering
\begin{tabular}{cc}
\epsfig{file={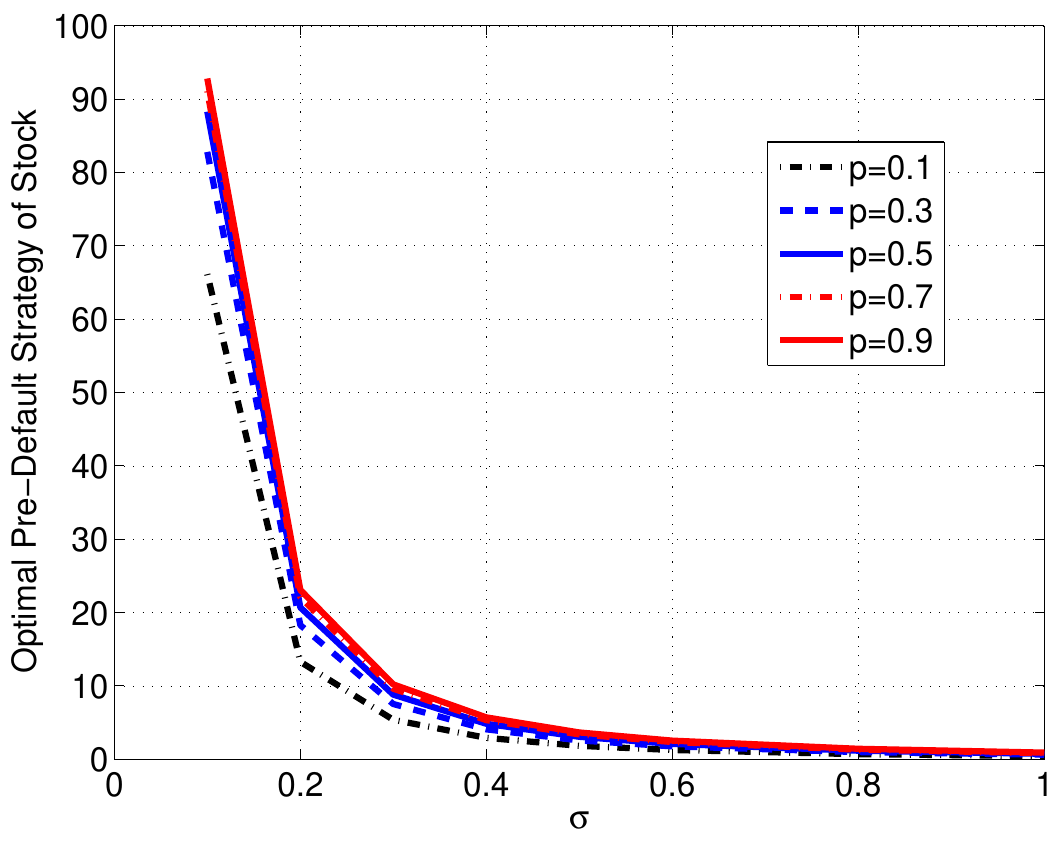},width=0.4\linewidth,clip=}
\epsfig{file={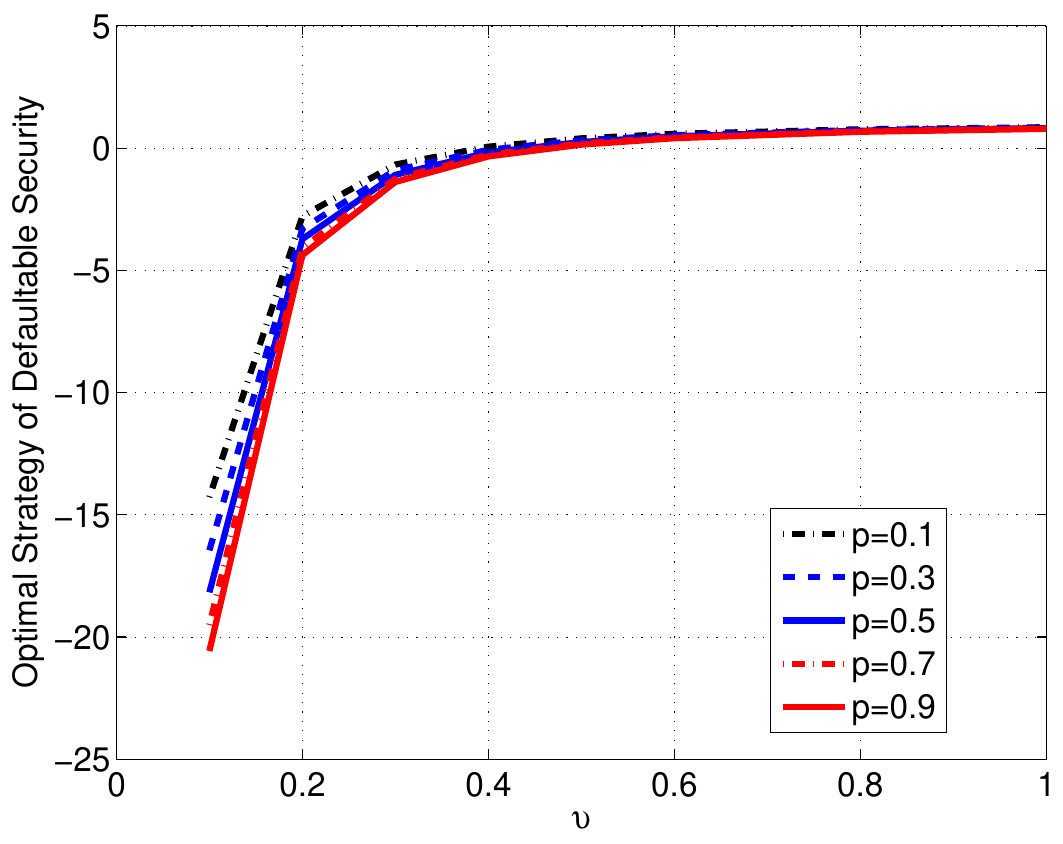},width=0.4\linewidth,clip=} \\
\end{tabular}
\caption{The left panel reports the dependence of the optimal stock investment strategy on $\sigma$. The right panel reports the dependence of
 the optimal investment strategy in the vulnerable security  on $\upsilon$. We set $t=0$. In the right panel, we set  $h_1 = 0.05$ and $h_2=0.01$.}
\label{fig:stockbondsnu}
\end{figure}

\section{{Conclusions}} \label{sec:conclusions}
We have studied the optimal investment problem of a power investor in an economy consisting of a
defaultable security, {a} stock, and {a} money market account. The price processes of these
securities are assumed to have drift coefficients and default intensities modulated by a hidden Markov chain.
We have reduced the partially observed stochastic control problem to a risk sensitive one, where the state
is given by the filtered regime probabilities. The conditioning filtration, generated by the prices
of the stock and of the defaultable security, and by the indicator of default occurrence, is
driven both by a Brownian component and by a pure jump martingale. The filter has been used to derive
the HJB partial differential equation corresponding to the risk sensitive control problem. We have split the latter into a pre-default and {a} post-default dynamic
programming {subproblem}. The HJB PDE corresponding to the post-default value function can be transformed to a linear
parabolic PDE, for which existence and uniqueness of a classical solution can be guaranteed. By contrast, the
HJB PDE corresponding to the pre-default value function has exponential nonlinearity and quadratic gradient growth.
We have provided a detailed mathematical analysis of such PDE and established the existence of a classical solution with
$\mathcal{C}^{2,\alpha}_P$ regularity. We have then proven verification theorems establishing the correspondence between the PDE solutions
and the value functions of the control problem. Our study has been complemented with a thorough numerical analysis illustrating the role of regime
uncertainty, default risk, and price volatilities on the optimal allocation decisions and value functions.

\subsection*{Acknowledgments}
The authors gratefully acknowledge two anonymous reviewers and Wolfgang Runggaldier for providing constructive and insightful comments which improved significantly the quality of the manuscript.
Agostino Capponi would also like to thank Ramon Van Handel for very useful discussions and insights provided in the original model setup.

\appendix

\section{Proofs related to Section \ref{sec:utilmax}}\label{app:utilmax}

\begin{lemma}\label{lem:filqi}
Let
\begin{equation}
q_t^i = \Ex^{\hat{\Px}} \left[L_t {\idc_{\{X_t=e_i\}}}  \bigg| \mathcal{G}_t^I \right]. 
\label{eq:qtidef}
\end{equation}
{Then, the} dynamics of {$(q_t^i)_{t\geq{}0}$, $i=1,\dots,N$}, under the measure $\hat{\Px}$, is given by the following system of stochastic differential equations (SDE){:}
\begin{align}
 dq_t^i &= {\sum_{\ell=1}^{N}\varpi_{\ell,i}(t)q^{\ell}_{t}} dt + q_t^i {\CaddBlue Q^{\top}(t,e_i,\pi_t)} {\Sigma_{Y}} d \hat{W}_t +
 {q_{t^{-}}^i} (h_i-1) d\hat{\xi}_t - \gamma {\CaddBlue \eta(t,e_i,\pi_t)} q_t^i dt{,} \label{eq:dynqti} \\
        \nonumber  q_0^i &= p_0^i.
\label{eq:dynqti}
\end{align}
\end{lemma}

\medskip
\noindent
\proof
Let us introduce the following notation
\[
    H_{t}^{i}:={\bf 1}_{\{X_{t}=e_{i}\}}.
\]
Note that $X_{t}=(H^{1}_{t},\dots,H^{N}_{t})^{\top}$ and, from (\ref{eq:MCsemim}),
\begin{equation}\label{eq:MCsemimb}
    H^{i}_{t}=H_{0}^{i}+\int_{0}^{t} \sum_{\ell=1}^{N}\varpi_{\ell,i}(s)H^{\ell}_{s} ds + \CRed{\varphi_{i}(t)}.
\end{equation}
From Eq.~\eqref{eq:lteq} and~\eqref{eq:stocexpdyn}, we deduce that, under ${\hat{\Px}}$,
$$
dL_t = L_{t-} (h_t-1) d\hat{\xi}_t + L_t {\CaddBlue Q^{\top}(t,X_t,\pi_t)} dY_t - L_t \gamma {\eta(t,X_t,\pi_t)}dt
$$
which yields that
$$
[L, {H^{i}}]_t = \int_0^t L_{s-} {\CaddBlue Q^{\top}(s,X_s,\pi_s)} d[Y, {H^{i}}]_s + \int_0^t L_{s-} (h_s-1) d[\hat{\xi}, {H^{i}} ]_s.
$$
As $(Y_t)_{t\geq0}$ and $(H_t)_{t\geq0}$ are independent of $(X_t)_{t\geq0}$ (and, hence, of $H^{i}$), under ${\hat{\Px}}$ it holds (see also \cite{Wong}) that, ${\hat{\Px}}$ almost surely,
$$
    [Y, {H^{i}}]_s = [\hat{\xi}, {H^{i}} ]_s = 0,{\text{ for all }s\geq{}0.}
$$
Thus, applying It\^o{'}s formula, we obtain
\begin{eqnarray}
\nonumber L_t {H^{i}_{t}} &=& {H^{i}_{0}} + \int_0^t {H^{i}_{s^{-}}} dL_s + \int_0^t L_{s-} d{H_{s}^{i}} \\
\nonumber                      &=&  {H^{i}_{0}} + \int_0^t  {H^{i}_{s}} L_s {\CaddBlue Q^{\top}(s,X_s,\pi_s)} dY_s +  \int_0^t  {H^{i}_{s^{-}} L_{s^{-}} (h_{s^{-}} - 1)}
                               d\hat{\xi}_s \\
                      & &  - \int_0^t {H^{i}_{s}} L_s \gamma {\CaddBlue \eta(s,X_s,\pi_s)} ds
+ \int_0^t L_{s} {\sum_{\ell=1}^{N}\varpi_{\ell,i}(s)H^{\ell}_{s}} ds +
                               \int_0^t L_{s-} d{{\CRed \varphi_{i}}(s)}
\label{eq:dynfilter}
\end{eqnarray}
Since {$({\CRed \varphi_i(t)})_{t\geq{}0}$} is a $((\mathcal{F}_t^{{X}})_{t\geq{}0}, {\hat{\Px}})$-martingale, and {$\mathcal{G}_T^I$} is independent of {$\mathcal{F}_T^{X}$} under ${\hat{\Px}}$, we have that
$\Ex^{{\hat{\Px}}}\left[\int_0^t L_{s-} d{{\CRed \varphi_i(s)}} | {\mathcal{G}_t^I} \right] = 0$.
Therefore, taking $\G_t^I$ conditional expectations in Eq.~\eqref{eq:dynfilter}, we obtain
\begin{eqnarray}
\nonumber \Ex^{{\hat{\Px}}}\left[L_t {H^{i}_{t} |\mathcal{G}_t^I}\right] &=& 1 + \int_0^t \Ex^{{\hat{\Px}}} \left[L_s {H^{i}_{s}}  {\CaddBlue Q^{\top}(s,{e_i}, \pi_s)} | \G_s^I \right] dY_s
                                                          + \int_0^t \Ex^{{\hat{\Px}}} \left[L_{s^{-}} {H^{i}_{s^{-}}} ({h_{s^{-}}} - 1) | \G_s^I \right] d{\hat{\xi}_s} \\
                                                 & & - \int_0^t  \Ex^{{\hat{\Px}}} \left[{H^{i}_{s}}  L_s \gamma {\CaddBlue \eta(s,e_i, \pi_s)} | \G_s^I \right] ds + \int_0^t  \Ex^{{\hat{\Px}}} \left[ {\sum_{\ell=1}^{N}\varpi_{\ell,i}(s)L_{s} H^{\ell}_{s}} | \G_s^I \right]  ds,
\label{eq:Pcondexp}
\end{eqnarray}
where we have used that, if $\phi_t$ is {$\Gx$-predictable} then (see, for instance, \cite{Wong}, {Ch. 7,} Lemma 3.2)
\begin{eqnarray}
\nonumber \Ex^{{\hat{\Px}}} \left[\int_0^t \phi_s L_{s-} dY_s | \G_t^I \right] &=& \int_0^t \Ex^{{\hat{\Px}}}\left[\phi_s L_{s-} | \G_s^I \right] dY_s \\
\nonumber \Ex^{{\hat{\Px}}} \left[\int_0^t \phi_s L_{s-} d\hat{\xi}_s | \G_t^I \right] &=& \int_0^t \Ex^{{\hat{\Px}}}\left[\phi_s L_{s-} | \G_s^I \right] d\hat{\xi}_s \\
\nonumber  \Ex^{{\hat{\Px}}} \left[\int_0^t \phi_s L_{s-} ds | \G_t^I \right] &=& \int_0^t \Ex^{{\hat{\Px}}}\left[\phi_s L_{s-} | \G_s^I \right] ds
\end{eqnarray}
Observing that under ${\hat{\Px}}$ $dY_t = {\Sigma_{Y}} d \hat{W}_t$, using that {\CaddBlue $Q(t,{e_i},\pi_t)$} and {\CaddBlue $\eta(t,e_i, \pi_t)$} are {$(\mathcal{G}_t^I)_{t\geq{}0}$} adapted, and that the Markov-chain generator $A(t)$ is deterministic, we obtain Eq.~\eqref{eq:dynqti} upon taking the differential of Eq.~\eqref{eq:Pcondexp}.
\endproof

\begin{lemma}\label{lem:qLp}
The following identities hold
\begin{eqnarray}
q_t^i &=& \hat{L}_t p_t^i
\label{eq:relqti}\\
p_t^i &=& \frac{q_t^i}{\sum_{j=1}^N q_t^j}
\label{eq:qtisumq}
\end{eqnarray}
where $q_t^i$, $\hat{L}_t$, and $p_t^i$ are defined, respectively, by~\eqref{eq:qtidef},~\eqref{eq:hatLt}, and~\eqref{eq:filtprob}.
\end{lemma}


\medskip
\noindent
\proof
We start establishing the relation~\eqref{eq:relqti} by comparing the dynamics of $q_t^i$ and of ${\hat{L}_t p_t^i}$. The dynamics of $q_t^i$ is known from Lemma \ref{lem:filqi} and given in Eq.~\eqref{eq:dynqti}. Next, we derive the dynamics of $\hat{L}_t p_t^i$.
We have
\begin{equation*}
d (\hat{L}_t p_t^i) = \hat{L}_{{t^{-}}} dp_t^i + p_{{t^{-}}}^i d \hat{L}_t + d\left[\hat{L},p^{i} \right]_t.
\label{eq:prodltp}
\end{equation*}
From Eq.~\eqref{eq:ltdyn} and~\eqref{eq:dynpti}, we obtain
\begin{align}
d\left[\hat{L},p^i \right]_t &= p_t^i \hat{L}_t {\Red \hat{\loga}^{\top}(t,p_t)} \Sigma_Y \Sigma_Y^{-1} \left({\CaddBlue \loga(t,e_i) - {\Red \hat{\loga}}(t,p_t)}  \right) dt + p_t^i \hat{L}_t \gamma \pi_t^{\top} \left( {\CaddBlue \loga(t,e_i) - {\Red \hat{\loga}}(t,p_t)} \right) dt \nonumber \\
&\quad+ \left(\hat{h}_{{t^{-}}} -1 \right) \frac{h_i - \hat{h}_{{t^{-}}}}{\hat{h}_{{t^{-}}}} \hat{L}_{{t^{-}}} p_{{t^{-}}}^i d H_t.
\label{eq:firstvar}
\end{align}
Using {the above equations, along with \eqref{eq:dynpti}, we obtain
\begin{eqnarray}
\label{eq:fulldyn}
d \left(\hat{L}_t p_t^i \right) &=& \hat{L}_t \left(\sum_{\ell=1}^{N}\varpi_{\ell,i}(t) p^{\ell}_{t} dt \right) +
                                                                    \hat{L}_t p_t^i \left({\CaddBlue \loga^{\top}(t,e_i) - {\Red \hat{\loga}}^{\top}(t,p_t)} \right) \left(\Sigma_Y \Sigma^{\top}_Y \right)^{-1} \left(dY_t - {\CaddBlue {\Red \hat{\loga}(t,p_t)}} dt \right) \\
\nonumber &+& \hat{L}_{{t^{-}}} p_{{t^{-}}}^i \frac{h_i - \hat{h}_{{t^{-}}}}{\hat{h}_{{t^{-}}}} \left(d H_t - \hat{h}_{{t^{-}}}{\bar{H}_{t^{-}}}dt \right) + p_t^i \hat{L}_t {\Red \hat{Q}^{\top}(t,p_t,\pi_t)} dY_t - p_t^i {\hat{L}_t \gamma {\Red \hat{\eta}(t,p_t,\pi_t)}}dt \\
\nonumber &+& p_{{t^{-}}}^i \hat{L}_{{t^{-}}} (\hat{h}_{{t^{-}}} - 1) (d H_t -  {\CaddBlue \bar{H}_{t^{-}}}dt) + (\hat{h}_{{t^{-}}}-1) \frac{h_i - \hat{h}_{{t^{-}}}}{\hat{h}_{{t^{-}}}} \hat{L}_{{t^{-}}} p_{{t^{-}}}^i d H_t \\
\nonumber &+& p_t^i \hat{L}_t {\Red \hat{\loga}^{\top}(t,p_t)} \left(\Sigma_Y \Sigma_Y^{\top} \right)^{-1} {\left(\loga(t,e_i) - \Red {\hat{\loga}}(t,p_t)  \right) dt} + \gamma p_t^i \hat{L}_t \pi_t^{\top} \left({\CaddBlue \loga(t,e_i) - {\Red \hat{\loga}}(t,p_t)} \right) dt.
\end{eqnarray}
Next, observe that
\begin{eqnarray}
\nonumber \hat{L}_t p_t^i \left({\CaddBlue \loga^{\top}(t,e_i) - {\Red \hat{\loga}}^{\top}(t,p_t)} \right) (\Sigma_Y \Sigma_Y^{\top})^{-1} (dY_t - {\Red \hat{\loga}}(t,p_t) dt) +
p_t^i \hat{L}_t {\Red \hat{Q}}^{\top}(t,p_t,\pi_t) dY_t &=& \\
\hat{L}_t p_t^i {\CaddBlue Q^{\top}(t,e_i,\pi_t)} dY_t - \hat{L}_t p_t^i \left({\CaddBlue \loga^{\top}(t,e_i)-{\Red \hat{\loga}}^{\top}(t,p_t)} \right) (\Sigma_Y \Sigma_Y^{\top})^{-1}
{\Red \hat{\loga}}(t,p_t) dt
\label{eq:ltpsim}
\end{eqnarray}
Moreover,
{\CaddBlue
\begin{equation}
\eta(t,e_i,\pi_t) - {\Red \hat{\eta}}(t,p_t,\pi_t) = \pi_t^{\top} \left({\Red \hat{\loga}(t,p_t)} -\loga(t,e_i)  \right).
\label{eq:etas}
\end{equation}
}
Using relations~\eqref{eq:ltpsim}, and~\eqref{eq:etas}, along with {straightforward simplifications}, we may simplify {Eq.~\eqref{eq:fulldyn}} to
\begin{equation}
d(\hat{L}_t p_t^i) = \left( \sum_{\ell=1}^{N}\varpi_{\ell,i}(t) \hat{L}_t p^{\ell}_{t} dt \right) +  \hat{L}_t p_t^i {\CaddBlue Q^{\top}(t,e_i,\pi_t)} dY_t - \gamma \hat{L}_t p_t^i {\CaddBlue \eta(t,e_i,\pi_t)}dt   + \hat{L}_{{t^{-}}} p_{{t^{-}}}^i (h_i-1) d\hat{\xi}_t.
\label{eq:dynLtpti}
\end{equation}
Using that $dY_t = \Sigma_Y d\hat{W}_t$, we have that the equality~\eqref{eq:relqti} holds via a direct comparison of equations~\eqref{eq:dynLtpti} and~\eqref{eq:dynqti}.

\noindent Next, we establish~\eqref{eq:qtisumq}. Using Eq.~\eqref{eq:relqti} and that $\sum_{i=1}^N p_t^i = 1$, we deduce that
$$
d \left(\sum_{i=1}^N q_t^i \right) = d \left(\sum_{i=1}^N \hat{L}_t p_t^i \right) = d\hat{L}_t
$$
hence obtaining that $\sum_{i=1}^N q_t^i  = \hat{L}_t$. Using again~\eqref{eq:relqti}, this gives
$$
p_t^i = \frac{q_{t}^i}{\hat{L}_t} = \frac{q_{t}^i}{\sum_{j=1}^N q_t^j}
$$
This completes the proof.
}
\endproof

\medskip

\noindent{\textbf{Proof of Proposition \ref{prop:equivform}.}}
\proof

\smallskip
\noindent
{\BGreen

Using Eq.~\eqref{eq:gammaeq},~\eqref{eq:qtidef}, and the relation~\eqref{eq:relqti} established in Lemma~\ref{lem:qLp} , we have that
\begin{eqnarray*}
\nonumber J(v,\pi,T) &=&\frac{1}{\gamma} \Ex^{\Px} \left[V_T^{\gamma} \right] = \frac{v^{\gamma}}{\gamma} \Ex^{\hat{\Px}} \left[L_T  \right] = \frac{v^{\gamma}}{\gamma} \Ex^{\hat{\Px}} \left[ \Ex^{\hat{\Px}} \left[L_T \big| \mathcal{G}_T^I \right] \right] \\
\nonumber &=& \frac{v^{\gamma}}{\gamma} \sum_{i=1}^N \Ex^{\hat{\Px}} \left[ \Ex^{\hat{\Px}} \left[L_T \idc_{{\{X_T=e_i\}}} \big| \mathcal{G}_T^I \right] \right] = \frac{v^{\gamma}}{\gamma} {\sum_{i=1}^N  \Ex^{\hat{\Px}} \left[q_T^i \right]} \\
&=& \frac{v^{\gamma}}{\gamma} \sum_{i=1}^N  \Ex^{\hat{\Px}} \left[\hat{L}_T p_T^i \right] =  \frac{v^{\gamma}}{\gamma}  \Ex^{\hat{\Px}} \left[\hat{L}_T \right],
\end{eqnarray*}
thus proving the statement.
}
\endproof

\section{Proofs related to Section \ref{sec:ver}} \label{app:verifications}
We start with a Lemma, which will be needed in the section where the verification theorem is proven.
\begin{lemma}\label{Rem:TryPrvRmnPs}
For any $T>0$ and $i\in\{1,\dots,N\}$, it holds that
\begin{enumerate}
\item[(1)] ${\Px}\left(p^i_t> 0,\, \textrm{for\ all }\ t\in [0,T)\right)=1$.
\item[(2)] ${\Px}\left(p^i_t< 1,\, \textrm{for\ all }\ t\in [0,T)\right)=1$.
\end{enumerate}
\end{lemma}

\medskip

\proof
\smallskip
\noindent
{\BGreen
Define $\varsigma = \inf \{t : p^i_t = 0\} \wedge T$.
}
If $p^i_t$ can hit zero, then $\Px(p^i_{\varsigma} = 0) > 0$. Recall that {\BGreen $p_{t}^i = \frac{q_{t}^i}{\sum_j q_t^j}$ from Eq.~\eqref{eq:qtisumq}, hence $p_{\varsigma}^i = \frac{q_{\varsigma}^i}{\sum_j q_\varsigma^j}$, where the equality
$$
q_{\varsigma}^i = \Ex^{\hat{\Px}}\left[L_{\varsigma} \idc_{X_{\varsigma} = e_i} \big| \mathcal{G}_{\varsigma}^I \right]
$$
is true by the optional projection property, see \cite{Rogers}.}
Define the two dimensional (observed) log-price process {$Y_t = (\log(S_t), {\log(P_t)})^{\top}$.} As $q_\varsigma^i = \Ex^{\hat{\Px}} \left[L_{\varsigma} 1_{X_{\varsigma}=e_i} | \mathcal{G}_{\varsigma}^I \right]$, and using that $L_{\varsigma}>0$, we can choose a modification
$g(Y,H,X_{\varsigma})$ of $\Ex^{{\hat\Px}}\left[L_{\varsigma} | \mathcal{G}_{\varsigma}^I, X_{\varsigma} \right]$ such that {$g>0$}, and, for each {$e_{i}$}, $g(Y,H,{e_{i}})$ is $G_{\varsigma}^I$-measurable.
By the tower property
\begin{equation*}
q_{\varsigma}^i = \Ex^{\hat{\Px}}[g(Y,H,X_{\varsigma}) \idc_{X_{\varsigma}=e_i} | \mathcal{G}_{\varsigma}^I] = g(Y,H,e_i) \hat{\Px}(X_{\varsigma}=e_i|\mathcal{G}_{\varsigma}^I) = g(Y,H,e_i) \Px(X_t=e_i)\big|_{t=\varsigma},
\end{equation*}
where the first equality follows because $\varsigma$ is $\mathcal{G}^I_{\varsigma}$-measurable and the {last two equalities} because $X$ is independent of {$\mathcal{G}^{I}$} under $\hat{\Px}$. As $\Px(X_t=e_i) >0$ and $g>0$, we get that {$q_{\varsigma}^{i}>0$ a.s, which contradicts that $\Px(p_{\varsigma}^{i}=0)>0$}. This proves the first statement in the Lemma. Next, we notice that
\begin{eqnarray*}
\Px\left(p^i_t = 0 ,\, \textrm{for\ some }\ t\in [0,T) \right) = 1- \Px\left(p^i_t > 0 ,\, \textrm{for\ all }\ t\in [0,T) \right) = 0,
\end{eqnarray*}
where the last equality follows from the first statement. This immediately yields the second statement.\endproof

\medskip

\noindent{\textbf{Proof of Eq.~(\ref{eq:finHJB0})}}
\proof
\smallskip
\noindent
Let us first analyze the first term in the sup of Eq.~\eqref{eq:HJBrewr}, i.e. ${\CadBlue \beta_{\gamma}^{\top}} \nabla_{\tilde{p}} w$. {\CadBlue For brevity, we use
$\beta_{\varpi} := \beta_{\varpi}(t,\tilde{p},0)$.}
By definition of {${\CadBlue \beta_{\gamma}}$}, and using the maximizer {$\pi_t:=\pi^{*}_{t}$} in~\eqref{eq:pimaximizer}, we have
\begin{eqnarray}
\nonumber {{\CadBlue \beta^{\top}_{\gamma}}={\CadBlue \beta_{\varpi}^{\top}} + \gamma {\pi^{\top}} \Sigma_Y \bar{\kappa}^{\top}} &=& {\CadBlue \beta_{\varpi}}^{\top}  + \frac{\gamma}{1-\gamma} \left(\Sigma_Y \bar{\kappa}^{\top} (\nabla_{\tilde{p}} \wpre)^{\top}  -  \Upsilon \right)^{\top} (\Sigma_Y^{\top} \Sigma_Y)^{-1} \Sigma_Y \bar{\kappa}^{\top} \\
 &=& {\CadBlue \beta_{\varpi}^{\top}} + \frac{\gamma}{1-\gamma} (\nabla_{\tilde{p}} \wpre) \bar{\kappa} \Sigma_Y^{\top} (\Sigma_Y^{\top} \Sigma_Y)^{-1} \Sigma_Y \bar{\kappa}^{\top} - \frac{\gamma}{1-\gamma} \Upsilon^{\top} (\Sigma_Y^{\top} \Sigma_Y)^{-1} \Sigma_Y \bar{\kappa}^{\top}
\label{eq:term1}
 \end{eqnarray}
Further, again using the expression for {$\pi=\pi^{*}$}, the second term in the sup is equal to
\begin{eqnarray}
\nonumber {-}\gamma \pi^{\top} \Upsilon &=& {-}\frac{\gamma}{1-\gamma} \left(- \Upsilon + \Sigma_Y \bar{\kappa}^{\top} {(\nabla_{\tilde{p}} \wpre)^{\top}} \right)^{\top} (\Sigma_Y^{\top} \Sigma_Y)^{-1} \Upsilon \\
                             &=& \frac{\gamma}{1-\gamma} \Upsilon^{\top} (\Sigma_Y^{\top} \Sigma_Y)^{-1} \Upsilon\, {-}\, \frac{\gamma}{1-\gamma} {(\nabla_{\tilde{p}} \wpre)} \bar{\kappa}  \Sigma_Y^{\top} (\Sigma_Y^{\top} \Sigma_Y)^{-1} {\Upsilon.}
\label{eq:term2}
\end{eqnarray}
The third term in the sup may be simplified as
\begin{align}
\nonumber &-\frac{1}{2} \frac{\gamma}{1-\gamma} \left( - \Upsilon + \Sigma_Y \bar{\kappa}^{\top} {(\nabla_{\tilde{p}} \wpre)^{\top}} \right)^{\top} (\Sigma_Y^{\top} \Sigma_Y)^{-1} (- \Upsilon + \Sigma_Y \bar{\kappa}^{\top} {(\nabla_{\tilde{p}} \wpre)^{\top}}) = \\
\nonumber &\quad-\frac{1}{2} \frac{\gamma}{1-\gamma}\Upsilon^{\top} (\Sigma_Y^{\top} \Sigma_Y)^{-1} \Upsilon  + \frac{1}{2} \frac{\gamma}{1-\gamma}\Upsilon^{\top}(\Sigma_Y^{\top} \Sigma_Y)^{-1}\Sigma_Y \bar{\kappa}^{\top} {(\nabla_{\tilde{p}} \wpre)^{\top}}   \\
&\quad+ \frac{1}{2} \frac{\gamma}{1-\gamma} {(\nabla_{\tilde{p}} \wpre)} \bar{\kappa} \Sigma_Y^{\top} (\Sigma_Y^{\top} \Sigma_Y)^{-1} \Upsilon - \frac{1}{2} \frac{\gamma}{1-\gamma} {(\nabla_{\tilde{p}} \wpre)} \bar{\kappa} \Sigma_Y^{\top}(\Sigma_Y^{\top} \Sigma_Y)^{-1} \Sigma_Y \bar{\kappa}^{\top} {(\nabla_{\tilde{p}} \bar{w})^{\top}.}
\label{eq:term3}
\end{align}
Using Eq.~\eqref{eq:term1},~\eqref{eq:term2}, and~\eqref{eq:term3}, we obtain that
\begin{eqnarray}
\nonumber & & \sup_{\pi} \bigg \{{\CadBlue \beta_{\gamma}^{\top}} {(\nabla_{\tilde{p}} \wpre)^{\top}} - \gamma \pi_t^{\top} \Upsilon - \frac{1}{2} \gamma (1-\gamma) \pi_t^{\top} \Sigma^{\top}_Y \Sigma_Y \pi_t \bigg \}  = \\
\nonumber & & {\CadBlue \beta_{\varpi}^{\top}} (\nabla_{\tilde{p}} \wpre)^{\top} + \frac{1}{2} \frac{\gamma}{1-\gamma} {(\nabla_{\tilde{p}} \wpre)} \bar{\kappa} \Sigma_Y^{\top} (\Sigma_Y^{\top} \Sigma_Y)^{-1} \Sigma_Y \bar{\kappa}^{\top} {(\nabla_{\tilde{p}} \wpre)^{\top}}  + \frac{1}{2} \frac{\gamma}{1-\gamma} \Upsilon^{\top} (\Sigma_Y^{\top} \Sigma_Y)^{-1} \Upsilon{- \frac{\gamma}{1-\gamma} (\nabla_{\tilde{p}} \wpre)} \bar{\kappa} \Sigma_Y^{-1} \Upsilon,
\end{eqnarray}
and therefore, after re-arrangement, we obtain Eq.~\eqref{eq:finHJB0}.
\endproof
\medskip

\noindent{\textbf{Proof of Theorem \ref{MntPstDftVal}.}}
\proof
\smallskip
\noindent
In order to ease the notational burden, throughout the proof we will write $\tilde{p}$ for $\tilde{p}^{\circ}$, $\tilde{p}_{s}$ for $\tilde{p}^{t}_{s}$, $\pi$ for $\pi^{t}$, $\tilde{\Px}$ for ${\tilde{\Px}^{t}}$, $\Px$ for ${\Px^{t}}$, $\tilde{W}$ for $\tilde{W}^{t}$, {\Blue $X$ for $X^{t}$,} and $\mathcal{G}_{s}^{I}$ for $\mathcal{G}_{s}^{t,I}$.
{Let us first remark that
\begin{equation}\label{Eq:AACTldp0}
	{\Px}\left(\tilde{p}_{s}\in \tilde{\Delta}_{N-1},\,t\leq s\leq T\right)=1.
\end{equation}
Indeed, set $\tilde{p}^{N}_{s}=1-\sum_{j=1}^{N-1}\tilde{p}^{j}_{s}$ and recall from Remark \ref{Exist2} that the process $\tilde{p}_{s}^{i}$ is given by
\begin{equation}\label{Eq:ArtRptildep}
   {\tilde{p}_s^i := {{\Px}}\left(X_{s} = e_i \big| \mathcal{G}_{s}^I\right), 
    \qquad (t\leq{}s\leq{}T,\; i=1,\dots,N).}
\end{equation}
Therefore, using Lemma \ref{Rem:TryPrvRmnPs}, we deduce that all the $\tilde{p}^{i}$, with $i=1,\dots,N$, remain positive in $[t,T]$, a.s., and, hence, (\ref{Eq:AACTldp0}) is satisfied.


Next, we prove that the feedback trading strategy $\widetilde{\pi}_{s}:=(\tilde{\pi}^{S}_{s},\tilde{\pi}^{P}_{s})^{\top}$, $\tilde{\pi}^{P}_{s} := 0$, is admissible; i.e.,}
\begin{equation}\label{NecCndEMT}
    \Ex^{\Px}\left[\exp\left(\frac{\sigma^{2}\gamma^{2}}{2}\int_{t}^{T}\left(\widetilde{\pi}^{S}\left(s,\tilde{p}_{s}\right)\right)^{2}ds\right)\right]<\infty.
\end{equation}
We have that (\ref{NecCndEMT}) follows from {Eq.~(\ref{Eq:AACTldp0}) and} the fact that $(\widetilde{\pi}^S(s,\tilde{p}))^{2}$ is uniformly bounded on $[0,T]\times \widetilde{\Delta}_{N-1}$. To see the latter property, note that 
\[
	\sup_{(s,\tilde{p})\in[0,T]\times \widetilde{\Delta}_{N-1}}\left(\widetilde{\pi}^S(s,\tilde{p})\right)^{2}\leq
	\frac{2}{\sigma^{4}(1-\gamma)^{2}}\sup_{(s,\tilde{p})\in[0,T]\times \widetilde{\Delta}_{N-1}}\left(\tilde{\mu}(\tilde{p})-r\right)^{2}
	+\frac{2}{\sigma^{{\CRed 2}}(1-\gamma)^{2}}
    \sup_{(s,\tilde{p})\in[0,T]\times \widetilde{\Delta}_{N-1}}\left({\nabla_{\tilde{p}}  \wpost(s,\tilde{p})\underline{\kappa}(\tilde{p})}  \right)^{2}.
\]
The first term on the right hand side is clearly bounded since $|\tilde{\mu}(\tilde{p})|\leq \max_{i}|\mu_{i}|$ for any $\tilde{p}\in \widetilde{\Delta}_{N-1}$. For the second term, using the definition of $\underline{\kappa}$ given in Eq.~(\ref{Eq:DfnUnderAlpha}), we have
\begin{equation}\label{BndNdLo}
    \sup_{(s,\tilde{p})\in[0,T]\times \widetilde{\Delta}_{N-1}}\left({\nabla_{\tilde{p}}  \wpost(s,\tilde{p})\underline{\kappa}(\tilde{p})}  \right)^{2}=\frac{1}{\sigma^{2}}\sup_{(s,\tilde{p})\in[0,T]\times \widetilde{\Delta}_{N-1}}\left(\sum_{j=1}^{N-1}\partial_{\tilde{p}^{j}}\wpost(s,\tilde{p})\tilde{p}^{j}\left(\mu_{j}-\sum_{i=1}^{N}\mu_{i}\tilde{p}^{i}\right)\right)^{2},
\end{equation}
where $\tilde{p}^{N}:=1-\sum_{i=1}^{N-1}\tilde{p}^{i}$. The last expression  is bounded since each $\partial_{\tilde{p}^{j}}\underline{w}(s,\tilde{p})$ is bounded on $[0,T]\times \widetilde{\Delta}_{N-1}$ in view of Lemma \ref{lem:classpost} and Remark \ref{rem:sol}, {\CRed where it is shown $\mathcal{C}_P^{2,\alpha}$ regularity for $\underline{w}(s,\tilde{p})$, hence bounded first and second order space derivatives on $\widetilde{\Delta}_{N-1}$.}

Now, fix an arbitrary feedback control $\pi_{s}^{S}:=\pi^{S}(s,\tilde{p}_{s})$ such that $(\pi^{S},\pi^{P})\in{\mathcal{A}}(t,T;\tilde{p},1)$, {where $\pi^{P}_{s}\equiv 0$ and ${\mathcal{A}}(t,T;\tilde{p},1)$ is defined as in Definition~\ref{def:admiss}},
and define the process
\begin{equation*}
    M_{s}^{\pi^{S}} :=  e^{-\gamma \int_t^s \underline{\eta}(u,\tilde{p}_u,\pi^{S}_u) du} e^{\underline{w}(s,\tilde{p}_s)}, \qquad (t\leq{}s\leq{}T),
\end{equation*}
where
\begin{align}\label{Eq:Dfnundereta}
    \underline{\eta}(u,\tilde{p},\pi^{S})=
    {\eta}(u,\tilde{p},(\pi^S,0)^{\top})=-r + {\pi^{S}}(r-\tilde{\mu}(\tilde{p})) +   \frac{1-\gamma}{2} \sigma^{2}\left(\pi^{S}\right)^{2}.
\end{align}
In what follows, we write for simplicity $M^{\pi}$ for $M^{\pi^{S}}$ and {$\pi$ for $\pi^{S}$}. Note that the process $\{M^{\pi}_{s}\}_{t\leq{}s\leq{}T}$ is uniformly bounded. Indeed, (\ref{Eq:Dfnundereta}) is convex in {$\pi^{S}$} and by minimizing it over {$\pi^{S}$}, it follows that, {for any $\tilde{p}\in \tilde{\Delta}_{N-1}$,}
\[
    - \underline{\eta}(t,\tilde{p},\pi)\leq  r +\frac{(\tilde{\mu}(\tilde{p})-r)^{2}}{2(1-\gamma)\sigma^{2}}
\leq  r +\frac{(\max_{i}\mu_{i}^{2}+r^{2})}{(1-\gamma)\sigma^{2}} < {\infty}.
\]
Therefore, since $\underline{w}\in C([0,T]\times \tilde{\Delta}_{N-1})$, there exists a constant $K<\infty$ for which
\begin{equation}\label{Eq:UBFM}
     M_{s}^{\pi}=e^{-\gamma \int_t^s \underline{\eta}(u,\tilde{p}_u,\pi_u) du} e^{\underline{w}(s,\tilde{p}_s)}\leq K e^{\gamma \|\underline{\eta}\|_{\infty} (T-t)}=:A<\infty.
\end{equation}
We prove the result through the following steps:

\smallskip
\noindent {\bf (i)}  Define the {process} $\calY_{s}=e^{\underline{w}(s,\tilde{p}_{s})}$. By It\^o's formula and the generator formula (\ref{Eq:SemiMrtDcm1}) with $f(s,\tilde{p})=e^{\underline{w}(s,\tilde{p})}$,
\begin{align*}
      M_{s}^{\pi} &= M_{t}^{\pi}+\int_{t}^{s}e^{-\gamma \int_t^u \underline{\eta}(r,\tilde{p}_r,\pi_r) dr} d \calY_{u}
     -\gamma\int_{t}^{s} \underline{\eta}(u,\tilde{p}_u,\pi_u) e^{-\gamma \int_t^u \underline{\eta}(r,\tilde{p}_r,\pi_r) dr}\calY_{u}du\\
     &=M_{t}^{\pi}+\int_{t}^{s} M_{u}^{\pi}\left({\frac{\pa \wpost}{\pa u}} + \frac{1}{2} \text{tr}(\underline{\kappa} \underline{\kappa}^{\top} D^2  \wpost) + \frac{1}{2}{(\nabla_{\tilde{p}} \wpost) \underline{\kappa} \underline{\kappa}^{\top} (\nabla_{\tilde{p}} \wpost)^{\top}} +{(\nabla_{\tilde{p}} \wpost)\underline{\beta}_{\gamma}}
     -\gamma\underline{\eta}  \right)du
     +\int_{t}^{s}M_{u}^{\pi} \nabla_{\tilde{p}} \wpost \,\underline{\kappa}\,d {\tilde{W}}^{(1)}_{u}.
\end{align*}
Using the expression of $\underline{\eta}$ in (\ref{Eq:Dfnundereta}) and some rearrangement, we may write $M^{\pi}$ as
\[
     M_{s}^{\pi} =M_{t}^{\pi}+\int_{t}^{s} M_{u}^{\pi}R(u,\tilde{p}_{u},\pi_{u})du
     +\int_{t}^{s}M_{u}^{\pi} \nabla_{\tilde{p}} \wpost \,\underline{\kappa}\,d {\tilde{W}}^{(1)}_{u}
\]
with
\begin{align}
R(u,\tilde{p},\pi)  = {\underline{w}}_u + \frac{1}{2} \text{tr}(\underline{\kappa} \underline{\kappa}^{\top} D^2  \wpost) + \frac{1}{2} {(\nabla_{\tilde{p}} \wpost) \underline{\kappa} \underline{\kappa}^{\top} (\nabla_{\tilde{p}} \wpost)^{\top}} + \gamma r  +{(\nabla_{\tilde{p}} w)\underline{\beta}_{\gamma}} - \gamma \pi (r- \tilde{\mu}(\tilde{p})) - \frac{{\sigma^2}}{2} \gamma (1-\gamma) \pi^2.
\label{eq:Requation}
\end{align}
Clearly, $R(u,\tilde{p},\pi)$ is a concave function in $\pi$ since  $R_{\pi\pi}=-\sigma^{2}\gamma(1-\gamma) < 0$.
If we maximize $R(u,\tilde{p},\pi)$ as a function of $\pi$ for each $(u,\tilde{p})$, we find that the optimum is given by (\ref{eq:optpi}). Upon substituting (\ref{eq:optpi}) into (\ref{eq:Requation}), we get that
\begin{align*}
\nonumber R(u,\tilde{p},\pi)&\leq R(u,\tilde{p},{\widetilde\pi^{S}(u,\tilde{p})})=\wpost_{u} + \frac{1}{2} \text{tr}(\underline{\kappa} \underline{\kappa}^{\top} D^2 \wpost) + \frac{1}{2(1-\gamma)} {(\nabla_{\tilde{p}} \wpost) \underline{\kappa}\, \underline{\kappa}^{\top} (\nabla_{\tilde{p}} \wpost)^{\top}}+ {(\nabla_{\tilde{p}} \wpost)\underline{\Phi}} + \underline{\Psi} = 0,
\end{align*}
where the last equality follows from Eq.~(\ref{Eq:PstDfltDP}). Therefore,
we get the inequality
\begin{align*}
    {\Ex^{\tilde{\Px}}}\left[M^{\pi}_{T}\right]&\leq M_{t}^{\pi} + {\Ex^{\tilde{\Px}}}\left[\int_{t}^{T}
    M_{u}^{\pi}  \nabla_{\tilde{p}} \wpost \,\underline{\kappa}\,d {\tilde{W}}^{(1)}_{u}
    \right],
\end{align*}
with equality if $\pi=\widetilde{\pi}^{S}$.
From (\ref{Eq:DfnUnderAlpha}), it is easy to check that $\sup_{\tilde{p}\in \tilde{\Delta}_{N-1}}\|\underline{\kappa}(\tilde{p})\|^{2}\leq 2\max_{i}\{\mu_{i}\}/\sigma$. Then, since the partial derivatives $\partial_{\tilde{p}^{j}}\underline{w}(s,\tilde{p})$ are uniformly bounded on $[0,T]\times \tilde{\Delta}_{N-1}$ (see also the argument after Eq.~(\ref{BndNdLo})), (\ref{Eq:UBFM}) implies that
\[
    {\sup_{t\leq u\leq T} \left|M_{u}^{\pi}\nabla_{\tilde{p}} \wpost \,\underline{\kappa}\right|^{2}\leq A \sup_{t\leq u\leq T}\|\underline{\kappa}(\tilde{p}_{u})\|^{2} \sup_{t\leq u\leq T}\|\nabla_{\tilde{p}} \wpost(u,\tilde{p}_{u})\|^{2} \leq B},
\]
for {some} {non-random} constant $B<\infty$. We conclude that
\begin{equation}\label{Eq:KInGP}
     \Ex^{\tilde{\Px}} \left[M^{\pi}_{T}\right]\leq M_{t}^{\pi}=e^{\underline{w}(t,\tilde{p}_{t})}=e^{\underline{w}(t,\tilde{p})},
\end{equation}
with equality if $\pi=\widetilde{\pi}^{S}$.


\smallskip
\noindent
{\bf (ii)}
For simplicity, let us write $\widetilde{\pi}_{s}:=\widetilde{\pi}^{S}(s,\tilde{p}_{s})$.
First, note that
from
the fact that we have equality in (\ref{Eq:KInGP}) when $\pi=\tilde{\pi}$,
\begin{equation}\label{FIdN2}
    {e^{\underline{w}(t,\tilde{p})}=
    \Ex^{\tilde{\Px}}\left[M^{\tilde\pi}_{T}\right]=\Ex^{\tilde{\Px}} \left[ e^{-\gamma \int_t^T \underline{\eta}(u,\tilde{p}_u,\tilde\pi_u) du} e^{\underline{w}(T,\tilde{p}_T)}\right]
    =\Ex^{\tilde{\Px}}\left[ e^{-\gamma \int_t^T \underline{\eta}(u,\tilde{p}_u,\tilde\pi_u) du} \right]}.
\end{equation}
Similarly,
for every feedback control $\pi_{s}=\pi(s,\tilde{p}_{s})$ such that $(\pi,0)\in\mathcal{A}(t,T;\tilde{p},1)$,
\begin{align*}
    {{\Ex^{\tilde{\Px}}\left[ e^{-\gamma \int_t^T \underline{\eta}(u,\tilde{p}_u,\pi_u) du} \right]=\Ex^{\tilde{\Px}} \left[M^{\pi}_{T}\right]
     \leq e^{\underline{w}(t,\tilde{p})}=\Ex^{{\tilde{\Px}}}\left[ e^{-\gamma \int_t^T \underline{\eta}(u,\tilde{p}_u,\tilde\pi_u) du} \right],}}
\end{align*}
where the {inequality in the previous equation} {comes} from (\ref{Eq:KInGP}) and the last equality {therein} follows from (\ref{FIdN2}). The previous relationships show the optimality of $\tilde{\pi}$ and prove the assertions (1) and (2).
\endproof

\medskip


\noindent{\textbf{Proof of Theorem \ref{t3}.}}
\proof

\smallskip
\noindent
For brevity, define the operator
  $$\mathcal{B}=\p_{t}+ \frac{1}{2} \text{tr}\left(\bar{\kappa} \bar{\kappa}^{\top} D^2\right)+ \nabla_{\tilde{p}} \bar{\Phi}$$
and denote by
  $$H(t,\tilde{p},u)=-\tilde{h}({\tilde{p}}) e^{\wpost \left(t, \frac{1}{\tilde{h}({\tilde{p}})} {\tilde{p}} \cdot h^{\pr}\right)}\frac{u^{\gamma}}{1-\gamma}, \qquad u\in\R_{+},$$
the non-linear term of the PDE \eqref{eq:finHJB2}. Notice that since $\tilde{h} > 0$ by
construction, then $H\le 0$. Moreover, $u\mapsto H(t,{\tilde{p}},u)$ is smooth and Lipschitz
continuous on $[\bar c,+\infty[$ for any $\bar c>0$, uniformly w.r.t. $(t,{\tilde{p}})$.
We set
\begin{equation*}
 \bar{\psi}_0(t,{\tilde{p}})= e^{c(T-t)},\qquad t\in[0,T],
\end{equation*}
where $c$ is a suitably large positive constant such that
\begin{equation}\label{and6}
 c u+H(t,\tilde{p},u)-\frac{{\bar\Psi(t,\tilde{p})}}{1-\gamma}u\ge 0,\qquad \text{ for any } (t,\tilde{p})\in (0,T)\times\tilde\Delta_{N-1} \text{ and } u\ge1.
\end{equation}
Then we define recursively the sequence $(\bar{\psi}_j)_{j \in \N}$ by
\begin{equation}\label{e2}
\begin{cases}
  \left(\mathcal{B}+\frac{\bar{\Psi}}{1-\gamma}\right) \bar{\psi}_j-\l\bar{\psi}_j=H(\cdot,\cdot,\bar{\psi}_{j-1})-\l \,\bar{\psi}_{j-1}, 
  \\
  \bar{\psi}_{j}(T,\cdot) =1,
 \end{cases}
\end{equation}
where $\l$ is the Lipschitz constant of $u\mapsto H(\cdot,\cdot,u)$ on $[\bar c,+\infty[$ and
$\bar c$ is the strictly positive constant defined as
\begin{equation}\label{and3}
 \bar c=e^{-\frac{T}{1-\gamma}\left\|\bar{\Psi}\right\|_{\infty}}.
\end{equation}
{\CRed
Let us recall that the {\it linear problem} (\ref{e2}) has a classical solution in $C^{2,\a}_P$ 
whose existence can be proven as in Lemma \ref{lem:classpost}, see also the following Remark~\ref{rem:sol}.
}
Next we prove by induction that
\begin{itemize}
  \item[i)] $(\bar{\psi}_{j})$ is a decreasing sequence, that is
\begin{equation}\label{and1}
  \bar{\psi}_{j+1}\le \bar{\psi}_{j},\qquad j\ge 0;
\end{equation}
  \item[ii)] $(\bar{\psi}_{j})$ is uniformly strictly positive and in particular
\begin{equation}\label{and2}
  \bar{\psi}_{j+1}\ge \bar c,\qquad j\ge 0,
\end{equation}
with $\bar c$ as in \eqref{and3}.
\end{itemize}
First, we observe that
\begin{equation}\label{and5}
  \bar{\psi}_{0}\ge 1,\qquad \text{and}\qquad \left(\mathcal{B}+\frac{\bar{\Psi}}{1-\gamma}\right)
  \bar{\psi}_0=\left(-c+\frac{\bar{\Psi}}{1-\gamma}\right)\bar{\psi}_0.
\end{equation}
Next we prove \eqref{and1}-\eqref{and2} for $j=0$: by \eqref{and5} and \eqref{and6} we have
\begin{equation}
  \begin{cases}
  \left(\mathcal{B}+\frac{\bar{\Psi}}{1-\gamma}-\l\right)\left(\bar{\psi}_1-\bar{\psi}_0\right)=
  H(\cdot,\cdot,\bar{\psi}_0)+c \bar{\psi}_0-\frac{\bar{\Psi}}{1-\gamma}\bar{\psi}_0\ge 0,
  \\
  \left(\bar{\psi}_1-\bar{\psi}_0\right)(T,{\tilde{p}}) =0.
 \end{cases}
\label{eq:Beq}
\end{equation}
where the inequality above follows from the fact that $c$ is chosen as in~\eqref{and6}, and
$\bar{\psi}_0 \geq 1$ as observed in Eq.~\eqref{and5}. Since the process $\tilde{p}_t$ never
reaches the boundary of the simplex by Lemma~\ref{Rem:TryPrvRmnPs}, it follows from the
Feynman-Kac representation theorem (or, equivalently, the maximum principle) that $\bar{\psi}_1\le
\bar{\psi}_0$: indeed we have
\begin{eqnarray}
\nonumber (\bar{\psi}_1-\bar{\psi}_0)(t,\tilde{p}) = {\CRed \Ex^{\tilde{\Px}}} \left[-\int_t^T e^{-\int_t^s \left(\frac{1}{1-\gamma}\bar{\Psi}(r,\tilde{p}_r) - \lambda \right) dr}
\left(H(s,\tilde{p}_s,\bar{\psi}_0) + c \bar{\psi}_0(s,\tilde{p}_s) - \frac{1}{1-\gamma} \bar{\Psi}(s,\tilde{p}_s) \bar{\psi}_0(s,\tilde{p}_s) \right) \bigg|
\tilde{p}_t=\tilde{p} \right] \leq 0 \\
\label{eq:Fkdiff}
\end{eqnarray}
where the last inequality follows directly from the inequality in~\eqref{eq:Beq}. This proves~\eqref{and1} when $j=0$. Using the recursive definition~\eqref{e2}, along with the fact that $H\le 0$, $\l>0$ and inequality~\eqref{eq:Fkdiff}, we obtain
\begin{equation}\label{and8}
  \left(\mathcal{B}+\frac{\bar{\Psi}}{1-\gamma}\right)\bar{\psi}_1=H(\cdot,\cdot,\bar{\psi}_0)+\l \left(\bar{\psi}_1-\bar{\psi}_0\right)\le
  0.
\end{equation}
Then \eqref{and2} with $j=0$ follows again from the Feynman-Kac theorem: indeed by \eqref{and8} we have
\begin{eqnarray}\label{and10}
 \nonumber \bar\psi_{1}(t,{\tilde{p}}) &=& {\Ex^{\tilde{\Px}}}\left[-\int_t^T e^{-\int_t^s \frac{1}{1-\gamma}\bar{\Psi}(r,\tilde{p}_r) dr} \left(H(s,\tilde{p}_s,\bar{\psi}_0)+\l \left(\bar{\psi}_1-\bar{\psi}_0\right)(s,\tilde{p}_s)\right) \bigg| \tilde{p}_t = \tilde{p} \right]  + {\CRed \Ex^{\tilde{\Px}}}\left[e^{\frac{1}{1-\gamma}\int_{t}^{T}\bar{\Psi}(s,\tilde{p}_s)ds} \bigg| \tilde{p}_t = \tilde{p} \right] \\
  & & \ge e^{-\frac{T}{1-\gamma}\left\|\bar{\Psi}\right\|_{\infty}},
\end{eqnarray}
where the last inequality follows from the positivity of the first expectation above guaranteed by~\eqref{and8}.

Next we assume the inductive hypothesis to hold,
\begin{equation}\label{and9}
 \bar c\le\bar\psi_{j}\le \bar\psi_{j-1}
\end{equation}
and prove \eqref{and1}-\eqref{and2}. Recalling that $\l$ is the Lipschitz constant of $u\mapsto
H(\cdot,\cdot,u)$ on $[\bar c,+\infty[$, by \eqref{and9} we have
  $$
  \begin{cases}
  \left(\mathcal{B}+\frac{\bar{\Psi}}{1-\gamma}-\l\right)\left(\bar{\psi}_{j+1}-\bar{\psi}_j\right)=
  H(\cdot,\cdot,\bar{\psi}_j)-H(\cdot,\cdot,\bar{\psi}_{j-1})-\l\left(\bar{\psi}_j-\bar{\psi}_{j-1}\right)\ge 0,
  \\
  \left(\bar{\psi}_{j+1}-\bar{\psi}_{j}\right)(T,{\tilde{p}}) =0.
 \end{cases}
 $$
Thus \eqref{and1} follows from the Feynman-Kac theorem using the same procedure as in~\eqref{eq:Beq} and~\eqref{eq:Fkdiff}. Moreover we have
\begin{equation}
  \left(\mathcal{B}+\frac{\bar{\Psi}}{1-\gamma}\right)\bar{\psi}_{j+1}=H(\cdot,\cdot,\bar{\psi}_j)+\l \left(\bar{\psi}_{j+1}-\bar{\psi}_j\right)\le 0,
\end{equation}
where the inequality above follows by \eqref{and1} and using that $H\le 0$ and $\l>0$. Then, as in \eqref{and10}, we have that
\eqref{and2} follows from the Feynman-Kac theorem.

In conclusion, for $j\in\N$, we have
\begin{equation}\label{e1}
 \bar c \leq \bar{\psi}_{j+1} \leq \bar{\psi}_j \leq \bar{\psi}_0.
\end{equation}
Now the thesis follows by proceeding as in the proof of Theorem 3.3 in
\cite{DifrancescoPascucciPolidoro2007}. Indeed let us denote by $\bar{\psi}$ the pointwise limit
of $(\bar{\psi}_{j})$ as $j\to+\infty$: since $\bar{\psi}_{j}$ is a solution of \eqref{e2} and by
the uniform estimate \eqref{e1}, we can apply standard a priori Morrey-Sobolev-type estimates
(see, Theorems 2.1 and 2.2 in \cite{DifrancescoPascucciPolidoro2007}) to conclude that, for any
$\a\in\,]0,1[$, $\|\bar{\psi}_{j}\|_{C_P^{1,\a}((0,T)\times\tilde\Delta_{N-1})}$ is bounded by a
constant only dependent on $\mathcal{B}$, $\a$ and $\l$. Hence by the classical Schauder interior
estimate (see, for instance, Theorem 2.3 in \cite{DifrancescoPascucciPolidoro2007}), we deduce
that
$\|\bar{\psi}_{j}\|_{C_P^{2,\a}((0,T)\times\tilde\Delta_{N-1}))}$ is bounded uniformly in $j\in\N$. 
It follows that $(\bar{\psi}_j)_{j\in\N}$ admits a subsequence (denoted by itself) that converges
in $C^{2,\a}$. Thus passing at limit in \eqref{e2} as $j\to\infty$, we have
 $$\left(\mathcal{B}+\frac{\bar{\Psi}}{1-\gamma}\right) \bar{\psi} = H(\cdot,\cdot,\bar{\psi}), \qquad\text{ in } (0,T)\times\tilde\Delta_{N-1},$$
and $\bar{\psi}(T,\cdot)=1$.

Finally, in order to prove that $\bar{\psi}\in C\left((0,T]\times\tilde\Delta_{N-1}\right)$, we
use the standard argument of barrier functions. We recall that $w$ is a barrier function for the
operator $\left(\mathcal{B}+\frac{\bar{\Psi}}{1-\gamma}\right)$, on the domain
$(0,T]\times\tilde\Delta_{N-1}$, at the point $(T,\bar{p})$ if $w\in C^{2} \left(V \cap
\left((0,T]\times\tilde\Delta_{N-1}\right)\right)$ where $V$ is a neighborhood of $(T,\bar{p})$
and we have
\begin{itemize}
  \item[i)] $\left(\mathcal{B}+\frac{\bar{\Psi}}{1-\gamma}\right) w \leq -1$ in $V \cap \left((0,T)\times\tilde\Delta_{N-1}\right)$;
  \item[ii)] $w > 0$ in $V \cap \left((0,T)\times\tilde\Delta_{N-1}\right) \setminus \{(T,\bar{p})\}$ and $w(T,\bar{p}) = 0$.
\end{itemize}
Next we fix $\bar p\in\tilde\Delta_{N-1}$: following \cite{Friedman} Chap.3 Sec.4, it is not
difficult to check that
    $$w(t,\tilde p)=\left(|\tilde p-\bar x|^{2}+c_{1}(T-t)\right)e^{c_{2} (T-t)},$$
is a barrier at $(T,\bar{p})$ provided that $c_{1},c_{2}$ are sufficiently large. Then we set
  $$v^{\pm}(t,\tilde p)=1\pm k w(t,\tilde p)$$
where $k$ is a suitably large positive constant, independent of $j$, such that
  $$\left(\mathcal{B}+\frac{\bar{\Psi}}{1-\gamma}\right)(\bar\psi_{j}-v^{+})\ge H(\cdot,\cdot,\bar\psi_{j-1})-\l\,(\bar\psi_{j-1} - \bar\psi_{j})-\frac{\bar{\Psi}}{1-\gamma}-k\left(\mathcal{B}+\frac{\bar{\Psi}}{1-\gamma}\right)w\ge 0,$$
and $\bar\psi_{j}\le v^{+}$ on $\p\left(V \cap \left((0,T)\times\tilde\Delta_{N-1}\right)\right)$. 
The maximum principle yields $\bar\psi_{j}\le v^{+}$ on $V \cap
\left((0,T)\times\tilde\Delta_{N-1}\right)$;
analogously we have $\bar\psi_{j}\ge v^{-}$ on $V \cap
\left((0,T)\times\tilde\Delta_{N-1}\right)$,
and letting $j\to\infty$ we get
  $$1-k w(t,\tilde p)\le \bar\psi(t,\tilde p)\le 1+k w(t,\tilde p), \qquad (t,\tilde p)\in V \cap
\left((0,T)\times\tilde\Delta_{N-1}\right).$$ Therefore we deduce that
  $$\lim_{(t,\tilde p)\to(T,\bar p)} \bar\psi(t,\tilde p)=1$$
which  concludes the proof.
\endproof

\medskip

\noindent{\textbf{Proof of Theorem \ref{MntPreDftVal2}}}
\proof

\smallskip
\noindent
As in the proof of Theorem~\ref{MntPstDftVal}, to ease the notational burden we will write $\tilde{p}$ for $\tilde{p}^{\circ}$, $\tilde{p}_{s}$ for $\tilde{p}^{t}_{s}$, $\pi$ for $\pi^{t}$, $\tilde{W}$ for {$\tilde{W}^{t}$}, $\tilde{\Px}$ for {$\tilde{\Px}^{t}$}, $\Px$ for {$\Px^{t}$}, and $\mathcal{G}_{s}^{I}$ for {$\mathcal{G}_{s}^{t,I}$}.
Similarly to the proof of the post default verification theorem, it is easy to see that
the trading strategy
${\widetilde{\pi}_{s}:=(\widetilde{\pi}^S_{s},\widetilde{\pi}^P_{s})^{\top}}=\left(\widetilde{\pi}^{S}(s,{\tilde{p}_{s^{-}},H^{t}_{s^{-}}}), {\widetilde{\pi}^{P}(s,{\tilde{p}_{s^{-}},H_{s^{-}}})}\right)^{\top}$, as defined from equations~(\ref{eq:optpiPrD1})-(\ref{eq:optpiPrD2}), is admissible; i.e., satisfies (\ref{KCnAdm}). This essentially follows from the condition (\ref{CndFora}) and the fact that both $\underline{w}(s,\tilde{p})$ and $\bar{w}(s,\tilde{p})$ belong to $\mathcal{C}_{P}^{2\alpha}$, {\CRed hence their first and second order space derivatives are bounded on $[0,T]\times\widetilde{\Delta}_{N-1}$.} {Here, it is also useful to recall that ${\Px}\left(\tilde{p}_{s}\in \tilde{\Delta}_{N-1},\,t\leq s\leq T\right)=1$ as shown in the proof of Theorem~\ref{MntPstDftVal}.}

Next, for
a fixed feedback control $\pi_{s}:=(\pi_{s}^{S},\pi^{P}_{s}):=(\pi^{S}(s,\tilde{p}_{s^{-}},H_{s^{-}}),\pi^{P}(s,\tilde{p}_{s^{-}},H_{s^{-}}))$ such that $(\pi^{S},\pi^{P})\in\bar{\mathcal{A}}(t,T;\tilde{p},0)$,
define the process
\begin{equation}
    M_{s}^{\pi} :=  e^{-\gamma \int_{t}^{s} {\tilde{\eta}(u,\tilde{p}_u,\pi_u)} du} e^{w(s,\tilde{p}_{s},H_{s})}, \qquad (t\leq{}s\leq{}T),
\end{equation}
where $w(s,\tilde{p},z):=(1-z)\bar{w}(s,\tilde{p})+z\underline{w}(s,\tilde{p})$ and $\tilde\eta$ is defined as in Eq.~(\ref{Eq:DfnSmptldeta}). Note that $\tilde\eta$ can be written as
\begin{align*}\nonumber
    {\tilde{\eta}(t,\tilde{p},\pi)}&=-r + \pi^{S}(r-\tilde{\mu}(\tilde{p})) +   \frac{1-\gamma}{2} \sigma^{2}(\pi^S)^{2}+\pi^P\left({\BRed r- {\tilde{a}(t,\tilde{p})}} \right)+   \frac{1-\gamma}{2} \upsilon^{2}(\pi^P)^{2},
\end{align*}
{and, thus,} $-\tilde{\eta}$ is concave in {\CRed $\pi$}. {This in turn implies that there exists a {nonrandom} constant $A<\infty$ such that}
\begin{equation}\label{Eq:UBFMPreDfl}
     0<M_{s}^{\pi}\leq A<\infty,\qquad {t\leq{}s\leq{}T},
\end{equation}
{since $\underline{w}, \bar{w}\in C\left([0,T]\times \tilde{\Delta}_{N-1}\right)$.}
We prove the result through the following two steps:

\smallskip
\noindent {\bf (i)}  Define the {processes} $\calY_{s}=e^{w(s,\tilde{p},H_{s})}$ and $\mathcal{U}_{s}=e^{-\gamma \int_{t}^{s} {\tilde{\eta}(u,\tilde{p}_u,\pi_u)} du}$. By It\^o's formula, the generator formula (\ref{Eq:SemiMrtDcm1}) with $f(s,\tilde{p},z)=e^{w(s,\tilde{p},z)}$, and the same arguments as those used to derive (\ref{Eq:ApplGenexpw}),
\begin{align*}
      M_{s}^{\pi} &= M_{t}^{\pi}+\int_{t}^{s}{\calU_{u^{-}}} d \calY_{u}
     -\gamma\int_{t}^{s} {\tilde{\eta}(u,\tilde{p}_u,\pi_u)} \,{\calU_{u}}\, \calY_{u}du\\
     &=M_{t}^{\pi}+\int_{t}^{s} M_{u}^{\pi}\bigg[\frac{\pa w}{\pa u} + \frac{1}{2} \text{tr}({\kappa} {\kappa}^{\top} D^2  w) + \frac{1}{2}{(\nabla_{\tilde{p}} w) {\kappa} {\kappa}^{\top} (\nabla_{\tilde{p}} w)^{\top}} +(\nabla_{\tilde{p}} w){\beta}_{\gamma}\\
     &\qquad\qquad\qquad\qquad+(1-H_{u}) \tilde{h}(\tilde{p}_{u})\left( e^{\underline{w} \left(u,\frac{1}{\tilde{h}(\tilde{p}_{u})}\tilde{p}_{u}\cdot h^{\pr}\right)-\bar{w} \left(u,\tilde{p}_{u}\right)}-1 \right) -\gamma\tilde{\eta} \bigg]du
     +\mathcal{M}^{c}_{s}+\mathcal{M}^{d}_{s},
\end{align*}
where
\begin{equation}\label{Eq:DfnLcMart}
    \mathcal{M}^{c}_{s}:=\int_t^s M_{u}^{\pi}\nabla_{\tilde{p}}w \, {{\kappa}(u,\tilde{p}_u)} d \tilde{W}_u, \quad \mathcal{M}^{d}_{s}:=\int_{t}^{s} \calU_{u^{-}} \left( e^{\underline{w} (u,\frac{1}{\tilde{h}(\tilde{p}_{u^{-}})}\tilde{p}_{u^{-}}\cdot h^{\pr})}-e^{\bar{w} (u,\tilde{p}_{u^{-}})} \right) {d\tilde{\xi}_u}.
\end{equation}
Using the expression of ${\eta}$ in Eq.~(\ref{Eq:DfnSmptldeta}), and similar arguments to those used to derive (\ref{Eq:DfnSmptldeta}), we may write $M^{\pi}$ as
\begin{align*}
     M_{s}^{\pi} &=M_{t}^{\pi}+\int_{t}^{s} M_{u}^{\pi}R(u,\tilde{p}_{u},\pi_{u},H_{u})du+\mathcal{M}^{c}_{s}+\mathcal{M}^{d}_{s}
\end{align*}
with
\begin{align}\nonumber
R(u,\tilde{p},\pi,z)  &=  \frac{\pa w}{\pa u} + \frac{1}{2} \text{tr}({\kappa} {\kappa}^{\top} D^2 w) + \frac{1}{2} {(\nabla_{\tilde{p}} w ){\kappa} {\kappa}^{\top} ({\nabla_{\tilde{p}}} w)^{\top}} + \gamma r +
(1-z) \tilde{h}(\tilde{p})\left[ e^{\underline{w} \left(u,\frac{1}{\tilde{h}(\tilde{p})}\tilde{p}\cdot h^{\pr}\right)-\bar{w} \left(u,\tilde{p}\right)}-1 \right]  \\
& \quad  + z\left((\nabla_{\tilde{p}} \underline{w}) {\beta}_{\gamma} - \gamma \pi^S (r-\tilde{\mu}(\tilde{p}))-  \frac{\gamma(1-\gamma)}{2} \sigma^{2}(\pi^S)^{2}\right) \label{eq:RequationPreDflt}\\
&\quad +(1-z) \left((\nabla_{\tilde{p}} \bar{w}) {\beta}_{\gamma} -\gamma \pi^P ({\BRed r - {\CadBlue \tilde{a}(t,\tilde{p})}}) -   \frac{\gamma(1-\gamma)}{2} \upsilon^{2}(\pi^P)^{2}\right)
\nonumber
\end{align}
Clearly, $R(u,\tilde{p},\pi,z)$ is a concave function in $\pi$  for each $(u,\tilde{p},z)$. Furthermore, this function reaches its maximum at $\tilde{\pi}(u,\tilde{p},z)=(\tilde{\pi}^{S}(u,\tilde{p},z),\tilde{\pi}^{P}(u,\tilde{p},z))$ as defined in the statement of the theorem. Upon substituting this maximum into (\ref{eq:RequationPreDflt}) and rearrangements similar to those leading to (\ref{Eq:PstDfltDP}) and (\ref{eq:finHJB0}) (depending on whether $z=1$ or $z=0$), we get
\begin{align*}
\nonumber R(u,\tilde{p},\pi,z)&\leq R(u,\tilde{p},{\widetilde\pi(u,\tilde{p},z)},z)= 0,
\end{align*}
in light of the corresponding equations  (\ref{Eq:PstDfltDP}) and (\ref{eq:finHJB0}).
Therefore,
{we get the inequality
\begin{align*}
    \Ex^{{\tilde{\Px}}} \left[M^{\pi}_{T}\right]&\leq M_{t}^{\pi} +\Ex^{{\tilde{\Px}}} \left[\mathcal{M}^{c}_{T}+\mathcal{M}^{d}_{T}\right],
\end{align*}
with} equality if $\pi=\widetilde{\pi}$. Note that {$\Ex^{{\tilde{\Px}}}\left[\mathcal{M}^{c}_{T}\right]=0$ since it is possible to find a nonrandom constant $B$ such that
	\[
    \sup_{t\leq u\leq T} \left|M_{u}^{\pi}\nabla_{\tilde{p}}w \,\kappa(u,\tilde{p}_{u})\right|^{2}\leq A \sup_{t\leq u\leq T}\|{\kappa}(u,\tilde{p}_{u})\|^{2} \sup_{t\leq u\leq T}\|\nabla_{\tilde{p}} w(u,\tilde{p}_{u})\|^{2} \leq B,
\]
in view of (\ref{Eq:UBFMPreDfl}) and the fact that the partial derivatives of $\underline{w}$ and $\bar{w}$ are uniformly bounded on $[0,T]\times\widetilde{\Delta}_{N-1}$. {\CRed The latter statement follows from the fact that both $\underline{w}$ and $\bar{w}$ are $\mathcal{C}_P^{2,\alpha}$ on $\widetilde{\Delta}_{N-1}$ in light of Lemma \ref{lem:classpost} and Theorem \ref{t3}}.
}
To deal with $\mathcal{M}^{d}$, note that since $\underline{w},\bar{w}\in  C(([0,T]\times \tilde{\Delta}_{N-1})$ and {$\{\calU_{s}\}_{t\leq{}s\leq{}T}$} is uniformly bounded {(due to the fact that $-\tilde{\eta}$ is concave)}, we have that the integrand of the second integral in (\ref{Eq:DfnLcMart}) is uniformly bounded and, thus,  {$\Ex^{\tilde{\Px}} \left[\mathcal{M}^{d}_{T}\right]=0$} as well. The two previous facts, together with the initial conditions $H_{t}=0$ and $\tilde{p}_{t}=\tilde{p}$, lead to
\begin{equation}\label{Eq:KInGPPreDfl}
     {{\Ex^{\CRed \tilde{\Px}}\left[M^{\pi}_{T}\right]\leq M_{t}^{\pi}=e^{{w}(t,\tilde{p}_{t},H_{t})}=e^{{w}(t,\tilde{p},0)}}=e^{\bar{w}(t,\tilde{p})}},
\end{equation}
with equality if $\pi=\widetilde{\pi}$.


\smallskip
\noindent{\bf (ii)}
The rest of the proof is similar to the post default case. Concretely,
using the fact that we have equality in (\ref{Eq:KInGPPreDfl}) when $\pi=\tilde{\pi}$,
\begin{equation}\label{FIdN2PreD}
    e^{\bar{w}(t,\tilde{p})}=
    \Ex^{\tilde{\Px}} \left[M^{\tilde\pi}_{T}\right]=\Ex^{\tilde{\Px}}\left[ e^{-\gamma \int_{t}^{T} {\tilde{\eta}(u,\tilde{p}_u,\tilde\pi_u)} du} e^{w(T,\tilde{p}_{T},H_{T})}\right]
    =\Ex^{\tilde{\Px}} \left[ e^{-\gamma \int_{t}^{T} {\tilde{\eta}(u,\tilde{p}_u,\tilde\pi_u)} du}\right],
\end{equation}
since $w(T,\tilde{p}_{T},H_{T}):=(1-H_{T})\bar{w}(T,\tilde{p}_{T})+H_{T}\underline{w}(T,\tilde{p}_{T})\equiv 0$.
Also, {from (\ref{Eq:KInGPPreDfl}),
for every feedback control $\pi_{s}=\pi(s,\tilde{p}_{s},H_{s})\in\mathcal{A}(t,T;\tilde{p},0)$,
\begin{equation*}
    {\Ex^{\tilde{\Px}}\left[ e^{-\gamma \int_{t}^{T}  {\tilde{\eta}(u,\tilde{p}_u,\pi_u)} du} \right]=\Ex^{\CRed \tilde{\Px}}\left[M^{\pi}_{T}\right]
\leq{}
     M_{t}^{\pi}
     =e^{\bar{w}(t,\tilde{p})}=\Ex^{\tilde{\Px}}\left[ e^{-\gamma \int_{t}^{T}  {\tilde{\eta}(u,\tilde{p}_u,\tilde\pi_u)} du}\right],}
\end{equation*}
where 
the} last equality above follows from (\ref{FIdN2PreD}). This proves the assertions (1) and (2).
\endproof

\end{document}